\documentclass{aa}  

\usepackage{graphicx}
\usepackage{txfonts}
\newcommand{\re}{$R_e$}
\newcommand{\Lsig}{$L-\sigma$}
\newcommand{\Lsiga}{$L=L_0\sigma^{\alpha}$}

\newcommand{\Lsigb}{$L=L'_0\sigma^{\beta}$}
\newcommand{\Lsigbtempo}{$L=L'_{0}(t) \sigma^{\beta(t)}$}                   

\newcommand{\MRa}{$R_e$-$M_s$}

\newcommand{\Ie}{$I_e$}

\newcommand{\IeRe}{$I_e-R_e$}
\newcommand{\IeSig}{$I_e - \sigma$}

\newcommand{\FP}{$\log(R_e)= a \log(\sigma) + b \log(\langle I\rangle_e) + c$}

\newcommand{\kmsM}{km\, sec$^{-1}$\, Mpc$^{-1}$}

\begin{document}

   \title{The role of dry mergers in shaping the scaling relations  of galaxies}


   \author{Mauro D'Onofrio\fnmsep\thanks{Corresponding author Email: mauro.donofrio@unipd.it}
          \inst{1}
          \and
          Cesare Chiosi
          \inst{1}
          \and
          Francesco Brevi
          \inst{1}
          }

   \institute{Department of Physics and Astronomy, University of Padua
              Vicolo Osservatorio 3, I35122, Padua, Italy\\
              \email{mauro.donofrio@unipd.it, cesare.chiosi@unipd.it, francesco.brevi@studenti.unipd.it}
             }

   \date{Received November, 28th, 2024; Accepted January, 16th, 2025}
 
  \abstract
   {In the context of the hierarchical formation of galaxies,  we investigated the role played by mergers in shaping the scaling relations  of galaxies, that is the projections of their Fundamental Plane onto  the \IeRe, \IeSig, \MRa\ and \Lsig\ planes. To this aim, based on the scalar Virial Theorem, we developed a simple theory of multiple dry mergers to read both the large scale simulations and the companion scaling relations.}
   {The aim was to compare the results of this approach with the observational data and with two of the most recent and detailed numerical cosmo-hydro-dynamical simulations, that is Illustris-TNG and EAGLE (Evolution and Assembly of GaLaxies and their Environments).}
   {We derived the above scaling relations for the galaxies of the MaNGA (Mapping Nearby Galaxies at APO) and WINGS (Wide-field Imaging of Nearby Galaxy-Clusters Survey) databases and compared them with the observational data, the numerical simulations, and the results  of our simple theory of dry mergers.}
   {The multiple dry merging mechanism  is able to explain all the main characteristics of the observed scaling relations of galaxies, such as slopes, scatters, curvatures and zones of exclusion. The distribution of galaxies in these planes is continuously changing across time because of the merging activity and other physical processes, such as star formation,  quenching, energy feedback, and so forth.}
   {The simple merger theory presented here yields the correct distribution of galaxies in the main scaling relations at all cosmic epochs. The precision is comparable with that obtained by the modern cosmo-hydro-dynamical simulations, with the advantage of providing a rapid exploratory response on the consequences engendered  by different physical effects.}

   \keywords{galaxies formation and evolution --
                galaxies' structure --
                scaling relations
               }

   \maketitle
%

\section{Introduction}

In recent years several studies have addressed the problems posed by the observed scaling relationships (ScRs) of galaxies. It is clear today that the information encrypted in the ScRs is of fundamental importance to understand the physical processes occurred during the evolution of galaxies. The formation, structure, and evolution of galaxies depend on several phenomena (mergers, stripping, star formation events, feedback, etc.) that in most cases leave imprints on the observed galaxies' distribution in different ScRs. The analysis of the ScRs is therefore of great importance to reconstruct the history of galaxies and the evolution of our Universe.

Many physical parameters of galaxies are mutually correlated, in particular those defining the Virial Theorem (VT) and  the Fundamental Plane (FP) (\FP). The mass correlates with the radius (\MRa), the luminosity with the velocity dispersion (\Lsig), the effective surface brightness with the radius (\IeRe),  and others. In log units these relations are often linear, but significant deviations are visible in particular when the distributions of massive and dwarf galaxies are compared. 
In addition, when the sample of galaxies includes objects of very different luminosity, the ScRs  variations of the mean  slopes  and zones of exclusion (regions empty of galaxies) are visible.

Up to now, most of the efforts of the astronomical community have been dedicated to the analysis of the peculiar characteristics of each ScR. For example, many studies addressed the problem posed by the slope, zero-point, curvature and scatter \cite[among many others, see e.g.,][]{OegerleHessel91, Bernardietal2003, Shen_etal_2003, Courteau_etal_2007, Desroches_etal_2007, Sanders_2010, Nigoche-Netro_etal_2011, HydeBernardi2009, Montero-Dorta_etal_2016, Samir_etal_2020, Donofrio_etal_2024, Quenneville2024}. Other studies highlighted the role played by the environment, feedback effects, dissipation, mergers, Dark Matter halos, galaxy morphology and dynamics, and by cosmology \cite[see e.g.,][]{Bender_etal_1992, Burstein_etal_1995, Chiosi_Carraro_2002, Boylan-Kolchin2006, Damjanov_etal_2009, Fan_etal_2010, Bernardi_etal_2011, Shankar2013, Agertz_Kravtsov2015, Krajnovic2018, SanchezAlmeida2020, Chiosi_Donofrio_Piovan_2023}. A different approach, followed by many researchers, was that of comparing the results of numerical simulations with the observed galaxies in various ScRs \cite[see e.g.,][]{Lange_etal_2016, Ferrero_etal_2017, Furlong_etal_2017, Pillepichetal2018a, Pillepichetal2018b, Geneletal2018, Huertas-Companyetal2019, Cannarozzo2023}.

The approach followed here is somewhat  different from the previous ones and finds its roots on a series of papers \citep{Donofrioetal2017, Donofrioetal2019, Donofrioetal2020, DonofrioChiosi2021, Donofrio_Chiosi_2022, Donofrio_Chiosi_2023a, Donofrio_Chiosi_2023b}, that are based on the assumption that  instead of  separately analyzing  the single ScRs, more significant information can be extracted  by simultaneously looking in a self-consistent way at all the mutually related ScRs.
The novelty of this approach is to look at the behavior of each galaxy during its evolution, taking into account  that both luminosity and mass (and consequently all the other related parameters) continuously change for many reasons (star formation, mass acquisition/depletion by mergers or stripping events, natural aging of their stellar populations, and so forth). 

In the paradigmatic hierarchical context, repeated  mergers among galaxies occur and change the properties of these latter. At each merger, the dynamical structure and  equilibrium of a galaxy is perturbed, while  for most of the remaining time the galaxy is in mechanical equilibrium, that is they obey the VT. {Based on these grounds, \citet{Donofrioetal2017} first generalized the luminosity-velocity dispersion relation by explicitly introducing the time dependence. The new relationship is 

$$
 L(t)  = L'_0(t) \sigma(t)^{\beta(t)}, $$
\noindent
where $t$ is the time and $\sigma$ the velocity dispersion. The proportionality coefficient $L'_0$ and the exponent $\beta$ are functions of time and can vary from galaxy to galaxy. } Then, by coupling this law with the VT, which is a function of $M_s$, $R_e$ and $\sigma$, they obtained a system of equations in the unknowns $\log L'_0$ and $\beta$ whose coefficients are function of the variables characterizing a galaxy ($M_s$, $R_e$, $L$, $\sigma$, and $I_e$). 

The new empirical relation, although formally equivalent to the Faber-Jackson (FJ) relation for early-type galaxies (ETGs) \citep{FaberJackson1976}, has a profoundly different physical meaning: $\beta$ and $L'_0$ are time-dependent parameters that can vary considerably from galaxy to galaxy, according to the mass assembly history and the evolution of the stellar content of each object.
These parameters are found to be good indicators of a galaxy's  mass accretion history, star formation, and evolutionary processes, offering an immediate insight into its current stage of evolution. Indeed the $\beta$ parameter determines the direction of motion of a galaxy in the space of parameters at the base of the VT.

Adopting this new perspective, it was possible to simultaneously explain the tilt of the FP and the observed distributions of galaxies in the FP projections and, at the same time, to understand the real nature of the FJ relation. 

Having demonstrated that the ScRs observed today originate from the motion in the parameter space of each individual  galaxy during its evolution, what is still  lacking is to understand which physical phenomena drive the change of shape  of the ScRs in the course of time. In order to reach this goal we proceeded here in two parallel ways: i) First, we compared the ScRs related to the VT (the $I_e-R_e$, $M_s-R_e$ and $L-\sigma$ relations) with the most recent simulations of model galaxies in cosmological context, such as Illustris-TNG \citep{Vogelsberger_2014a, Vogelsberger_2014b} and EAGLE (Evolution and Assembly of GaLaxies and their Environments) \citep{Schaye_etal_2015, McAlpine_etal_2016}. The comparison was extended to $z\sim 1$ where observations are nowadays available in sufficient number to allow a meaningful analysis, and to even larger  redshifts to see what the simulations would predict. ii) Then, to cast light on the effects of mergers, in which different masses and energies are involved, on the various ScRs, we 
set up and present here a simple analytical model of mergers able to mimic the results of the large scale simulations and to predict the path of galaxies in the parameter space in the course of  time. This approach is based on the idea that galaxies in isolation are in virial equilibrium and that a merger is a perturbation of this state of short duration so that a new equilibrium condition is soon recovered. On this ground, the formation and evolution of a galaxy can be conceived as that of an isolated object whose mass and radius time to time are increased by a merger likely causing additional star formation.  
Starting from the idea and the formalism  developed by \cite{Naab_etal_2009} about the material falling on an already formed object during a single merger, we generalized their formalism to the case of a series of successive dry mergers among galaxies of different masses. We will see that this simple idea is able to catch the observed distribution of galaxies in the ScRs, making the dry merging phenomenon the principal mechanism shaping the ScRs observed today. We will also see that this approach is consistent with the results obtained in our previous  studies \citep[see, e.g.][]{Donofrioetal2017, Donofrioetal2020, DonofrioChiosi2021, Donofrio_Chiosi_2022, Donofrio_Chiosi_2023b}. The advantage, with respect to the  sophisticated numerical simulations, is in the possibility of obtaining new results on a short time and at no cost, whenever one may want to explore different solutions, such as different initial conditions, different efficiencies of mergers, different laws and modes of star formation, and so forth.

The paper is structured as follow. Sec. \ref{sec:2} presents the observational databases used for our comparisons of the structural parameters with the predictions of theoretical models. Sec. \ref{sec:3} introduces the EAGLE and Illustris-TNG simulations used in our analysis of the ScRs. Sec. \ref{sec:4} shows the comparison of simulations with the observational data at $z\sim0$. Sec. \ref{sec:5} gives an idea of the predicted distribution of galaxies in the ScRs at high redshift according to simulations. Sec. \ref{sec:6} presents the mean behavior of galaxies across time predicted by extensive numerical simulations. Sec. \ref{sec:7} is dedicated to presenting our simple analytical model of dry galaxy mergers and to a quick comparison with the observational data and the model galaxies produced by numerical simulations. In Sec. \ref{sec:8} we compare the observational data with the hydro-dynamical simulations. Sec. \ref{sec:9} contains a detailed comparison of the results obtained from our analytical models and the observational data and hydro-dynamical simulations. Sect. \ref{sec:9_beta_theory} summarizes the  $\beta-L_0'$ theory of \citet[][and references therein]{Donofrio_Chiosi_2024} and compares the results derived from the  application of it to the analytical model and observational data. Finally, Sec. \ref{sec:10_concl} draws some conclusions. 

\section{The observational data}\label{sec:2}
Three galaxy samples have been used in this work. The first one is the same adopted in our previous studies on the subject \citep[see,][]{Donofrio_Chiosi_2022,Donofrio_Chiosi_2023a}, that is the data at redshift z$\sim 0$ taken from the Wide-field Imaging of Nearby Galaxy-Clusters Survey (WINGS) and Omega-WINGS databases
\citep{Fasanoetal2006,Varela2009,Cava2009,Valentinuzzi2009,Moretti2014,Donofrio2014,Gullieuszik2015,Morettietal2017,Cariddietal2018,Bivianoetal2017}. {The whole WINGS photometric sample contains more than 32000 galaxies of all morphological types. }
The sub-sample also used here contains only ETGs, generally with stellar mass $M_s > 10^9 \, M_\odot$.
It has 270 galaxies, with measurements of their stellar masses, star formation rates (measured by \citet{Fritzetal2007}), morphologies (given by MORPHOT \citep{Fasanoetal2012}), effective radii and total luminosities {measured by GASPHOT \citep{Pignatelli, Donofrio2014}} and velocity dispersions, {collected from the literature}. 

The second galaxy sample is that of MaNGA (Mapping Nearby Galaxies at APO) survey \citep{Smee_etal_2013, Bundy_etal_2015, Drory_etal_2015, Blanton_etal_2017}. MaNGA obtained spectral and photometric measurements of nearby galaxies of different morphological types ($\sim10000$ objects across $\sim2700$ sky square degree), thanks to the “integral field units” (IFUs) of the Apache Point Observatory (APO). MaNGA is a byproduct of the SDSS survey and was designed to study the history of present day galaxies.

MaNGA provided a lot of galaxy parameters: morphology, stellar rotation velocity and velocity dispersion, magnitudes and radii, mean stellar ages and star formation histories, stellar metallicity, element abundance ratios, stellar mass surface density, ionized gas velocity, star formation rate, dust extinction, etc.. The galaxies were selected to span a stellar mass interval of nearly three orders of magnitude and no selection was made on size, inclination, morphology and environment, so this sample is fully representative of the local galaxy population and can be compared with the modern hydro-dynamical simulations like EAGLE and Illustris-TNG, which, however, do not provide information on the morphological types of their model galaxies.
The present sample contains information for 10220 galaxies of all morphological types.

  \begin{figure}        
   \centering
   \includegraphics[scale=0.40]{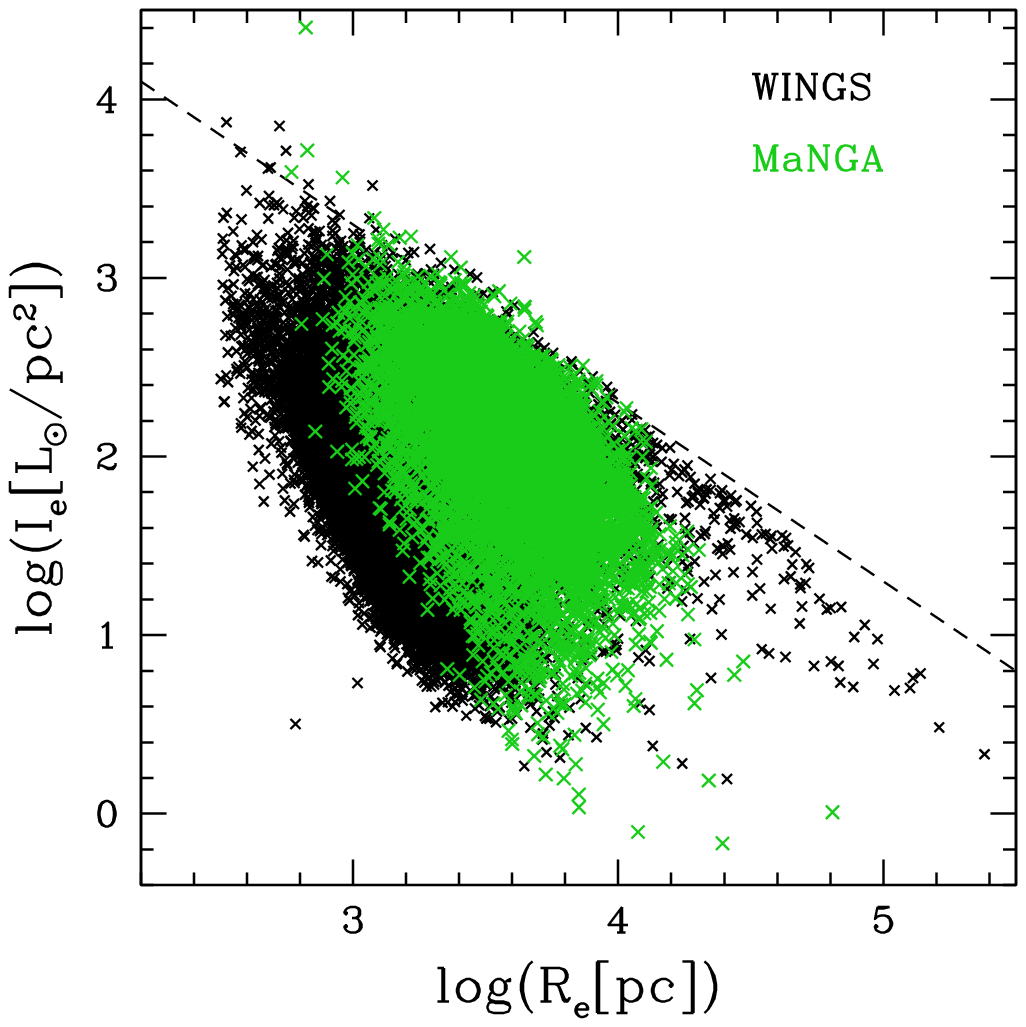}
   \caption{Comparison of the \IeRe\ plane for the WINGS and MaNGA data.
   The black symbols mark the whole WINGS dataset, while the green ones indicate the MaNGA data. All morphological types are included for both datasets.}
              \label{Fig:0}
    \end{figure}

{To get a qualitative idea of the ScRs we would expect from the  two datasets we present here the \IeRe\ plane as an example. ETGs and LTGs are in general well mixed in this plot, but for the ETGs which in general have larger radii \re\. The two datasets nicely overlap, with the exception of the largest ETGs that are not observed in MaNGA. Probably this is due to the fact that WINGS includes galaxies in clusters containing a number of very bright objects.}

At higher redshift ($z\sim 1$), the third sample in use is the Large Early Galaxy Astrophysics Census (LEGA-C) database. This sample contains $\sim 4000$ galaxies with measured parameters for objects at $z>0.6$ \citep{VanderWel2021}. Details on these measurements can be found in the cited paper and references therein.

{In the general, the parameters 
$L$, $M_s$, $R_e$, $\sigma$, $I_e$ are surely affected by some uncertainty, which for each object is estimated to amount to about  10 to 20\% at maximum. Since we are always using variables in logarithmic scale, uncertainties of this entity would correspond error bars from $2\times 0.042$ to $2\times 0.079$. Therefore, when presenting large samples of data with parametes encompassing wide intervals, error bars of this length would simply add some small noise to the general trends. For this reason, the data are always plotted neglecting the errors bars. }

\section{The EAGLE and Illustris-TNG data}\label{sec:3}

The simulations of galaxy formation and evolution used here are EAGLE and Illustris-TNG.

EAGLE \citep{Schaye_etal_2015, McAlpine_etal_2016} is a large-scale hydro-dynamical simulation of the Lambda-Cold Dark Matter universe ($\Lambda$-CDM). The project aims at examining the formation of galaxies and the co-evolution with the gaseous environment. In this simulation, great care was put in the treatment of feedback from massive stars and in the presence of accreting black holes.
EAGLE provides many galaxy parameters. Those used here are: the galaxy mass and luminosity, the mean effective half mass radius, the mean velocity dispersion of the star particles and the star formation rate.

Illustris-TNG \citep{Springeletal2018,Nelsonetal2018,Pillepichetal2018b} is the new version of Illustris-1. The new simulation seems to produce results in better agreement with the observations \citep{Nelsonetal2018, Rodriguez-Gomezetal2019}. In Illustris-TNG the structural parameters of galaxies, such as colors, magnitudes and radii, are in quite good agreement with the available data at low redshift \citep{Rodriguez-Gomezetal2019}. Here we used the TNG100 dataset. The extracted structural parameters are the same as the EAGLE one.

For both EAGLE and TNG100 simulations, the main progenitor branches of the merger trees were considered to follow the evolution of the same objects along time. The more massive galaxy has been always selected as the progenitor of each object going back to high redshift, from z=0 to z=4, in steps of 1.

Only those object that were already formed at z=4, with a stellar mass higher than the mass resolution of the simulations ($M_{res} = 1.4\times 10^6 M_\odot$) at all redshifts, and have a stellar mass $M_s > 10^{8.5}$ at z=0,  were considered.
To have data sets that are easier to handle and at the same time to keep the statistical robustness, only a maximum of 150 objects for each 0.2 dex of stellar mass at z=0 were considered.
At the end, a total of 2400 and 2200 objects were taken into consideration for EAGLE and TNG100 simulations, respectively.

With these data we want  to demonstrate that (i) these two hydro-dynamical simulations reproduce quite well the main characteristics of the observed ScRs at z=0; ii) our simple mathematical models of galaxy  structure are good proxies of both observational data and numerical N-body hydro-dynamical simulations.

\section{The ScRs at low redshift}\label{sec:4}

\begin{figure*}        
   \centering
   \includegraphics[scale=0.40]{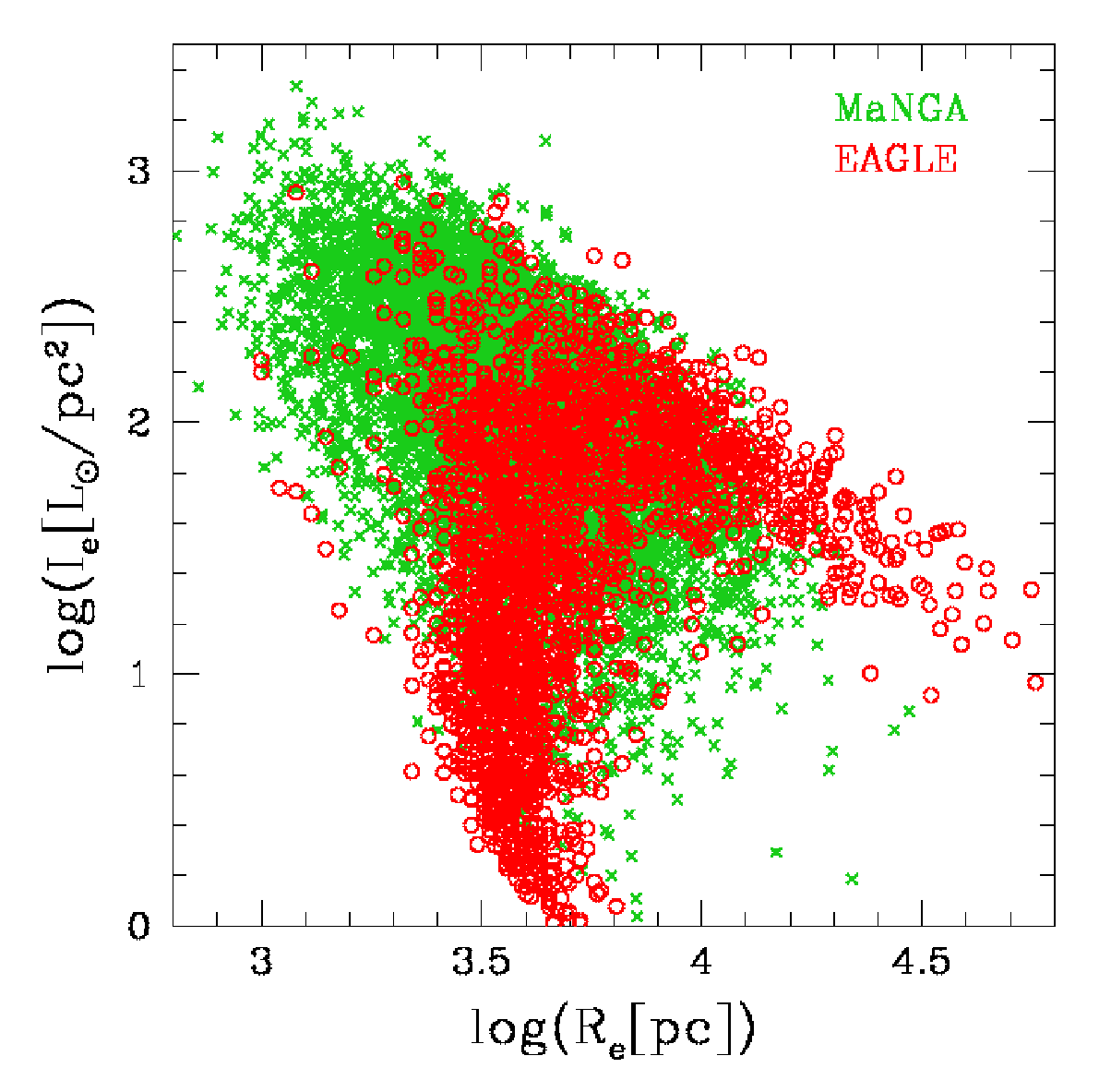}
   \includegraphics[scale=0.40]{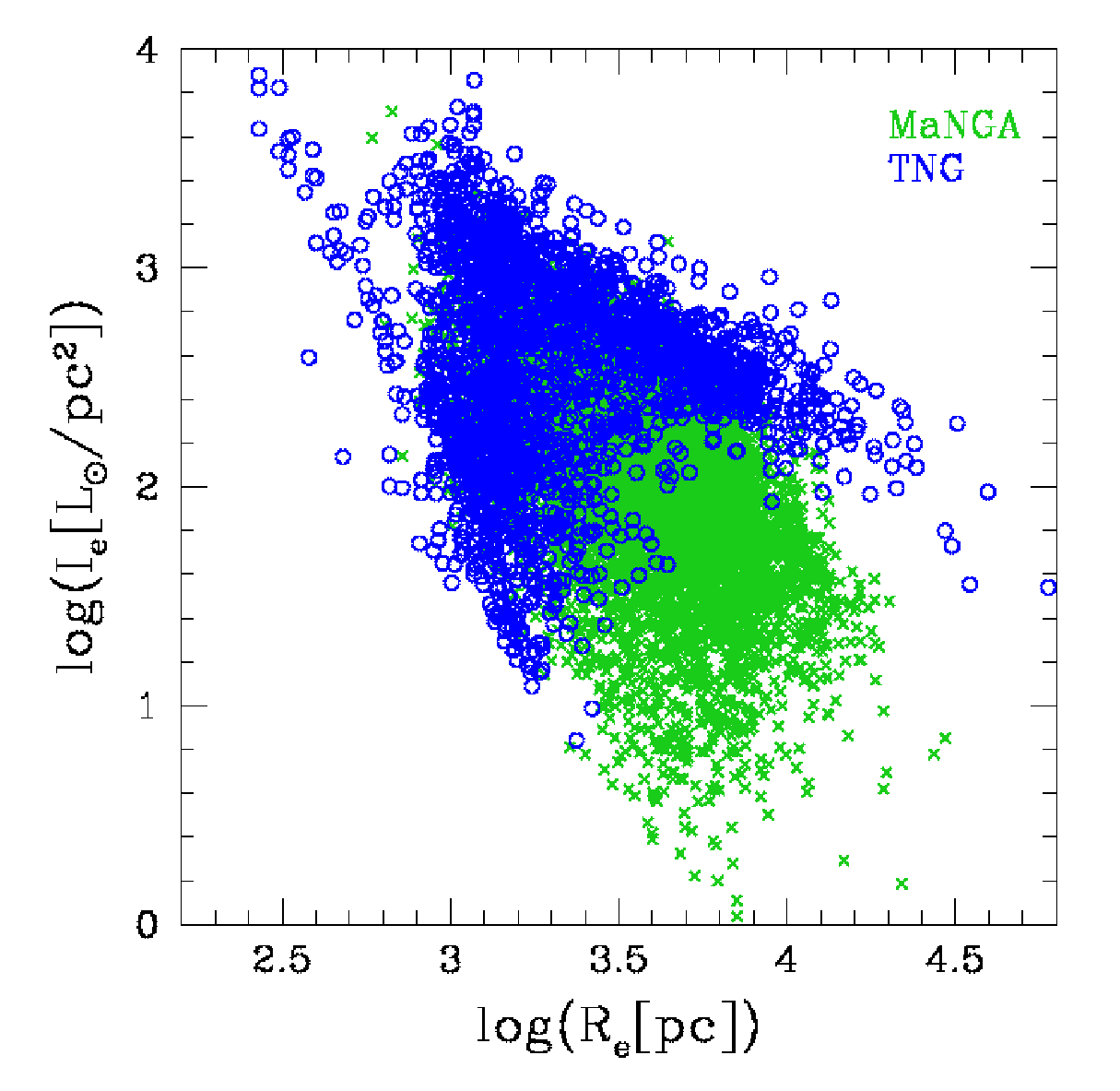}
   \caption{Comparison of the \IeRe\ plane with the MaNGA data and  simulations: EAGLE in the left panel and TNG100 in the right panel. The green crosses mark the MaNGA data for all morphological types, while the red circles are for EAGLE data and  the blue circles are for TNG100.}
              \label{Fig:1}
    \end{figure*}

Figure \ref{Fig:1} shows the \IeRe\ plane of the MaNGA database with superposed the results of the EAGLE (left panel) and Illustris-TNG100 (right panel) simulations. It should be remarked that the half light radius provided by the MaNGA data is somewhat different from the half mass radius given by simulations. 
Nevertheless, it is clear that both simulations reproduce many of the features observed in the \IeRe\ plane. This is the case, for example, of the $\Lambda$-shaped distribution of galaxies observed in this plane, in which the more massive galaxies are distributed along a tail  with a slope $\sim-1$, while the dwarf galaxies spread out on a cloud that extends along the  vertical direction. In our plot all morphological types are shown  for  both the observational data and simulations. In the case of the MaNGA data, the addition of the spiral galaxies produces a more uniform distribution of galaxies in the \IeRe\ plane in which the $\Lambda$-shape is less visible.

We note that the EAGLE radii are systematically larger than the TNG ones. This produces the higher surface brightness observed for the TNG data.

The lack of objects at low radii and very low surface brightness, that appears as a boundary, is due to selection effects. Only galaxies above a given luminosity are present in the observational data as well as in simulations.

On top right, we remark the presence of the Zone of Exclusion (ZoE). This boundary has a slope close to $-1$, a value predicted by the VT. No galaxies reside beyond this limit (at least at z=0, the present time). The origin of this boundary is still a matter of debate {\citep[see e.g.][]{Bender_etal_1992, Secco2001, Donofrioetal2006}}. 

\begin{figure*}   
   \centering
   \includegraphics[scale=0.40]{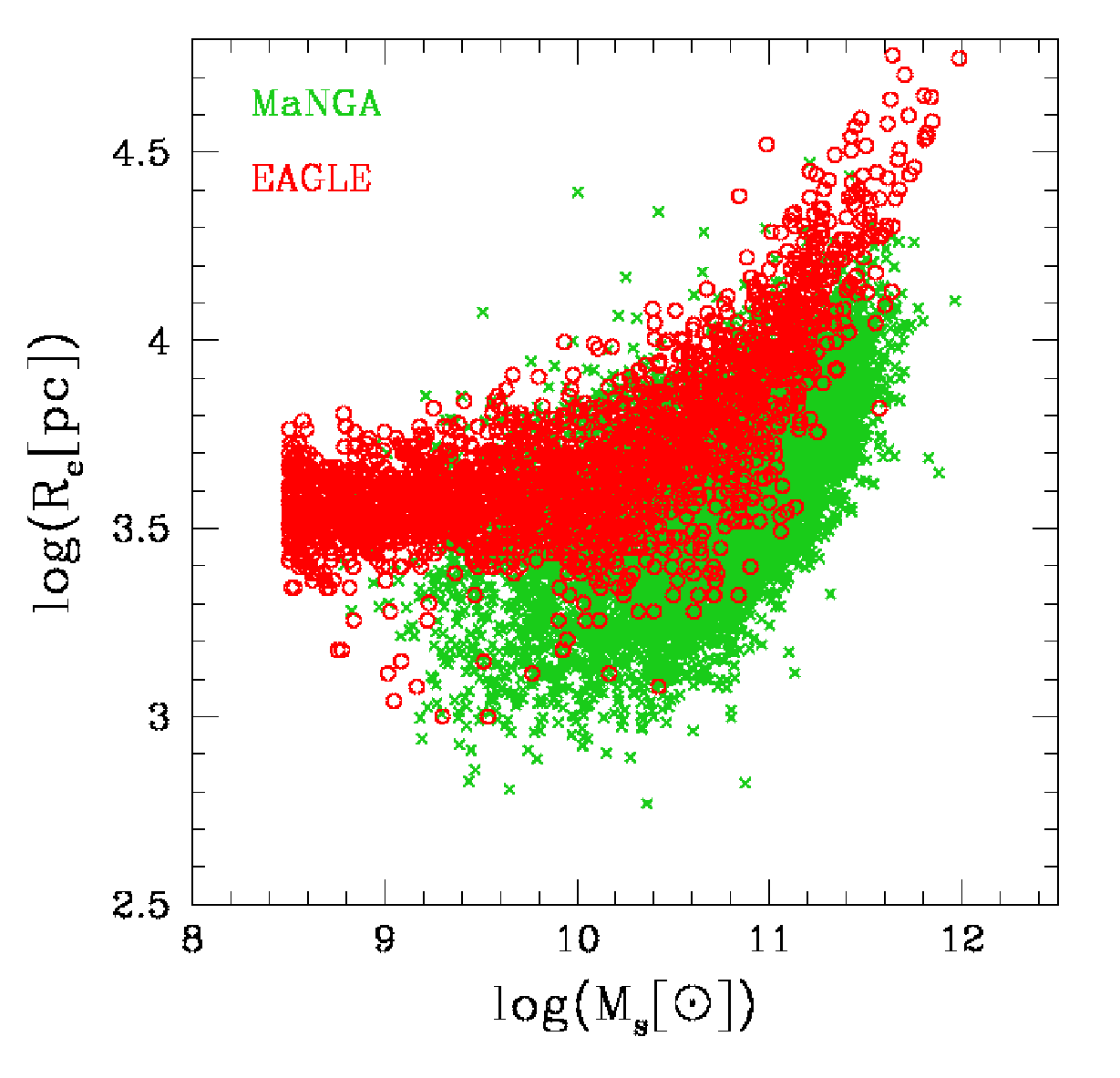}
   \includegraphics[scale=0.40]{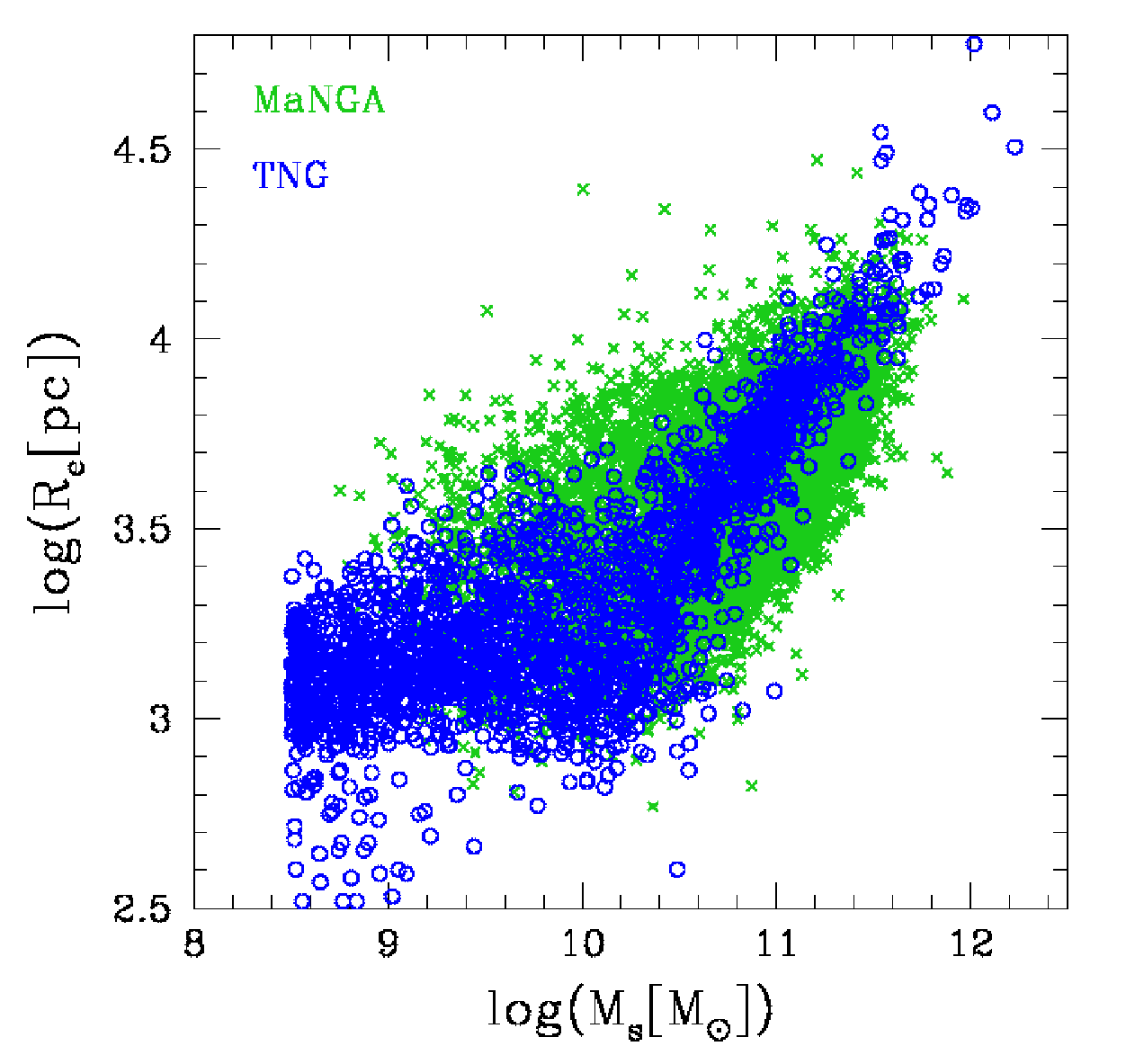}
   \caption{Comparison of the \MRa\ plane with the MaNGA data and simulations.  In the left panel the EAGLE models, in the right panel the TNG100 models. The color code and symbols are  as before.}
              \label{Fig:2}
    \end{figure*}

Fig. \ref{Fig:2} shows the \MRa\ plane for the same data-sets. Here again simulations are able to reproduce the curvature of the distribution at the high masses and the zone of exclusion on the right side with slope $\sim 1$. Only the scatter in radius is not fully matched. The observational data gives a much larger distribution of \re.
As for the previous plot this occurs when the spiral galaxies are added to the distribution.

In this plot it is better visible the systematic difference in the half-mass radius between EAGLE and TNG100. It is beyond the aims of this study to  analyze in detail   why this occurs. We limit ourselves  to note that  the physical prescriptions  used in the two simulations give rise to systematically different radii (most likely due to  feedback effects).

The simulations also predict  radii of the low mass galaxies (those with $M_s<10^9 M_\odot$) that are not present in our observational database. The flattening of the distribution at low mass galaxies  is clearly visible both in the observational data and theoretical simulations. In contrast, the tail formed by the massive galaxies is well matched by the models.

\begin{figure*}    
   \centering
   \includegraphics[scale=0.40]{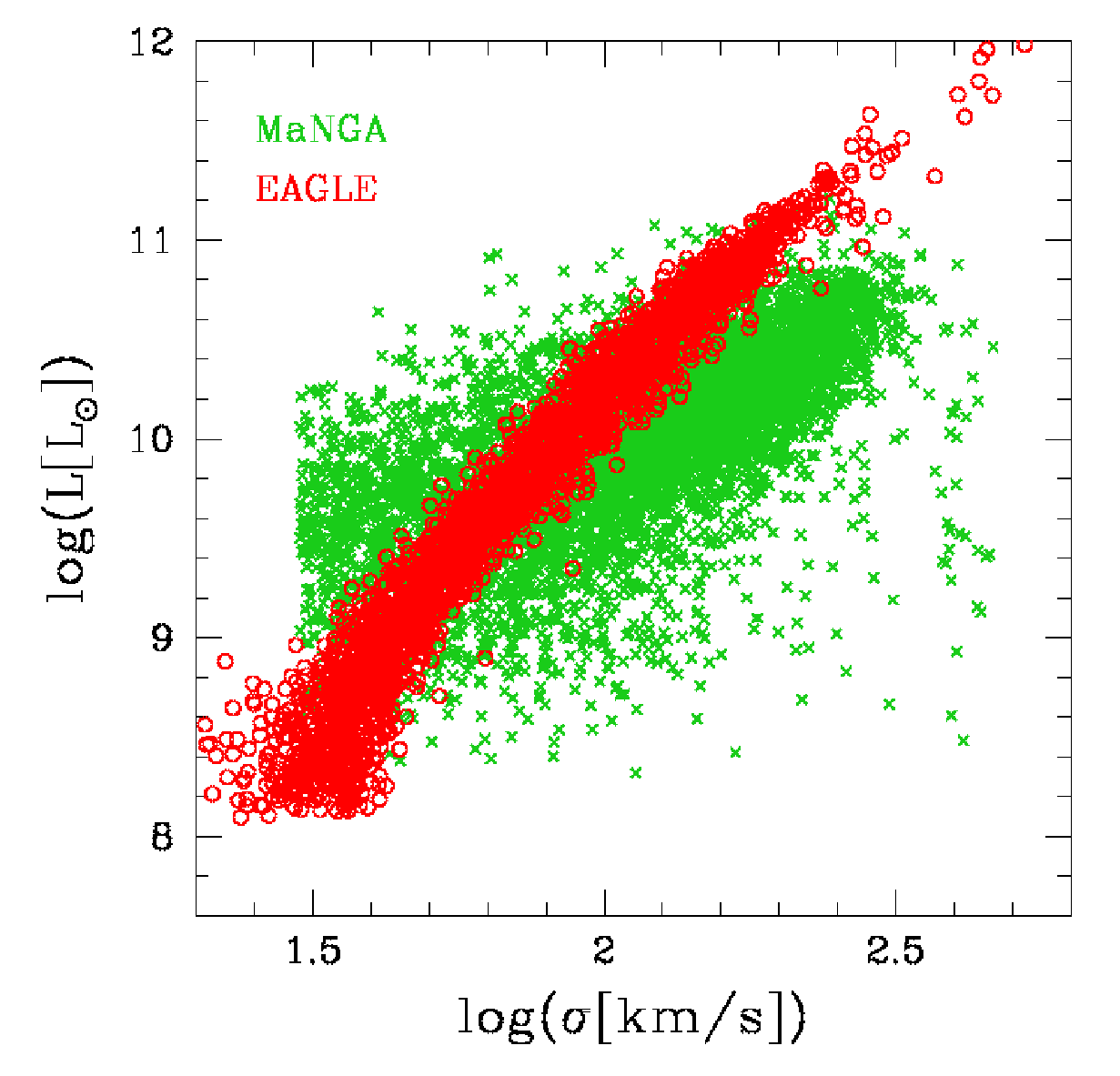}
   \includegraphics[scale=0.40]{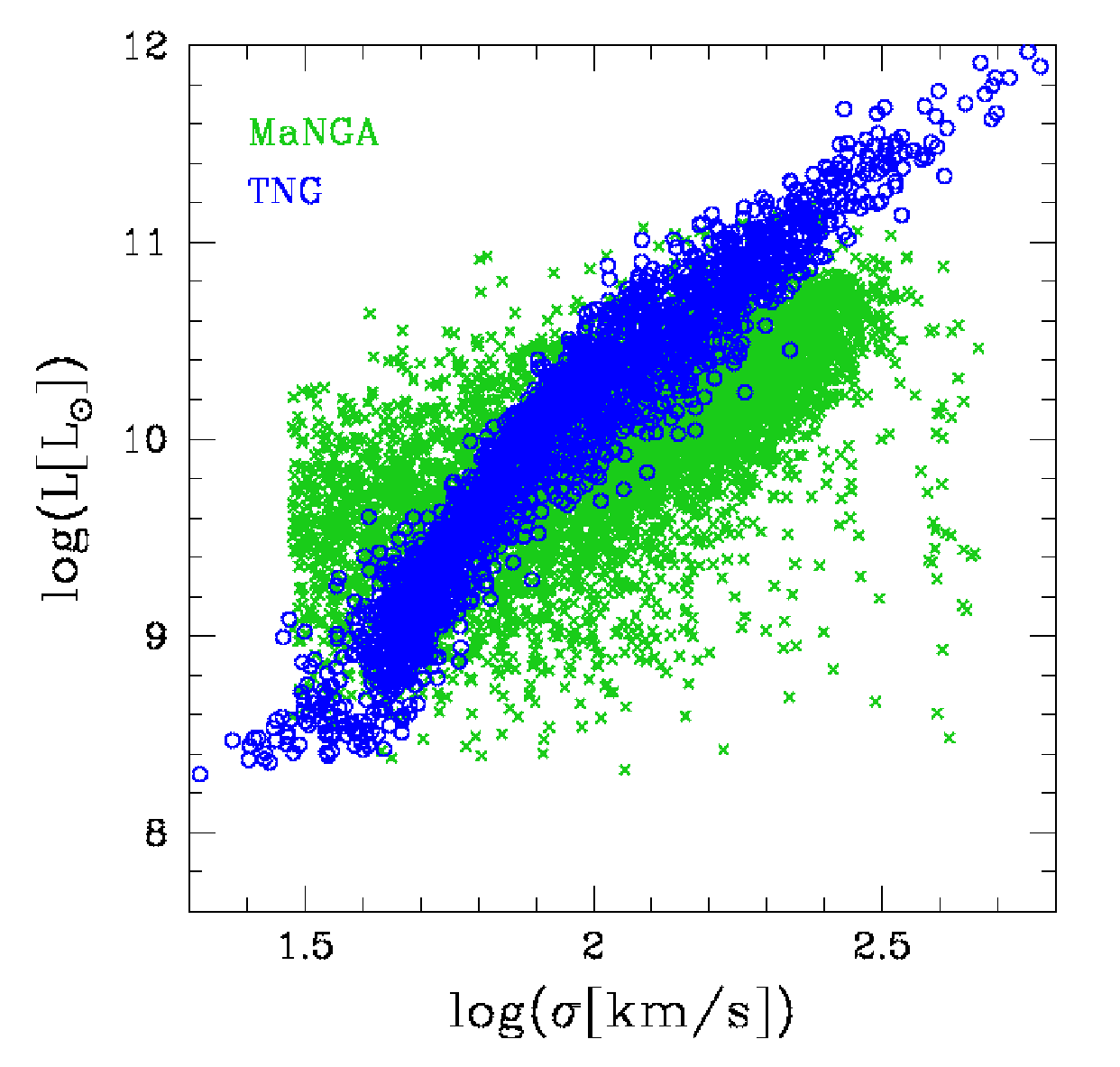}
   \caption{Comparison of the \Lsig\ plane with MaNGA data and simulations. In the left panel the EAGLE models, in the right panel the TNG100 models. Color code and symbols as before.}
              \label{Fig:3}
    \end{figure*}

Fig. \ref{Fig:3} shows the \Lsig\ plane. It is clear that in this case both sets of simulations do not fully reproduce the slope and the scatter in luminosity at each value of the velocity dispersion $\sigma$. A number of explanations can be found for this. It should be recalled for example that the MaNGA data derives $\sigma$ from the IFU. It follows that the velocity dispersion depends on the aperture of the fiber elements with respect to the galaxy dimension and the orientation of the galaxy.
On the theoretical side, simulations provide the mean velocity dispersion of the star particles within the half mass radius. The two measurements are far from being identical. Despite this,  a clear correlation of the two parameters is evident for both data samples.

The slope of the theoretical relation is slightly higher with respect to that of the data. Observations indicate that the galaxies of low luminosity progressively depart from the \Lsig\ relation and are distributed along a horizontal cloud of points. On the other hand, simulations predict a \Lsig\ relation holding  over the whole range of magnitudes with nearly constant  scatter everywhere.
Clearly the MaNGA data indicates that the addition of the spiral galaxies in this diagram, originally drawn  only for pressure supported systems like E and S0 galaxies, yields a less sharp and scattered distribution. The question arises, why simulations of all types of objects, yield such a narrow distribution. Probably, this is due to  the definition of velocity dispersion made for these systems.

As a final remark, we note that the model galaxies predict a slightly curved distribution that is not visible in the observational data.
It is not the aim of this work to discuss the possible explanations for the observed differences between observations and simulations. The qualitative comparison made here wants only to demonstrate that the present hydro-dynamical simulations are able to reproduce many of the main features of the observed distributions in all the ScRs generated by the FP projections. Examining these issues in more detail is left to other studies, more focused  on understanding the limits and drawbacks affecting  the present-day simulations of galaxies.

In other words, the physical prescriptions  adopted in the theoretical  models  catch the main behavior of galaxies in all ScRs. Perfect overlapping is, however, still lacking and requires some more work on several basic parameters that define a galaxy, for example a better description of the  luminosity profile at base of the half-mass core and half-mass radius and differences from galaxy to galaxy.

\section{Going up toward high redshift}\label{sec:5}

\begin{figure*}   
   \centering
   \includegraphics[angle=-90,scale=0.93]{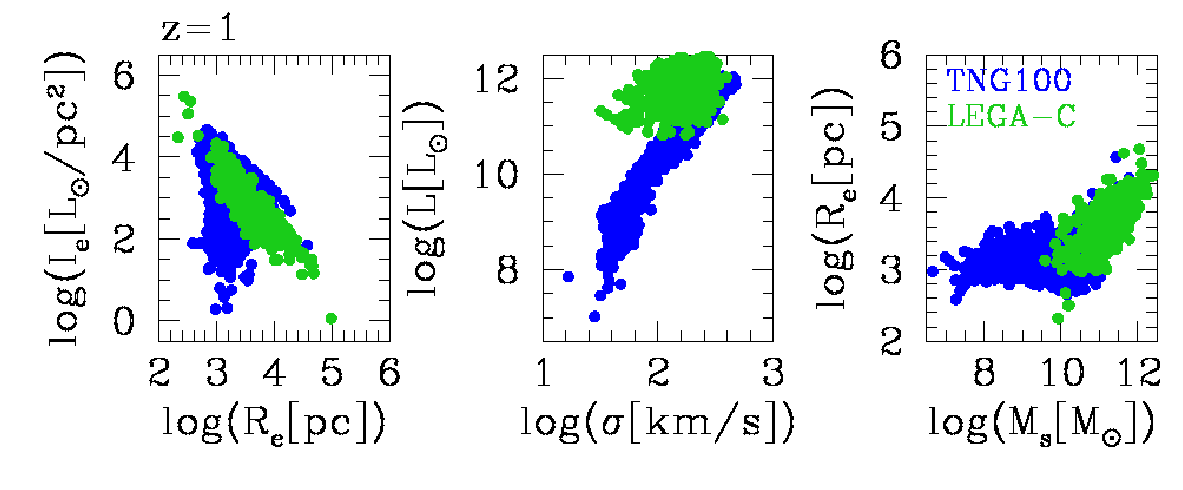}
   \caption{Comparison of the \IeRe, \Lsig\ and \MRa\ planes for the LEGA-C  data (green dots) and the TNG100 simulations (blue dots). Data and models refer to galaxies observed and/or predicted at redshift $z=1$.}
              \label{Fig:4}
    \end{figure*}

At z$\sim1$ the  sample of galaxies provided by the LEGA-C survey contains only the brightest objects. Fig. \ref{Fig:4} shows the comparison with the TNG100 data at the same redshift. {The comparison with EAGLE is very similar and we decided not to plot it here.} Again both simulations catch the main features of the ScRs. However, the \Lsig\ relation is only partially reproduced. Observations seem to indicate a much flatter distribution with respect to simulations. The reasons are likely the same as before. The data gives only the central velocity dispersion while the simulations yield the mean value of it inside the half-mass radius. In addition, a number of problems surely affect the measurement of $\sigma$ at  high redshifts. Nevertheless, the bulk of the distribution for the massive model galaxies well agrees with the observational data.

On the basis of the comparison of the structural parameters at z=0 and z$\sim1$, we may argue that the results of theoretical simulations provide a quite good descriptions of reality and therefore we can proceed to examine what they would predict for the ScRs at higher redshifts.

\begin{figure*}    
   \centering
   \includegraphics[scale=0.45]{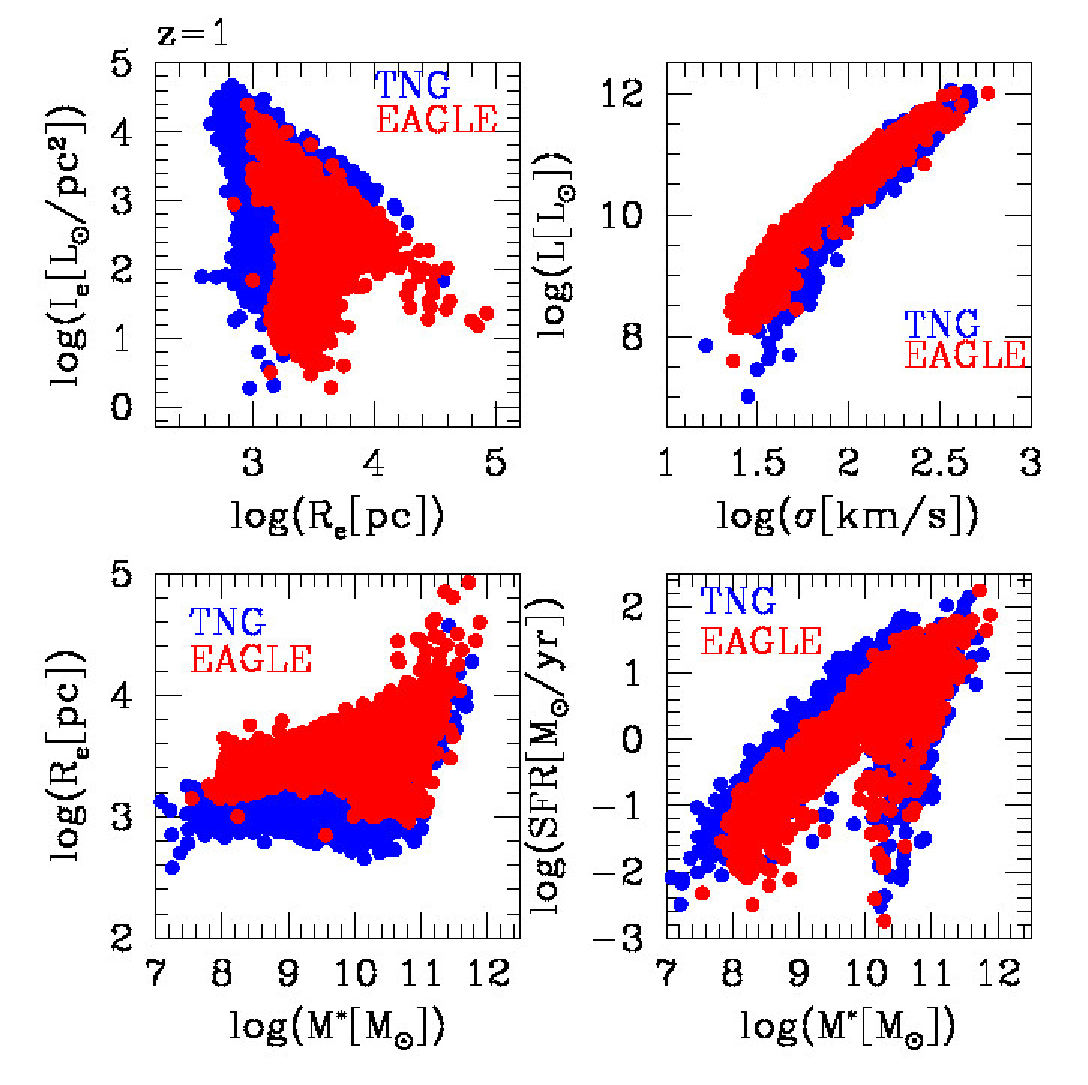}
   \includegraphics[scale=0.45]{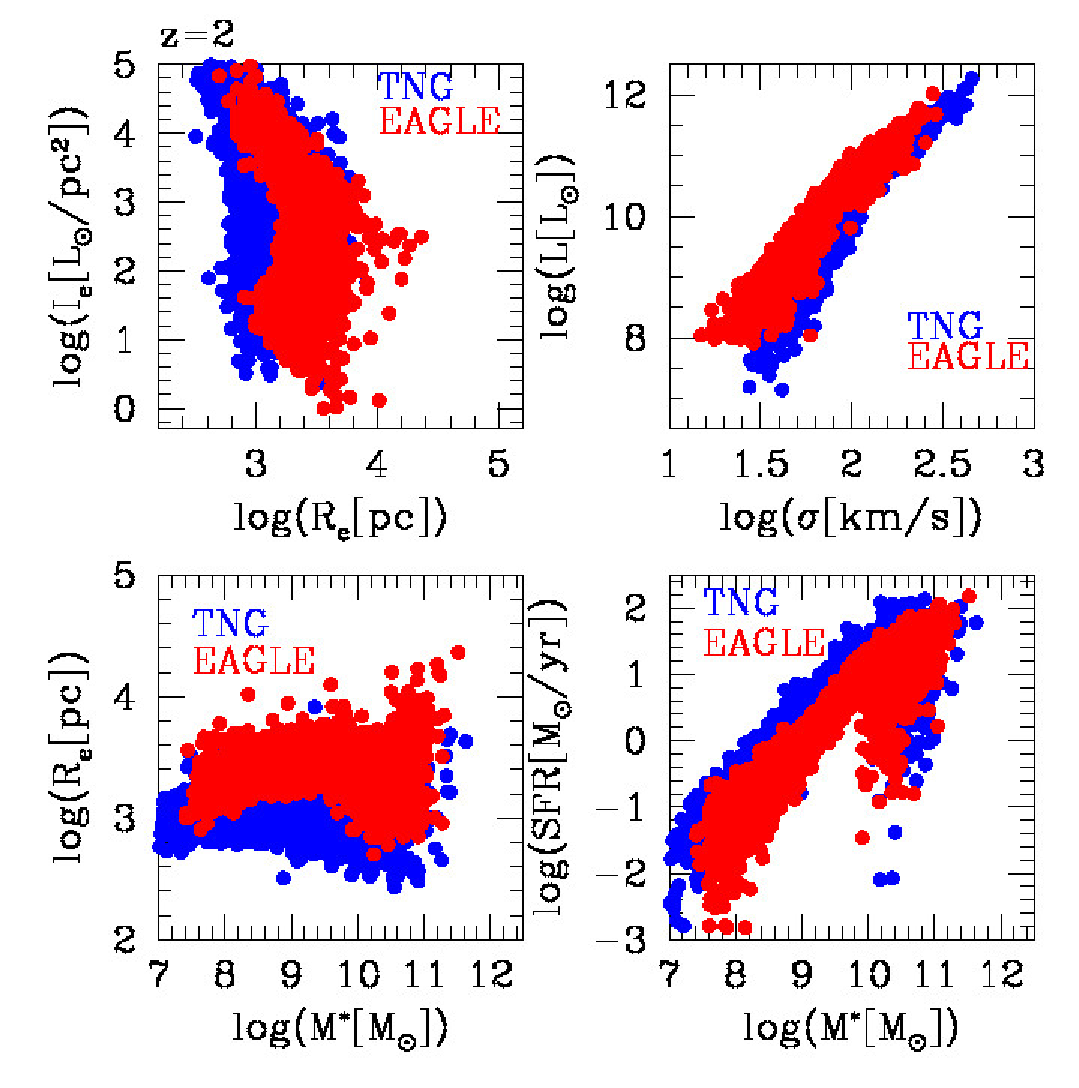}
   \caption{Comparison between EAGLE (red dots) and TNG (blue dots) at $z$=1 (the group of four panels at the left) and $z$=2 (the group pf four panels at the right) in four ScRs: the \IeRe\ plane (top left), the \IeSig\ plane (top right), the \MRa\ plane (bottom left) and the SFR-$M_s$ plane (bottom right).}
              \label{Fig:5}
    \end{figure*}

\begin{figure*}     
   \centering
   \includegraphics[scale=0.45]{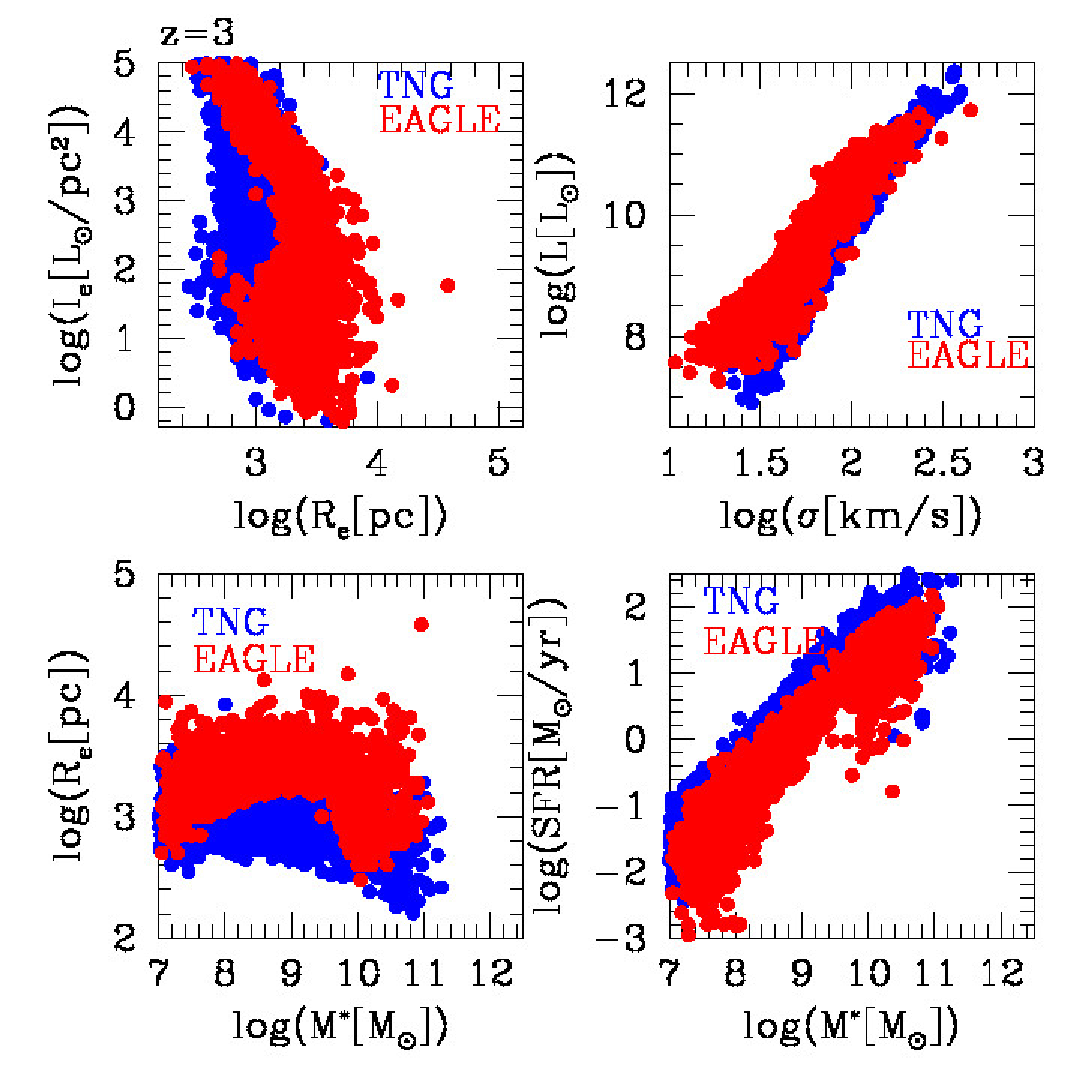}
   \includegraphics[scale=0.45]{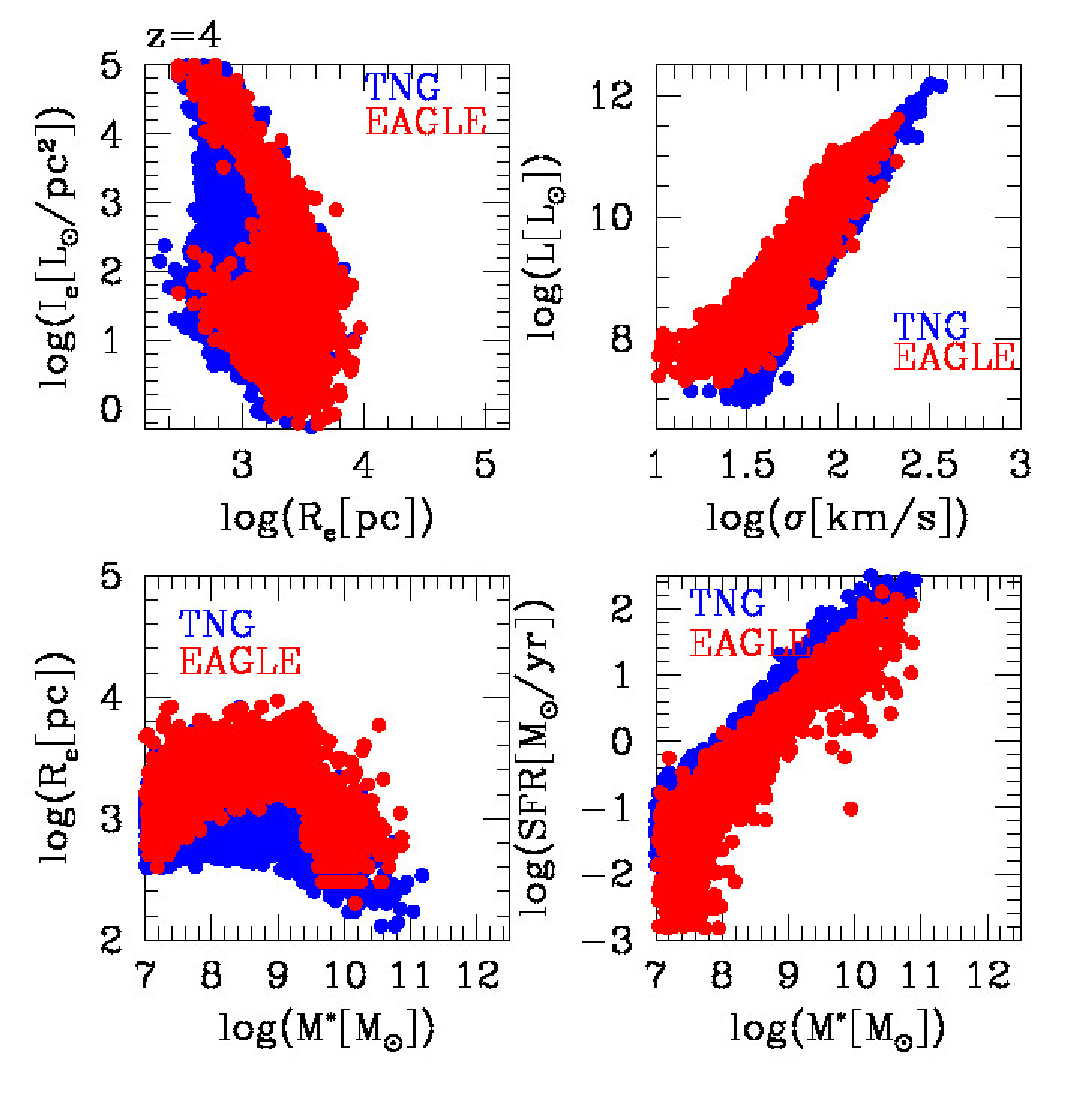}
   \caption{Comparison between EAGLE and TNG at $z=3$ (left panel) and $z=4$ (right panel). Symbols and diagrams as in Fig. \ref{Fig:5}.}
              \label{Fig:6}
    \end{figure*}

Figs. \ref{Fig:5} and \ref{Fig:6} show the predictions of EAGLE and Illustris-TNG for the ScRs at different redshifts, from z=1 to z=4. There is a substantial agreement between the two samples of models. Both predict that the effective half-mass radius decreases at large redshift, in particular that of massive galaxies. Consequently, the surface brightness increases together with the luminosity and the  tails in the \IeRe\ and \MRa\ planes disappear.

At high redshift both simulations predict  a \MRa\ relation curved toward an opposite direction with respect to that observed at z=0:  massive galaxies are all very small in size  and bright in luminosity.

The relation between the star formation rate (SFR) and the mass of galaxies is very similar, with the number of objects in the phase of SF quenching and/or already quiescent, that increases going towards the present epoch.

The \Lsig\ relation is progressively less curved going up in redshift and the scatter remains always approximately the same.

\section{The mean evolutionary  path of galaxies in simulations}\label{sec:6}

\begin{figure*}     
   \centering
   \includegraphics[scale=0.45]{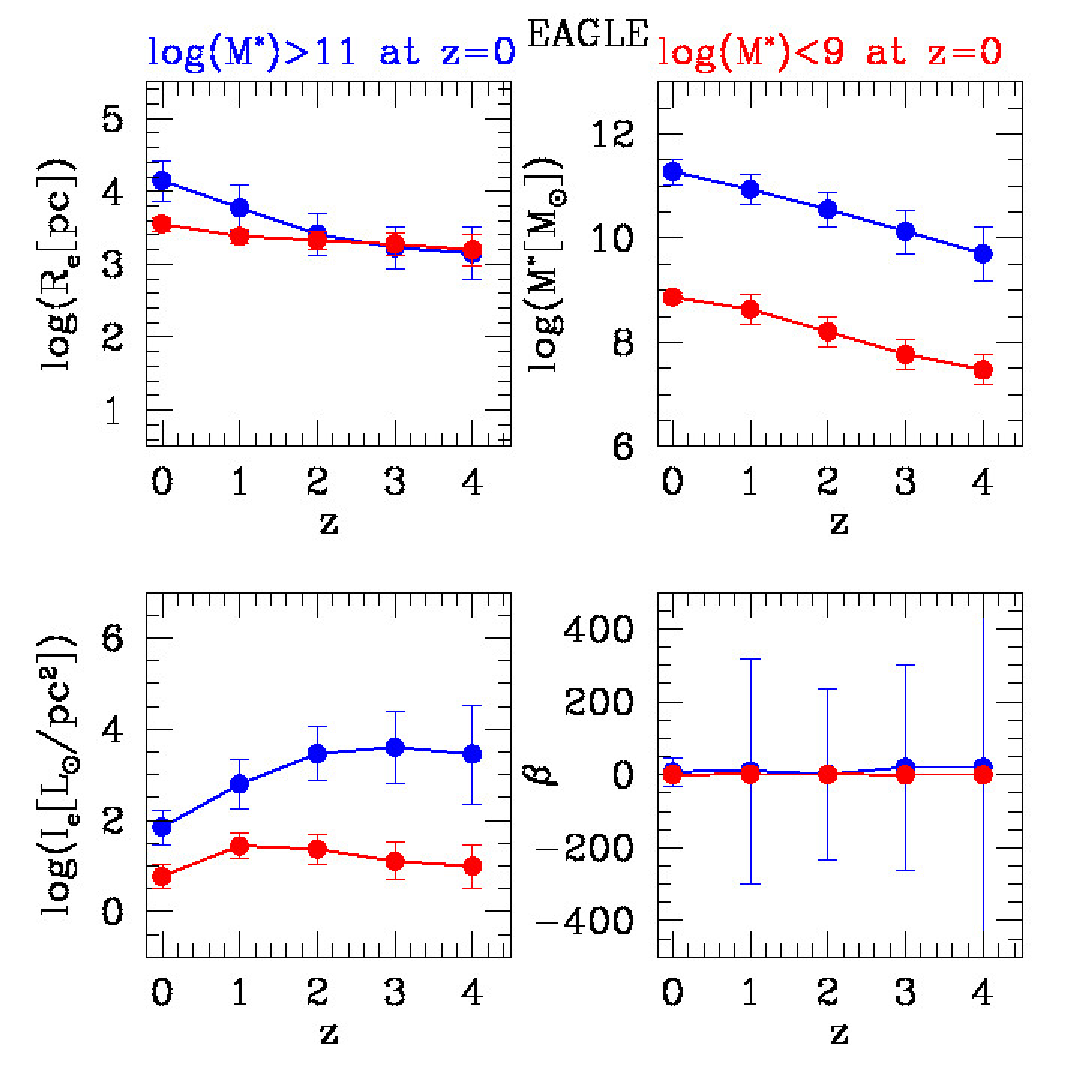}
   \includegraphics[scale=0.45]{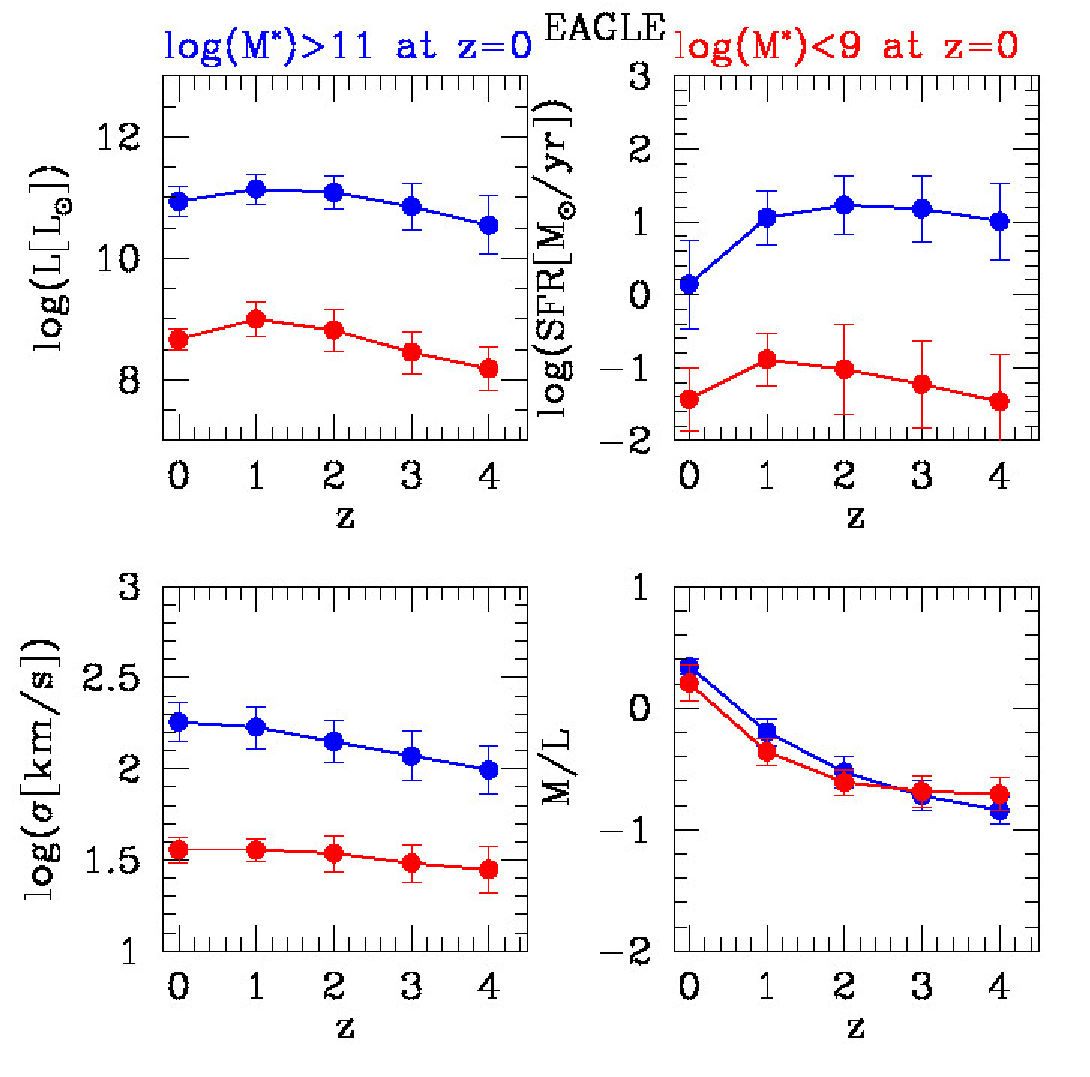}
   \caption{The mean evolutionary path of galaxies in the EAGLE simulations. The blue dots and lines give the mean path of the galaxies with today masses larger than $10^{11} M_\odot$, while the red dots and lines those for the masses lower than $10^{9} M_\odot$.
   The variables \re, $M_s$, \Ie\ and $\beta$ are plotted against the redshift $z$.}
              \label{Fig:7}
    \end{figure*}

\begin{figure*}      
   \centering
   \includegraphics[scale=0.45]{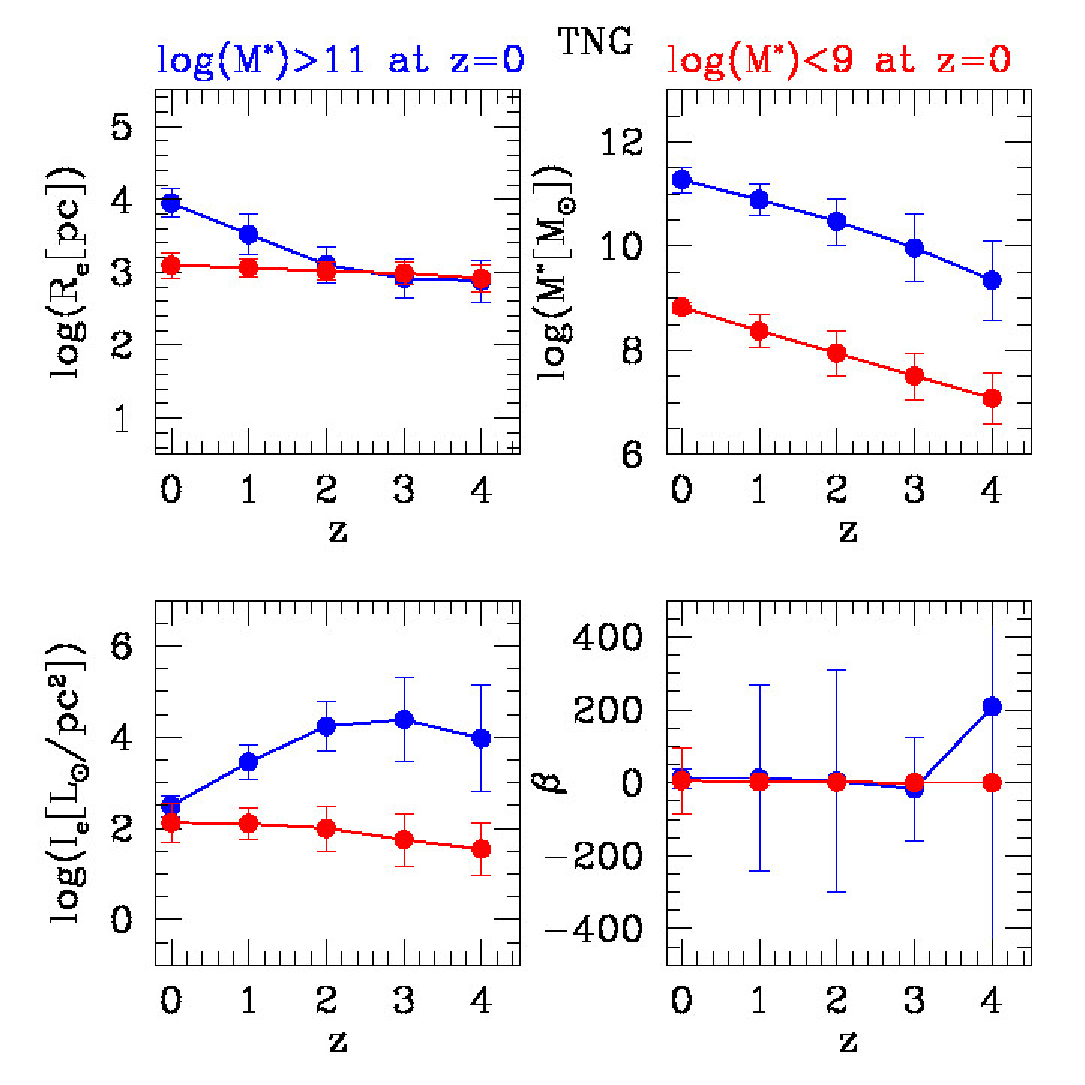}
   \includegraphics[scale=0.45]{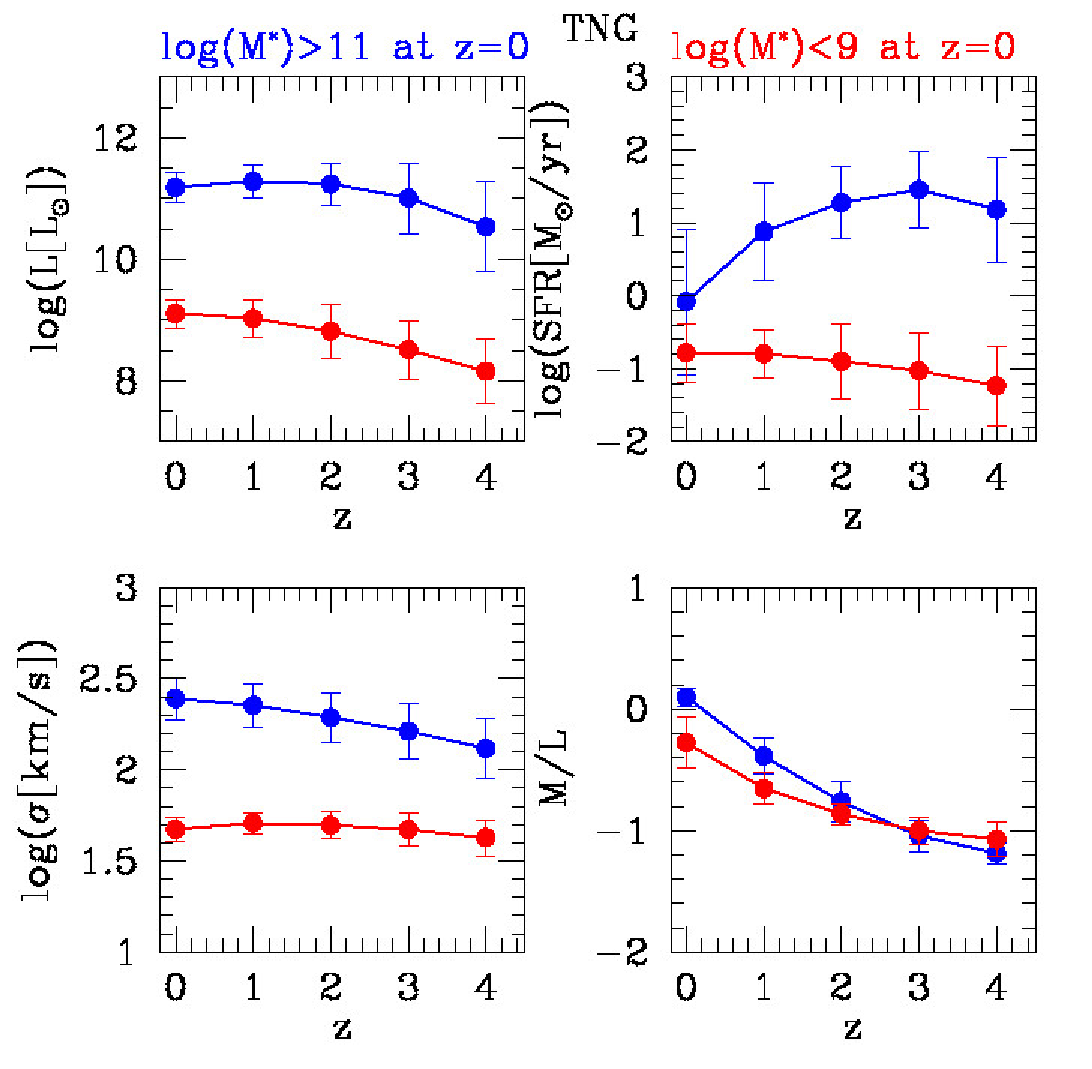}
   \caption{The mean evolutionary path of galaxies in the TNG simulations. Colors, symbols and diagrams as in Fig. \ref{Fig:7}.}
              \label{Fig:8}
    \end{figure*}

Figs. \ref{Fig:7} and \ref{Fig:8} give an idea of the mean evolutionary change in the structural parameters experienced by galaxies of different masses up to the present time (z=0). Objects with present mass larger than $10^{11} M_\odot$ and those with masses lower than $10^{9} M_\odot$ are shown in the figures with different colors. Fig. \ref{Fig:7} refers to the EAGLE simulation, while Fig. \ref{Fig:8} to TNG100. Both simulations show that the massive galaxies increase in radius with time, while the low massive objects essentially maintain the same radius. The mass increases with time, approximately with the same rate. The surface brightness decreases, in particular for the more massive galaxies. The same is true for the SFR, that decreases towards the present epoch. The total luminosity first increases and then starts to decrease. The $M/L$ ratio always increases towards the present time. The stellar velocity dispersion increases for the massive galaxies while it remains approximately the same for the low mass objects. The parameter $\beta$ defined by \citet[][see also below]{DonofrioChiosi2021} presents a large scatter around zero for the massive galaxies, while it remains always close to zero for the less massive ones.

The galaxies that today have masses within the limits we have considered exhibit properties intermediate in between  those we have described depending on their mass.

The galaxy parameters that undergo the largest variations across time are the half-mass radius and the SFR. We will see, with the aid simple models of mergers, that this is due to the combined effects of dry mergers and quenching of the stellar populations.

 \section{The role of mergers in shaping the ScRs of galaxies: a simple model of dry mergers} \label{sec:7}
 
In the first part of the paper, we have compared
the observational data to theoretical models calculated in the hierarchical scenario in which mergers among galaxies are the paradigm of their formation and evolution. Since data and theory were in satisfactory agreement,  we have implicitly  assumed that the hierarchical view was the right frame to work with and neglected to explore and highlight the role played by the merger mechanism in shaping the various scaling relations followed by galaxies. It goes without saying that the large scale simulations of galaxy formation and evolution in cosmological context are so heavy to plan, perform, and interpret that changing the physical ingredients to explore their effect on the ScRs of galaxies is out of reach. So trying to develop a simple analytical model of mergers as a smart proxy of the large scale numerical simulations is a meaningful and hopefully rewarding effort.
 
In this section we present a simple analytical model able to describe and predict the growth in radius as a consequence of the growth in mass by mergers together with the variations in luminosity, surface brightness, and velocity dispersion. 

The model strictly stands on the formalism  developed 
by \citet{Naab_etal_2009} for the case of a single object in virial equilibrium on which other objects of assigned total mass and energies fall in and merge together. Assuming energy conservation, the new system is also supposed to reach the virial equilibrium. The model we propose generalizes the formalism to the situation in which many successive mergers can occur. In addition to mass and radius the new model also considers the variations in luminosity and specific intensity.

\textsf{The Naab et al. (2009) case.} Let us start with the initial system characterized by mass $M_i$, half-mass radius $R_{e,i}$, velocity dispersion $\sigma_i$, luminosity $L_i$, specific intensity $I_{e,i}$. The kinetic ($K_i$),  gravitational ($W_i$),  and total ($E_i$) energies of the  initial  object are 
 
 $$ K_i = \frac{1}{2} M_i \sigma_i^2, \quad  W_i = - \frac{G M_i ^2}{ R_{e,i}}, \quad  {\rm and} \quad  E_i = K_i + W_i$$
 
 \noindent
 The virial equilibrium is expressed by $2 K_i + W_i = 0$ from which one derives the two identities  
 
 $$ K_i = - \frac{W_i}{2}  \quad {\rm and} \quad  E_i =  \frac{W_i}{2}. $$
 
 \noindent
 Suppose now that the system {\it i}  merges  another system {\it a} with  parameters $M_a$, $R_{a,i}$, $\sigma_a$, $L_a$,  $I_{a,i}$, $K_a$,  $W_a$,  and $E_a$, and that the virial condition is rapidly re-obtained after the merging. Finally, introduce the ratios  $\eta= M_a/M_i$ and $\epsilon = \sigma_a^2 /\sigma_i^2$. The total energy of the system is now $E_f = E_i + E_a$. After the merger event, the composite system will soon recover the virial equilibrium so that we can write
 
 \begin{equation}
  E_f = - \frac{1}{2} M_i \sigma_i^2 (1 + \epsilon \eta) = - \frac{1}{2} M_f \sigma_f^2 
 \end{equation}

 \noindent
 where $M_f = M_i + M_a = (1+ \eta) M_i$. It follows from it that 
 
 \begin{eqnarray} 
  \frac{M_f}{M_i}                &=& (1+ \eta)  \label{A_single_1} \\
  \frac{\sigma_f^2}{\sigma_i^2}  &=& \frac{1+ \eta \epsilon}{1+\eta}  \label{A_single_2}\\
  \frac{R_{e,f}}{R_{e,i}}        &=& \frac{(1+\eta)^2}{1+ \epsilon \eta} \label{A_single_3}  \\
  \frac{\rho_{f}}{\rho{i}}       &=& \frac{(1+\eta \epsilon )^3}{(1+ \eta)^5} \label{A_single_4} 
 \end{eqnarray}

 \noindent
 Depending on the parameters $\eta$ and $\epsilon$ the increase of the radius is different. If the mass is simply doubled $M_f = 2 M_i$,  $\eta =1$ and hence $\epsilon = 1$, the radius $R_{e,f} = 2 R_{e,i}$, and the velocity dispersion $\sigma_f = \sigma_i$.
  
\textsf{Extension to several mergers}. The formalism can be easily extended to the case of many successive  different mergers. 
For the sake of simplicity, we start considering an initial object characterized by the parameters $M_i$, $R_{e,i}$, $\sigma_i$, $L_i$, $I_{e,i}$ in suitable units and capturing other objects  characterized by the parameters $M_a$, $R_{e,a}$, $\sigma_a$, $L_a$ and $I_{e,a}$ in the same units. In this step of the reasoning, the luminosity and specific intensity are not included. The number of mergers is given by $N$ (integer). The captured objects are all the same and are small compared to the initial object.  Define also the ratios $\eta = M_a/M_i$  and $\epsilon = \sigma_a^2 / \sigma i ^2$. The final total mass is $M_f = M_i + N M_a$. The initial galaxy, the captured object, and the final object are all in virial equilibrium. Following the reasoning of the two-body case, we arrive to the expressions
 
 \begin{eqnarray}
 E_f &=& - \frac{1}{2} M_i \sigma_i^2 (1 + N \epsilon \eta)  \\
  E_f &=& - \frac{1}{2} M_f \sigma_f^2   
  \end{eqnarray}
 
 \noindent
 and the following scaling laws for the mass $M_f$, $\sigma_f^2$, $R_{e,f}$, and $\rho_f$ as functions of the corresponding values of the initial galaxy
 
 \begin{eqnarray}
 \frac{M_f}{M_i} &=& (1 + N \eta) \label{A_many_1} \\
  \frac{\sigma_f^2}{\sigma_i^2} &=& \frac{(1+N \eta \epsilon)}{(1+ N \eta)} \label{A_many_2} \\
  \frac{R_{e,f}}{R_{e,i}}  &=& \frac{(1+N \eta)^2}{(1+ N \epsilon \eta)} \label{A_many_3}  \\
  \frac{\rho_{f}}{\rho{i}}  &=& \frac{(1+N \eta \epsilon )^3}{(1+ N \eta)^5} \label{A_many_4}  
  \end{eqnarray}
  
\textsf{Luminosity and specific Intensity}. To include in the scaling laws those for the luminosity and specific intensity and derive the total luminosity $L_f$ and the total specific intensity $I_{e,f}$ of the remnant galaxy, we need to check in advance  some preliminary aspects concerning the luminosity because of its complex nature, more complex than the case of the mass, radius, and velocity dispersion.  Also in this case, the luminosity  $L_i$ and specific intensity $I_{e,i}$ of the initial object are  taken as the reference values for the luminosity, $L_a$, and specific intensity  $I_{e,a}$ of  the captured objects.    

We start from the notion that  the luminosity of a galaxy is proportional to its stellar mass content, the intensity of star formation, and the natural time evolution of its stellar populations. Looking at the luminosity to mass ratio $L/M$, the elementary theory of stellar population and many simple infall-models in literature \citet[see for instance][and references therein]{Tantalo_2005,Chiosi_Donofrio_Piovan_2023, Donofrio_Chiosi_2023b,Donofrio_Chiosi_2022} show that at the beginning the ratio $M/L$ is low (the effect of the luminosity overwhelms that of the mass), then grows to some value in relation to the star formation activity, and  past the initial period dominated by star formation the ratio steadily increases with the age. 
The typical behavior of the $L/M_s$ and $M_s/L$ ratios as function of the age and of the $L/M_s$ ratio as function of the stellar mass $M_s$ in infall models is shown in the {left and right panels} of Fig. \ref{masslum}, respectively. In the top panel, we also compare the $L/M_s$ ratio to that of single stellar populations (SSPs) of suitable chemical composition (solar abundances in our case) and derive their best-fit lines. Galaxy models and SSPs have the same trend. In the bottom panel we compare the $L/M_s$ variations as a function of the stellar mass of two model galaxies ($10^6$ $M_\odot$ and $10^{12}$ $M_\odot$). The two curves are identical but shifted by a certain factor (the mass). The best fit of the relations yields: 

  \begin{figure*}    
   \centering
   \includegraphics[scale=0.4]{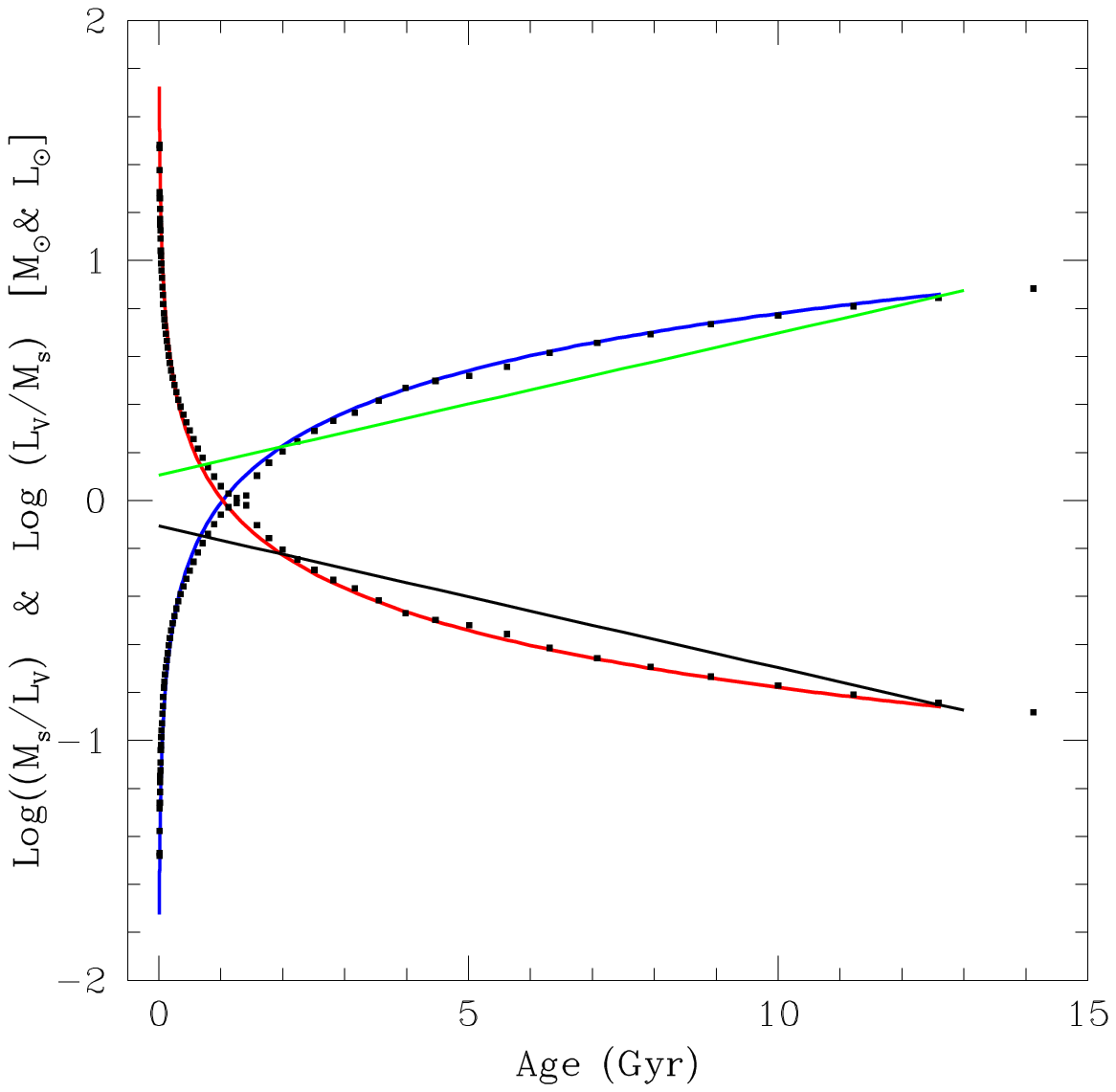}
   \includegraphics[scale=0.4]{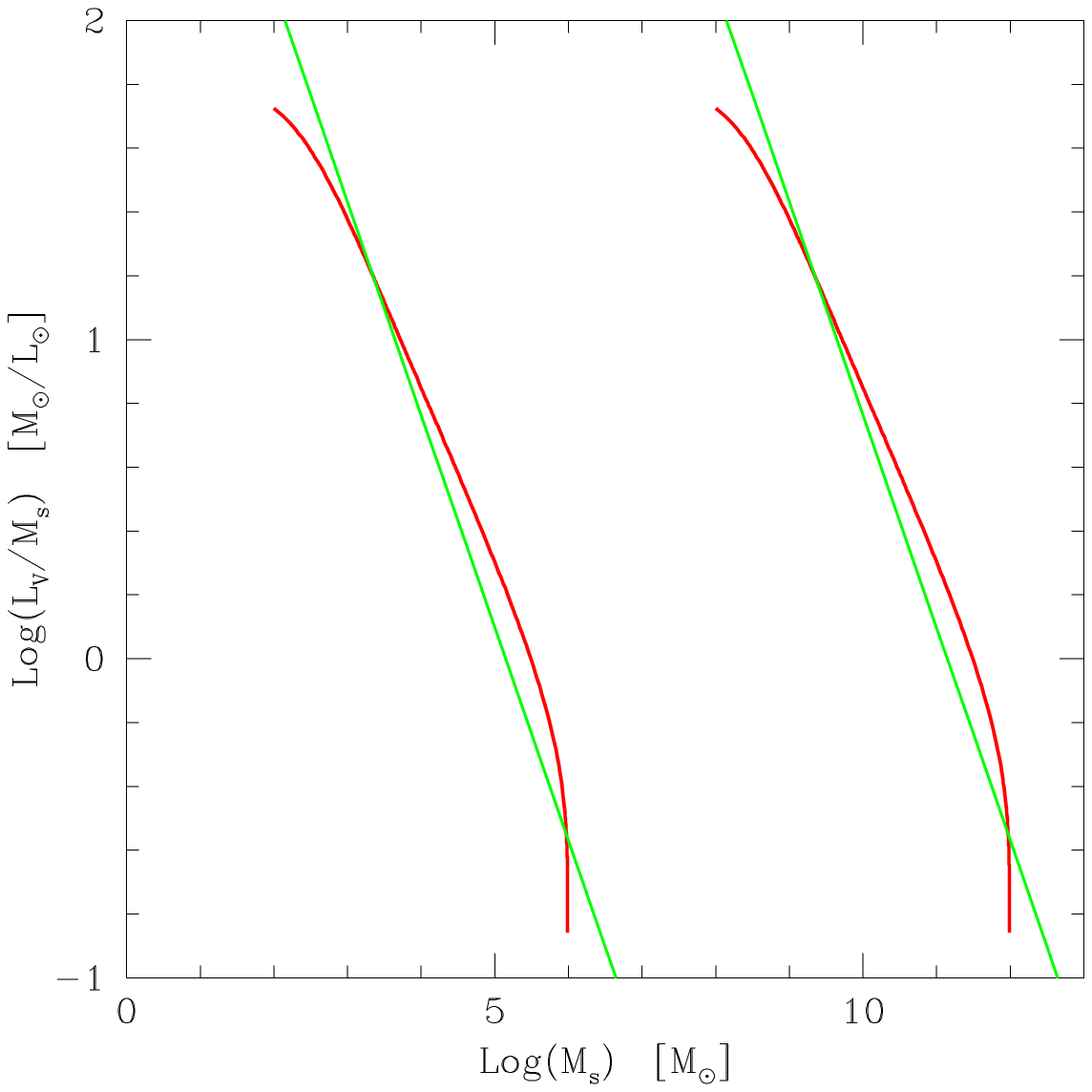}
   \caption{Mass to luminosity and luminosity to mass ratios as function of the age (left panel) and the luminosity to mass ratio as function of the stellar mass $M_s$ (right panel). Masses and luminosities are in solar units, the age is in Gyr. 
   \textsf{Left Panel}: In the top panel the solid red and blue lines are the $M_s/L_V$ and $L_v/M_s$ ratios, respectively, for an infall model of asymptotic mass of $10^{10} M_\odot$ calculated by the authors \citep[see][and references]{Chiosi_Donofrio_Piovan_2023},  while the black dots along each line are the same but for a SSP with solar chemical composition taken from the Padova library \citep{Tantalo_2005}. The straight lines are the best fit of the relations. \textsf{Right Panel}: the $M_s/L_V$ ratio is displayed for two infall models with different mass. For each case the stellar mass  increases along each curve from left to right. The two straight lines are the best fit of the data. See the text for more details.}
              \label{masslum}
    \end{figure*}

\begin{equation}
 \log \frac{L}{M_s} = -0.667 \log M_s  + K_{M_s},
\end{equation}

\noindent
where $K_{M_s}$ is 7.43 for the $10^{12}$ $M_\odot$ galaxy and 3.43 for the $10^{6}$ $M_\odot$) one. The intercept of the best-fit line is a function of the mass. Performing a linear interpolation and replacing the factor $K_{M_s}$, we obtain

\begin{equation}
\log \frac{L}{M_s} = -0.57 
\end{equation}

\noindent
and 

\begin{equation}
 \log L = \log M_s -0.57.
\end{equation}

\noindent
The factor -0.57 can always be expressed as the logarithm of a suitable number $-0.57 = Log A$. Therefore we may  write   
$$L = A M_s , $$
where $A \simeq 1/4$. Consequently,  in the derivation of the ratio $L_f / L_i$ the factor $A$ will cancel out.

\noindent
At this point we may wish to include also the effect of age on the luminosity  emitted by the stellar populations of a galaxy. From the relations shown in the top panel of Fig. \ref{masslum}, but for a short initial period, the luminosity of a single galaxy tends   to always decrease as a function of time.  Furthermore, looking at the captured objects of a certain mass there is no reason to say that they all have the same age and in turn the same luminosity. We may cope with this by introducing in the mass luminosity relation a factor $\delta$ varying from galaxy to galaxy within a certain range ($L = \delta A M_s$).  Therefore $\delta$ is  related to the luminosity per unit mass and is  also a function of time. 

With the present formalism, 
it is convenient to normalize the luminosity of the generic captured object to that of  the initial galaxy and write

  \begin{equation}
   \frac{L_f}{L_i} =  1 + \sum_j \frac{L_j}{L_i} = 1 + \sum_j \frac{\delta_j M_j}{\delta_i M_i}
  \end{equation}
posing $\theta_j = \delta_j / \delta_i$, we derive

\begin{equation}
\frac{L_f}{L_i} = (1 + \sum_j \theta_j \eta_j ).
\end{equation}

\noindent
Since  $\delta_j$ is an arbitrary quantity and $L_i$ is assigned,  we may always pose $\delta_i = 1$.

\noindent
If $\eta$ and $\theta$ are always the same we may write 

\begin{equation}
\frac {L_f}{L_i} = \frac {M_f}{M_i} = (1+ N \eta \theta) 
\end{equation}

\noindent
otherwise  the summation is needed.

\noindent 
If one is not interested to highlight the time dependence of the luminosity variation, the factor $\theta$ can be neglected by simply posing $\theta=1$.

\noindent
Since the specific intensity  is  given by $I_{e} = L /(2 \pi R_{e}^2)$,  the associated scaling relation is

\begin{equation}
\frac{I_{e,f} }{I_{e,i}} = \frac{ (1 + N \eta \theta) (1+ N\eta \epsilon )^2 }{(1+ N \eta)^4} .
\label{intensity_eq}
\end{equation}

\noindent
The last two relations complete the systems of equations from (\ref{A_single_1}) to (\ref{A_single_4})  and from (\ref{A_many_1}) to (\ref{A_many_4}).

Before proceeding further, we present in  Table \ref{Tab_1}  a few cases to illustrate the kind of effects we would expect on an initial object ($M_i$, $R_{e,i}$, $\sigma_i^2$ and $L_i$) after a series of mergers with an object with $M_a$, $R_{e,a}$, $\sigma_a$, and $L_a$) (always the same for simplicity). Finally,  the parameters $\eta$, $\epsilon$, and $\theta$ are supposed to be constant.  

   \begin{table*}
    \begin{center}
    \caption{Results from the analytical model of  mergers. Masses are in $M_\odot$, luminosities in $L_\odot$, radii in kpc, and specific intensities in $l_\odot/pc^2$.  } 
    \label{Tab_1}
    \begin{tabular}{|c r c c c c c c c c|}
     \hline
$M_i$    &  N   &$ \sigma_f^2/\sigma_i^2$ &    $R_{e,f}/R_{e,i}$ &$\rho_f/\rho_i$    & $L_f/L_i$ &  $I_{e,f}/I_{e,i}$ & $M_f$ & $\Delta log R_{ef}$ & $\Delta Log I_{ef}$\\
 \hline
\multicolumn{10}{|c|}{$\eta=0.1$, \qquad  $\epsilon=0.1$ \qquad $\theta=1$}\\
 \hline
  1E10 &   1 & 0.92E+00 & 0.12E+01 & 0.64E+00 & 0.11E+01 & 0.77E+00 & 0.11E+11 & 0.078 &-0.116\\
  1E10 &   3 & 0.79E+00 & 0.16E+01 & 0.29E+00 & 0.13E+01 & 0.48E+00 & 0.13E+11 & 0.215 &-0.316\\
  1E10 &   5 & 0.70E+00 & 0.21E+01 & 0.15E+00 & 0.15E+01 & 0.33E+00 & 0.15E+11 & 0.331 &-0.486\\
  1E10 &  10 & 0.55E+00 & 0.36E+01 & 0.42E-01 & 0.20E+01 & 0.15E+00 & 0.20E+11 & 0.561 &-0.820\\
  1E10 &  50 & 0.25E+00 & 0.24E+01 & 0.43E-03 & 0.60E+01 & 0.10E-01 & 0.60E+11 & 1.380 &-1.982\\
  1E10 & 100 & 0.18E+00 & 0.60E+02 & 0.50E-04 & 0.11E+02 & 0.30E-02 & 0.11E+12 & 1.782 &-2.522\\
  1E10 &1000 & 0.11E+00 & 0.93E+03 & 0.13E-06 & 0.10E+03 & 0.12E-03 & 0.10E+13 & 2.967 &-3.930\\
\hline
\end{tabular}
\end{center}
\end{table*}

\textsf{Including the kinetic energy of the merging object.}
For the sake of investigation, we  also took into account that the captured object with mass $M_a$, velocity dispersion $\sigma_a$, also  carries the kinetic energy of the relative motion  $K_a = (1/2) M_a v_a^2$. Part of this energy is supposed to be absorbed by the receiving object and the rest be  dissipated in some way. Also in this case, the mass $M_a$  and velocity $v_a^2$ are parameterized in terms of the initial mass $M_i$ and initial velocity dispersion $\sigma_i$ 
by the relation $M_a= \eta M_i$ and  $v_a^2 =  \lambda \sigma_i^2$. After the merger, the composite system recovers the virial condition. With the aid of  some simple algebraic manipulations, we obtain

 \begin{eqnarray}
 \frac{M_f}{M_i} &=& (1+N \eta) \label{B_many_lambda_1} \\
  \frac{\sigma_f^2}{\sigma_i^2} &=& \frac{(1+N \eta (\epsilon +\lambda)}{(1+ N \eta)} \label{B_many_lambda_2} \\
  \frac{R_{e,f}}{R_{e,i}}  &=& \frac{(1+N \eta)^2}{1+ N \eta (\epsilon + \lambda}    \label{B_many_lambda_3}\\
  \frac{\rho_{e,f}}{\rho{e,i}}  &=& \frac{(1+N \eta (\epsilon + \lambda )^3}{(1+ N \eta)^5}  \label{B_many_lambda_4} \\
  \frac{L_f}{ L_i} &=& (1+ N \eta \theta)   \label{B_many_lambda_5} \\
  \frac{I_{e,f} }{I_{e,i}} &=& \frac{ (1 + N \eta \theta) (1+ N\eta (\epsilon+\lambda) )^2 }{(1+ N \eta)^4} \label{B_many_lambda_6} 
  \end{eqnarray}
 
 \noindent
 Note that for $\lambda=0$ the previous case is recovered.

\textsf{Random variations of $\eta, \epsilon, \lambda$, and $\theta$.}  These relations can be further generalized to take into account the possibility that the parameters $\eta$, $\epsilon$, $\lambda$, and $\theta$  can slightly vary from one galaxy to another and the merger event. The  system of equations from (\ref{B_many_lambda_1}) to (\ref{B_many_lambda_6}) becomes 

 \begin{eqnarray}
 \frac{M_f}{M_i} &=& (1+ \sum_j\eta_j  )  \label{A_many_rand_1} \\
  \frac{\sigma_f^2}{\sigma_i^2} &=& \frac{(1+ \sum_j \eta_j \epsilon_j + \sum_j \eta_j \lambda_j)}{(1 + \sum_j \eta_j)} \label{A_many_rand_2} \\
  \frac{R_{e,f}}{R_{e,i}}  &=& \frac{(1+\sum_j \eta_j)^2}{1+ \sum_j \eta_j \epsilon_j  + \sum_j \eta_j \lambda_j}  \label{A_many_rand_3} \\
  \frac{\rho_{e,f}}{\rho{e,i}}  &=& \frac{(1+\sum_j \eta_j \epsilon_j + \sum \eta_j \lambda_j)^3}{(1+ \sum_j \eta_j)^5}  \label{A_many_rand_4}  \\
  \frac{L_f}{L_i} &=& (1+ \sum_j\eta_j \theta_j )  \label{A_many_rand_5} \\
  \frac{I_{e,f} }{I_{e,i}} &=& \frac{ (1 + \sum_j \eta_j \theta_j ) (1+  \sum_j \eta_j \epsilon_j + \sum_j \eta_j \lambda_j)^2 }{(1+ \sum_j \eta_j)^4} \label{A_many_rand_6}  
  \end{eqnarray}
 
 \noindent
where the suffix $j$ goes from 1 to $N$. In the summations  $\sum_j \eta_j$,  $\sum_j \epsilon_j$,  $\sum_j \eta_j \epsilon_j$, $\sum \eta_j \lambda_j$, $\sum \eta_j \theta_j$ the parameters $\eta_j$,  $\epsilon_j$,  $\lambda_j$, and $\theta_j$ are randomly  varied within suitable intervals by means of uniform distribution of random numbers between 0 and 1. Also in this case for $\lambda_j=0$ and  $\theta=1$ we recover the standard cases with no kinetic energy of the relative motion and no age effect on the luminosity. When considering the effect of random variations on the key parameters $\eta$, $\epsilon$,  $\lambda$,  and $\theta$  for each choice of the starting object we repeat the whole sequence  a number of times (typically 10) and retain all the results. In the following the models based on this formalism are referred to as Mod-A. 
   
\textsf{Changing the reference model step by step.} Before concluding this section, we like to present an alternative formulation of the problem which differs from the previous one from the initial hypothesis and yet leads to similar results. In the previous version the key hypothesis was: an initial object of mass $M_i$ that in all successive steps acquires a smaller object with  the same mass $M_a$, and energy smaller that those of the initial object. The mass $M_i$ and velocity dispersion $\sigma_i$ are always the reference values. 
  
In alternative we may suppose that at each merger the newly formed object becomes the reference system for the next episode. In other words, each merger is a two-galaxies event in which the parameters $\eta$, $\epsilon$ (and $\lambda$ and $\theta$) are referred to the last composed object.  These parameters must be continuously adjusted step by step to keep the perturbation small compared to the current object. This is achieved by simply dividing the parameters $\eta$, $\epsilon$, and  $\lambda$ by the current number of the step. The equations governing these models are 
  
 \begin{eqnarray}
  \frac{M_f}{M_i}               &=& (1 +  \eta) \label{Var_Ref_Gal_1} \\
  \frac{\sigma_f^2}{\sigma_i^2} &=& \frac{ (1 + \eta (\epsilon+\lambda) )}{(1+  \eta)} \label{Var_Ref_Gal_2} \\
  \frac{R_{e,f}}{R_{e,i}}       &=& \frac{(1 + \eta)^2}{ (1 +  \eta (\epsilon + \lambda) )}  \label{Var_Ref_Gal_3} \\
  \frac{\rho_{f}}{\rho{e,i}}    &=& \frac{(1 + \eta (\epsilon + \lambda) )^3}{(1+  \eta)^5} \label{Var_Ref_Gal_4}  \\
  \frac{L_f} {L_i}              &=& (1  +  \eta \theta)   \label{Var_Ref_Gal_5} \\
  \frac{I_{e,f} }{I_{e,i}}      &=& \frac{ (1 + \eta \theta) (1+ \eta (\epsilon+\lambda))^2 }{(1+  \eta)^4} \label{Var_Ref_Gal_6}  
  \end{eqnarray}

Given an initial set of values for $M_s$,  $R_e$, $\sigma $, $L$,  $I_e$, a sequence of mergers at increasing mass is made by $N$ steps indicated by $J = 1,... N$. In order to keep the perturbation small compared to the ever increasing mass we adopt the heuristic  prescription 
$\eta = \eta_r/J$, $\epsilon = \epsilon_r / J$, $\theta= \theta_r / J$  $\lambda= \lambda_r / J$,  where $\eta_r$, $\epsilon_r$, $\lambda_r$, and $\theta_r$ are the usual factors, randomly varied within suitable intervals. Models of this type are named Mod-B. 

{
\textsf{Flowchart of the models  and technicalities}. Starting from the \citet{Naab_etal_2009} idea and formalism, eqns. (\ref{A_single_1})  to (\ref{A_single_4}), we generalized it to the case of many mergers indicated by the parameter $N$, which can be any integer number from 1  to say 1000. This case is given by eqns. (\ref{A_many_1}) to (\ref{A_many_4}). The choice of $N$  is completely free, however some hint can be given by the ratio $M_f/M_i$ between the final mass we want to reach and the initial mass we started with. The parameters $\eta$ and $\epsilon$ are free but confined inside a suitable interval.  To follow the gradual variation of the galaxy properties we retain a number of intermediate models from $N$=1 to $N$=$N_{max}$, sometime we refer to these models as a ``sequence''. 

Following the same line of thought we derived the relation for $L_f/L_i$ and $I_{e,f}/I_{e,i}$. These equations can be added to  eqns. (\ref{A_single_1}) to (\ref{A_single_4}) to complete them with the expected variations of  luminosity and specific intensity, eqns. (\ref{A_many_1}) to (\ref{A_many_4}) .

Another important step is made by including the effect of the  kinetic energy carried by the merging object (parameter $\lambda$). This case corresponds to eqns. (\ref{B_many_lambda_1}) to (\ref{B_many_lambda_6}).

Finally, we suppose that the parameters $\eta$, $\epsilon$, $\lambda$ and $\theta$ instead of being rigorously constant passing from one model   to another, they may suffer  random fluctuations within some reasonable intervals of values. The new situation corresponds to eqns. (\ref{A_many_rand_1}) to (\ref{A_many_rand_6}). This is the most general case. 

All these models share the same basic feature, that is the merging object is always described in terms of the first merger of the whole sequence. This is true for the mass, the energy, and so forth. This type of model has been named Mod-A. 

A completely different point of view has been adopted for the last group of models, named Mod-B and given by eqns.(\ref{Var_Ref_Gal_1}) to (\ref{Var_Ref_Gal_6}). In this case, the initial model model is  varied along the sequence. In other words, we start with an initial galaxy characterized by the variables $M_i$, $R_i$, $\sigma_i$, etc. on which another galaxy, characterized by $M = \eta M_i$, $\sigma = \epsilon sigma_i$, etc., can merge. The step corresponds to $N$=1. The resulting object, with $M_f$, $\sigma_f$, etc., becomes the initial model for the successive step ($N$=2), and so forth at each step until a certain value of the mass or a maximum number of mergers have been reached. Along the sequence the parameters $\eta$, $\epsilon$, $\lambda$, and $\theta$ are suitably decreased so that the merging object is a small perturbation compared to the existing one. The parameters $\eta$, $\epsilon$, $\lambda$ and $\theta$ are also randomly varied within some suitable intervals passing from one step to the successive one. 
These models are the closed analytical simulation of real mergers.
}

\section{Results from the analytical merger models}\label{sec:8}
In this section we present some results of the  analytical model of mergers with the aim of testing their ability to approximate the data of real galaxies and/or the results of detailed numerical simulations. We  focus on three basic planes that can be set up with the physical quantities $L$, $M_s$, $R_e$, $\sigma$, and $I_e$: the planes in question are $I_e-R_e$, $L-\sigma$, and $R_e-M_s$. The analysis is carried out using both Mod-A and Mod-B.

The first step of the procedure is to set the initial seed from which a sequence of mergers is simulated. To this aim, we consider four models of galaxies whose evolution is calculated according to the infall formalism of \citet{Tantaloetal1998}. They are labeled according to the baryonic asymptotic mass $M_G= 10^6 M_\odot$,  $10^8 M_\odot$, $10^{10} M_\odot$, and $10^{12} M_\odot$, that is the baryonic mass reached at the present time.  The mass accretion time scale is $\tau=1$ Gyr. All other details of the models can be found in \citet{ Chiosi_Donofrio_Piovan_2023}. Suffice to recall here that they are supposed to start their evolution at redshift $z_f = 5$ in the $\Lambda-CDM$ Universe with parameters  $H_0 = 71$ \kmsM,  $\Omega_m = 0.27$, and $\Omega_\Lambda = 0.73$ age of the universe $T_u=13.67$ Gyr, age of the Universe at $z_f =5$  $T_{f} = 1.2$ Gyr, age of the galaxies at $z=0$  $T_G = 12.47$. For each sequence we mark six values of the redshift (4, 3, 2, 1, 0.5, and 0) at which the data for $M_s$, $R_e$, $\sigma$, $L_V$, $I_e$ and $M/L$   are recorded. They are the seeds of six sequences of  merger simulations. All data are listed in Table \ref{Tab_2}. 

A few comments on the data of Table \ref{Tab_2}  are mandatory. (i) First of all each model is labeled by the asymptotic baryonic mass (sum of the gas and stellar mass) at the present time. The asymptotic mass is indicated by $M_G(T_G)$ where $T_G$ is the present age of a galaxy. (ii) The baryonic mass starts from a small value at $T=0$, and increases to the final value $M_G(T_G)$. (iii) The mass of dark matter $M_D$ increases with the same rate in a proportion fixed by the ratio  $M_D/M_G \simeq  6$ set by the  adopted cosmological parameters \citep[see][]{Adams_2019}. (iv) The \citet{Fan_etal_2010} relationship is used to estimate the half-mass (luminosity) radius $R_e$.  This relation contains the ratio $M_s/M_D$ in addition to other  parameters. All of them  are largely  uncertain. Consequently,  this effective radius  $R_e$ is also highly uncertain. It may well be that the value for $R_e$  listed in Table \ref{Tab_2} is overestimated by a factor of 2. It follows from this that $I_e$ is underestimated by a factor of 4 and $\sigma$ is underestimated by a factor of $\sqrt{2}$. Drawing the relationships $\log I_e-\log R_e$, $\log L- \log \sigma$, and $\log R_e - \log M_s$,  those uncertainties should be kept in mind. The various relationships can be shifted horizontally and/or vertically by the factors -0.3  for $\log R_e$, +0.6 for $\log I_e$, and -0.15 for $\log \sigma$. Their slopes remain  unchanged. 

In each simulated sequence, a  number of steps are considered until a certain value of the final mass $M_f$ is reached. The length of each sequence is set by the parameter $N$ which may take the following values:
$N=1, 2, 3, 5, 10, 50, 100, 200, 600, 1000$. In general the minimum value for the parameter $\eta$ that determines the amount of mass  added at each step is $\eta=0.1$. The maximum value is 
$\eta=1.0$, while the typical choice is $\eta=0.2$-$0.3$.  Since the parameter $N$ can be as high as 1000, the final mass can be as high as 100 times or more than the initial value. In the case of seeds that are already massive, simulations with final masses exceeding those typical of massive galaxies, that is a few $10^{12}$, rarely $10^{13} M_\odot$, can be reached.  However, this is less of a problem because each sequence can be stopped at any value. The parameter $\epsilon$ is usually chosen from $0.1$ to $1.0$, typical choice is $\epsilon =0.2$ - $0.3$.  {The parameter 
$\lambda$ goes from $0.1$ to $0.5$.
A few cases were considered in which  $\epsilon = 0.01$ to $0.05$ and $\lambda$ goes from 0.1 to 0.5. Since the results are not of interest, they are not shown here for the  sake of brevity. } 
Finally, the parameter $\theta$ has the value $\theta=1$ when not in use, otherwise it is $1 \leq \theta  \leq 3$.  In any case, for all the models we are going to present, we indicate the interval in which each parameter is randomly varied.

\begin{figure*}      
   \centering
   {
   \includegraphics[scale=0.28]{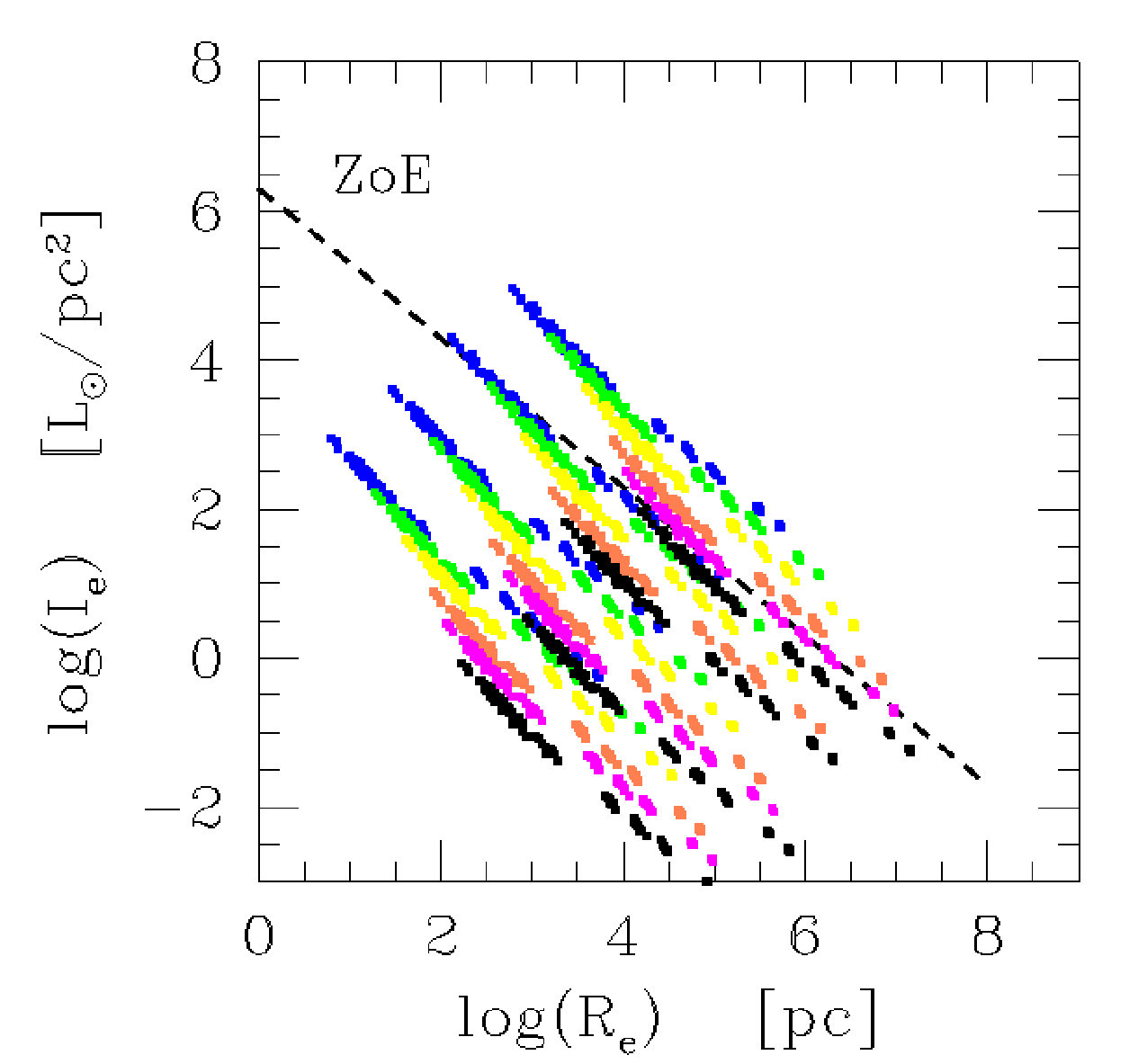}
   \includegraphics[scale=0.28]{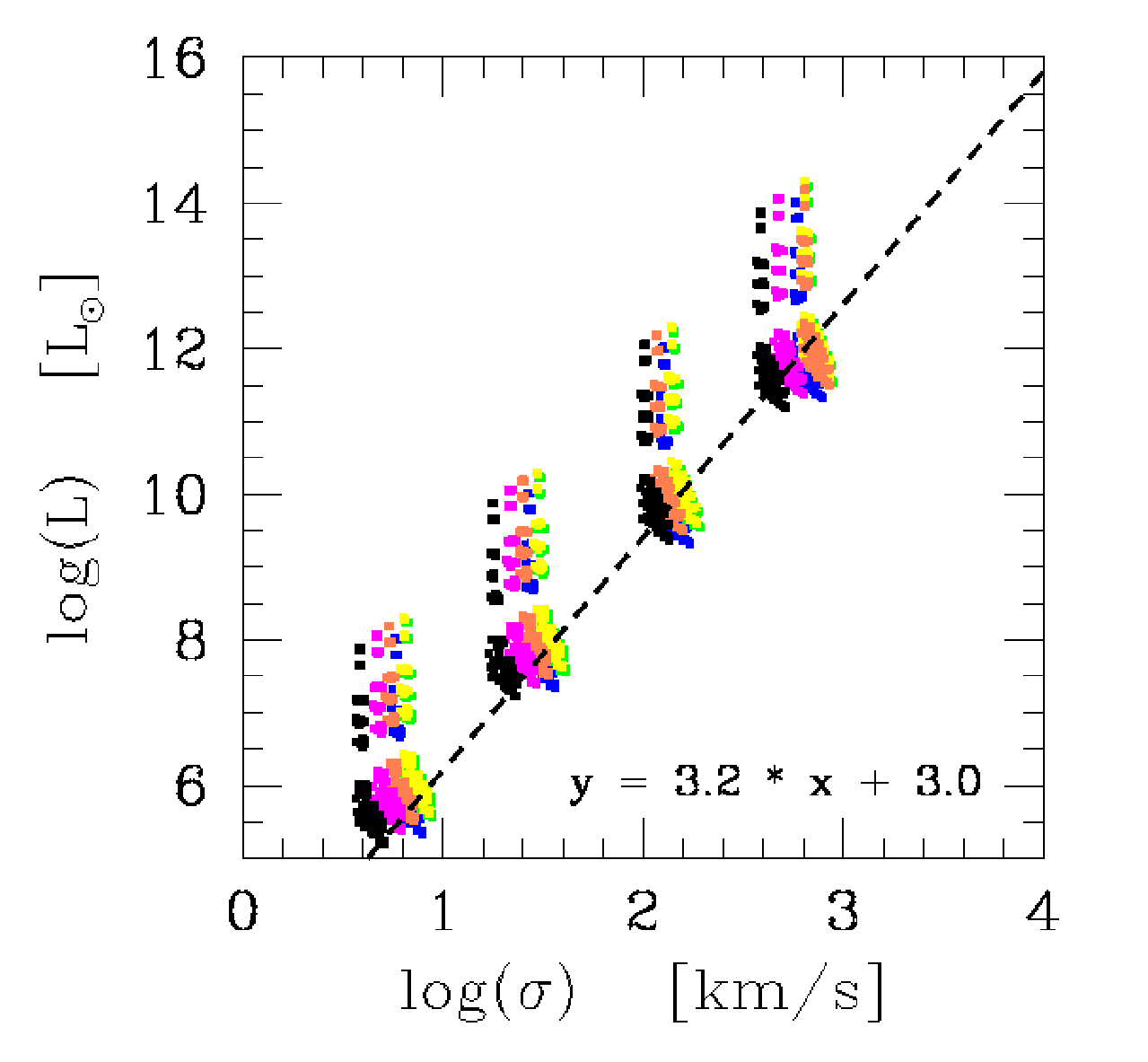} 
   \includegraphics[scale=0.28]{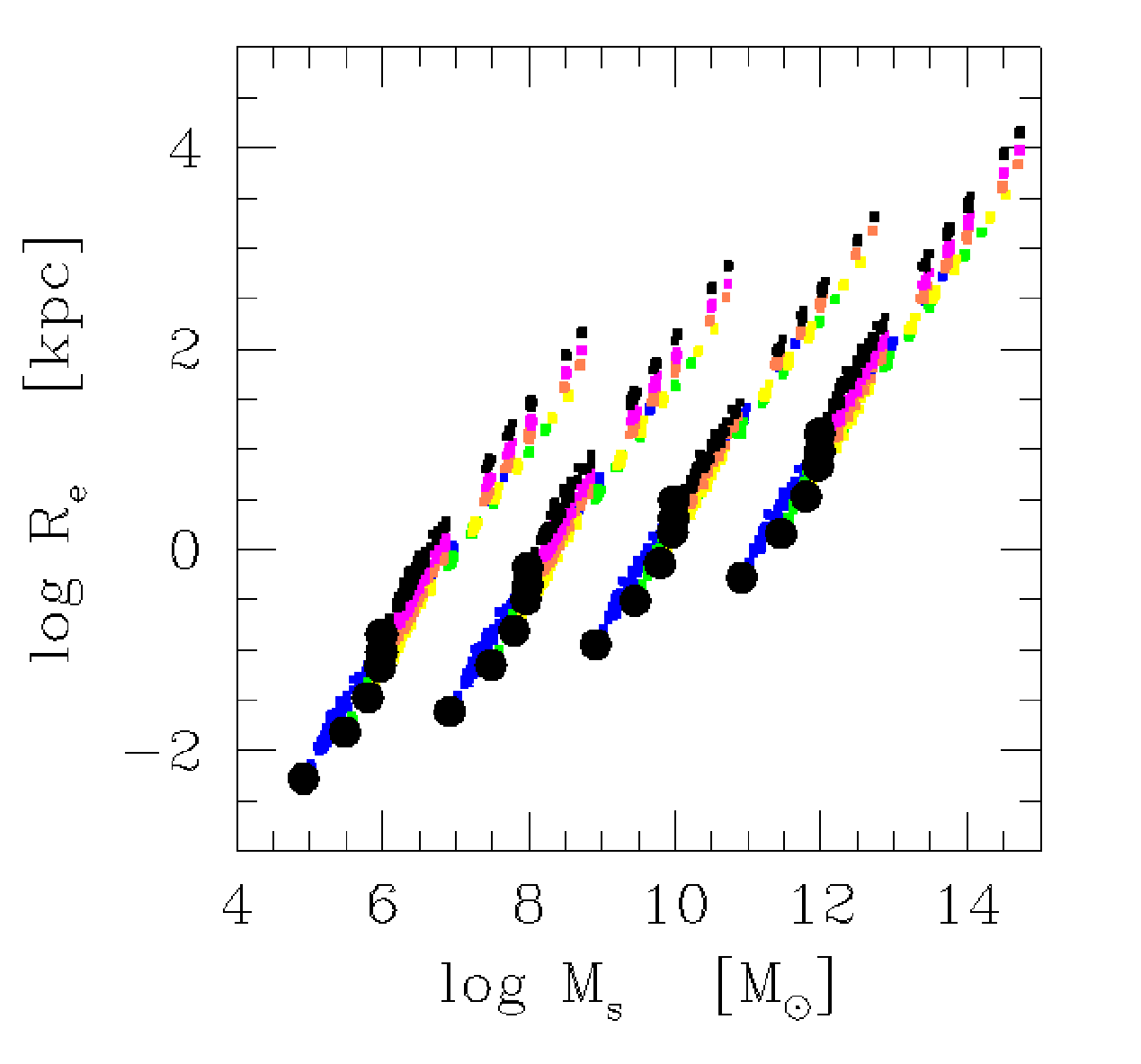}
   }
   \caption{Basic structural planes for Mod-A described by eqns. (\ref{A_many_rand_1}) to (\ref{A_many_rand_6}).  In this particular case the effect of the relative kinetic energy of the incoming object is not included ($\lambda = 0$) while the luminosity is simply proportional to the mass via the parameter $\eta$ and the luminosity effects are neglected ($\theta =1$). The parameters $\eta$ and $\epsilon$  are randomly varied in the intervals: $0.1 \leq \eta \leq 1.0 $,  $0.1 \leq \epsilon \leq 1.0$,  and finally each parameter is summed up N-times. In each panel there are four groups of lines. They correspond to the different infall models with mass $M_G$ out of which the initial seeds are taken. From the left to right $M_G = 10^6, 10^8, 10^{10}, 10^{12}\, M_\odot$. Furthermore, in each group there are  six dotted lines of different colors according to the age (i.e. redshift) at which the seeds are taken. The color code is as follows: blue (z=4), green (z=3), yellow (z=2), coral (z=1),  magenta (z=0.5), and black (z=0). All data for the seeds are listed in Table \ref{Tab_2}.   \textsf{Left Panel}: The $I_e$ versus $R_e$ plane. \textsf{Middle Panel}:  the $L_V$ versus $\sigma$ plane. The dashed line is the best fit limited to the models in four small clouds.  \textsf{Right Panel}: The $R_e$ versus $M_s$ plane. The filled circles are the original seed data.}
              \label{Mod1_3rel}
    \end{figure*}

\begin{figure*}     
   \centering
   {
   \includegraphics[scale=0.28]{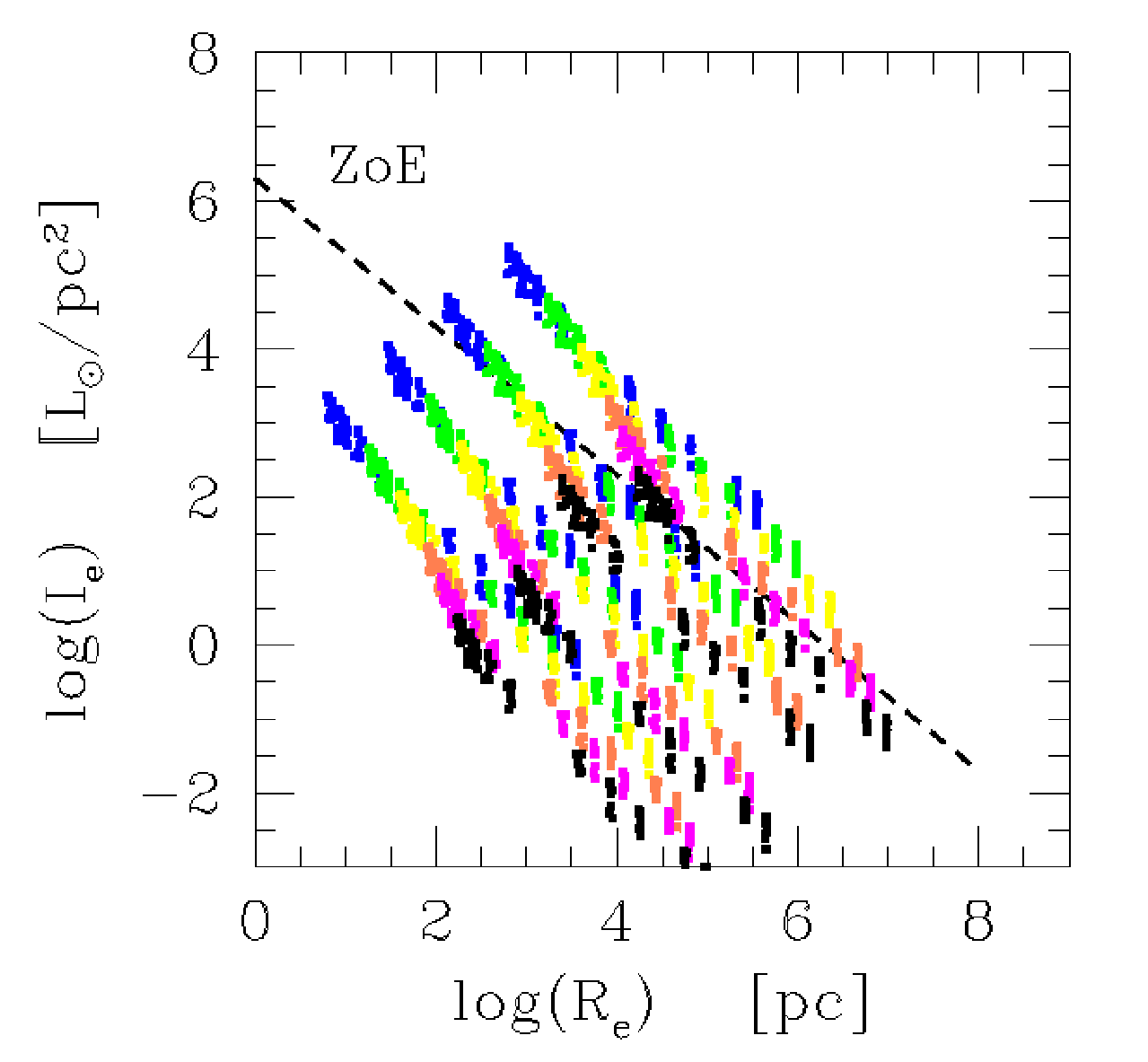}
   \includegraphics[scale=0.28]{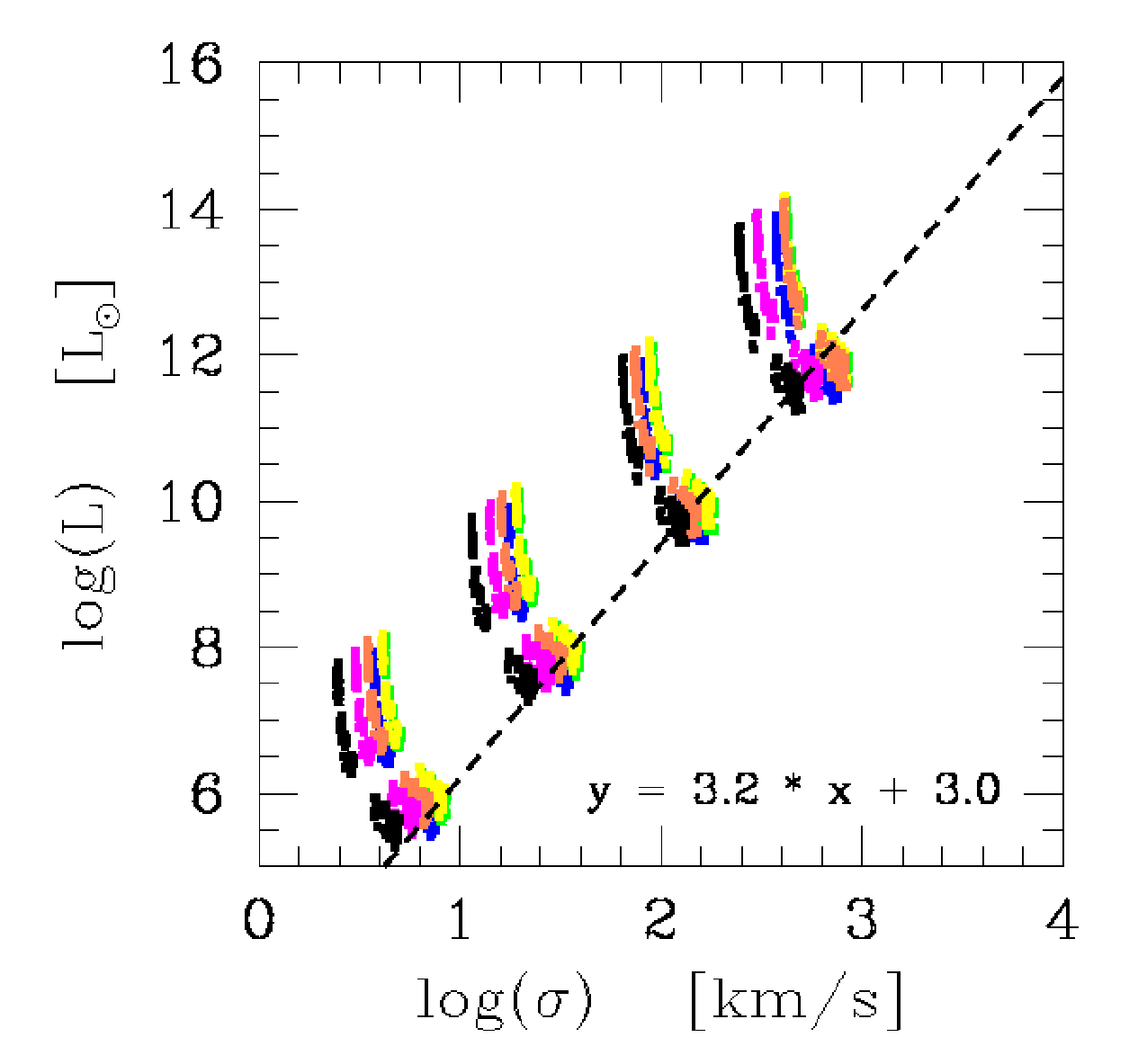} 
   \includegraphics[scale=0.28]{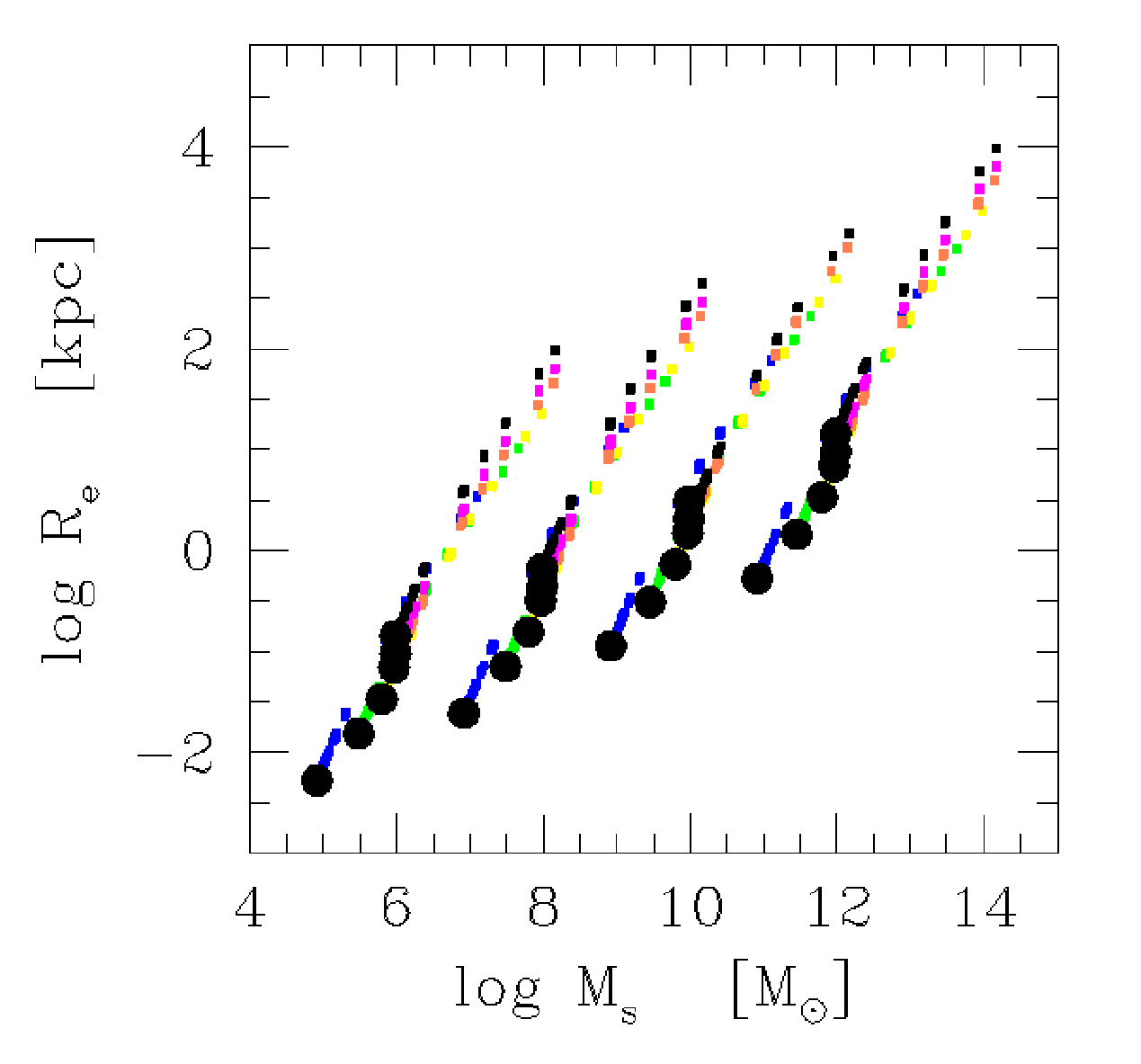}
   }
   \caption{Basic structural planes for Mod-A described by eqns. (\ref{A_many_rand_1}) to (\ref{A_many_rand_6}).  In this case, we include  the effect of the relative kinetic energy of the incoming object. The parameters $\eta$,  $\epsilon$, and $\lambda$ are randomly varied in the intervals $0.1 \leq \eta \leq 0.3 $ and $0.01 \leq \epsilon \leq 0.03$, $0.1 \leq \lambda \leq 0.3$, and summed up N-times. In addition to this, the expression for the luminosity contains the parameter $\theta $ which is randomly varied at each step in the interval $1 \leq \theta \leq 3$. 
   This parameter  is meant to simulate the effect of age of the luminosity of the incoming object. No integration on the number of steps $N$ is performed.  The meaning of the symbols and color code are as in Fig. \ref{Mod1_3rel}.  \textsf{Left Panel}: The $I_e$ versus $R_e$ plane. \textsf{Middle Panel}:  The $L_V$ versus $\sigma$ plane.  The dashed line is the best fit limited to the models in four small clouds.
   \textsf{Right Panel}: The $R_e$ versus $M_s$ plane. The filled circles are the original seed data.}
              \label{Mod2_3rel}
    \end{figure*}

\begin{figure*}         
   \centering
   {
   \includegraphics[scale=0.28]{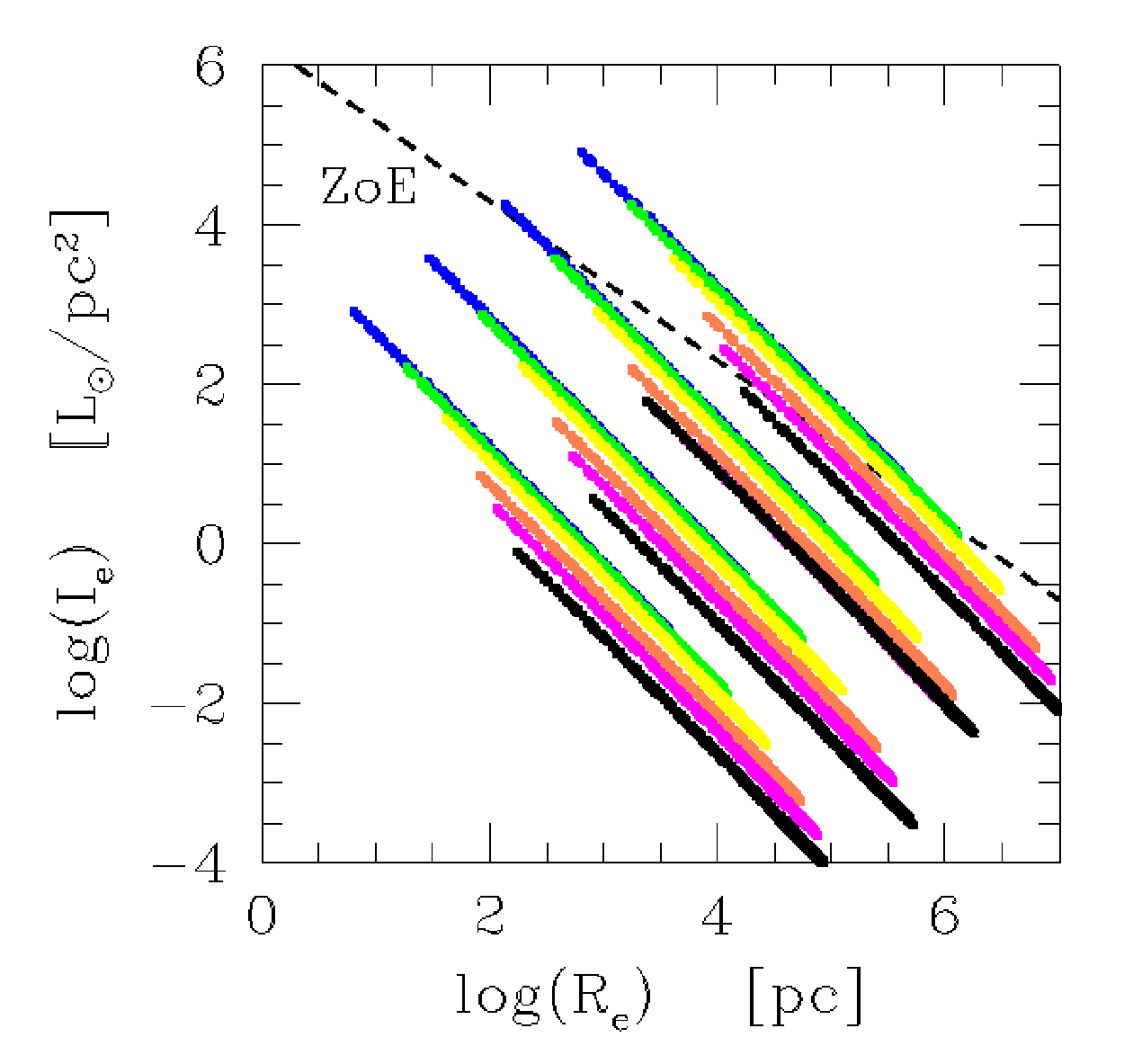}
   \includegraphics[scale=0.28]{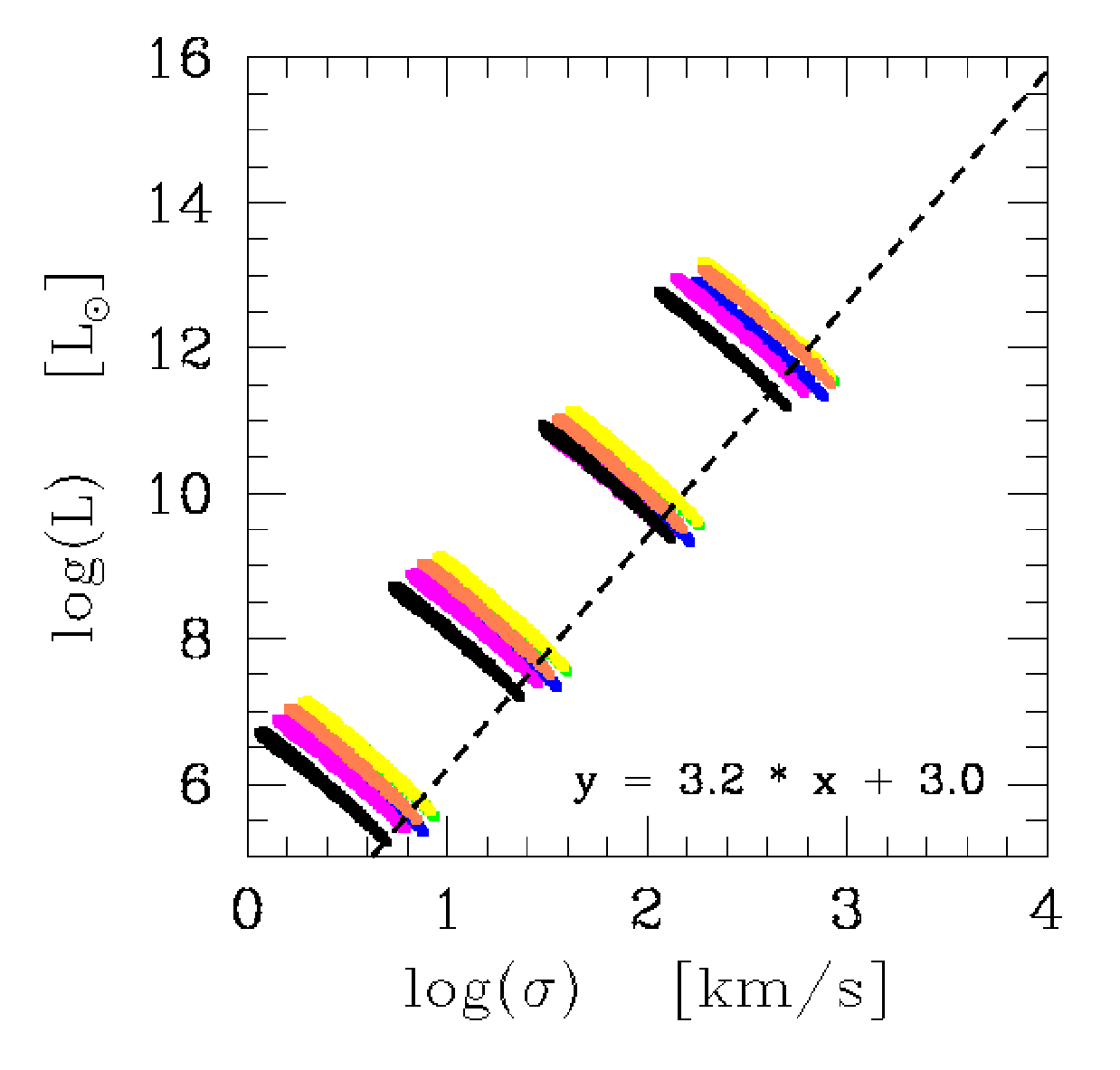} 
   \includegraphics[scale=0.28]{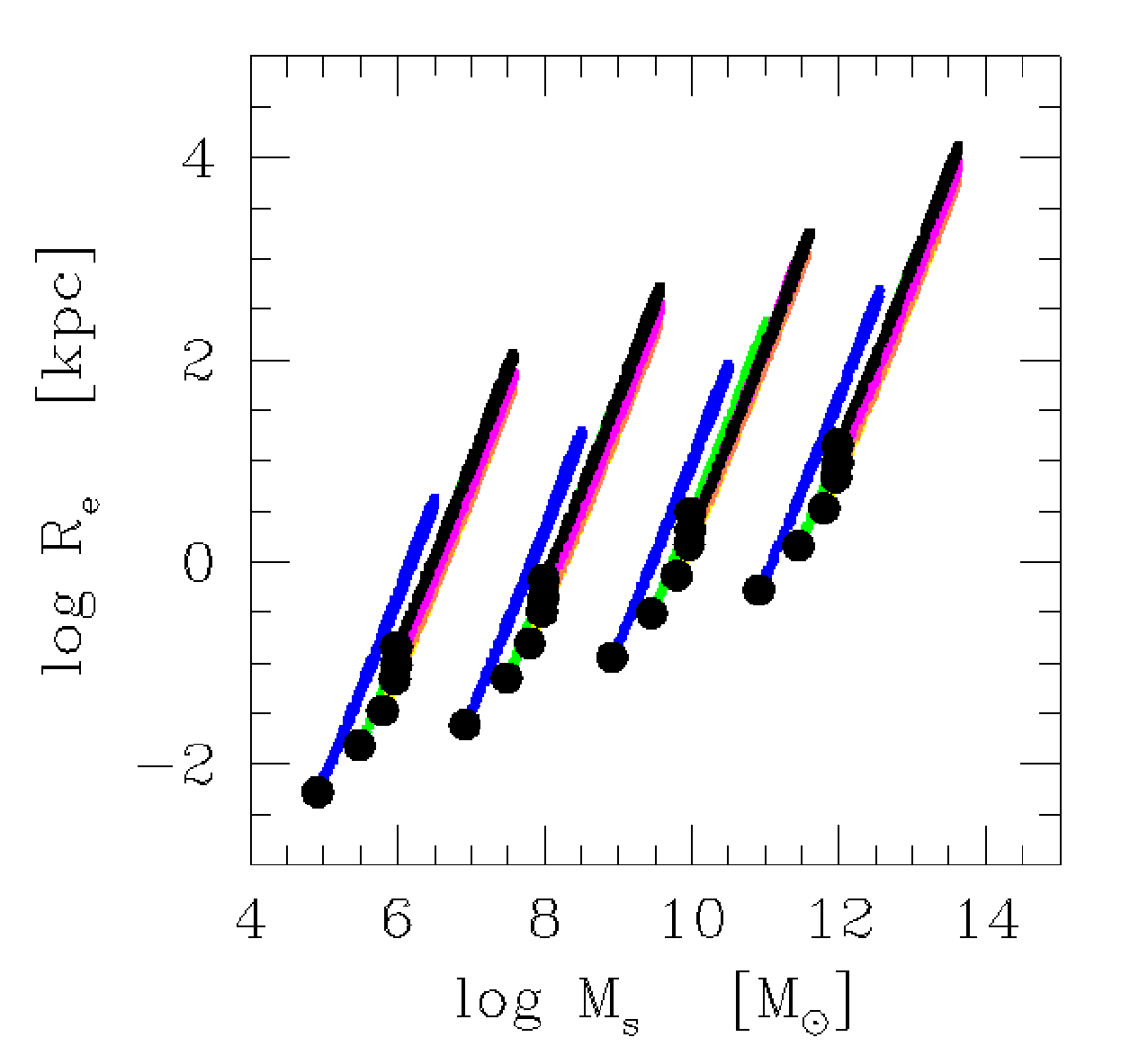}
   }
   \caption{Basic structural planes for Mod-B described by eqns. from (\ref{Var_Ref_Gal_1}) to (\ref{Var_Ref_Gal_6}).  In this particular case, we include  the effect of the relative kinetic energy of the incoming object. The parameters $\eta$,  $\epsilon$, and $\lambda$ are randomly varied in the intervals $0.1 \leq \eta \leq 0.3 $ and $0.1 \leq \epsilon \leq 0.3$ and summed up N-times. The parameter $\lambda$ is varied in the range  $0 \leq \lambda \leq 0.3$ and finally $\theta=1$ (only the effect of mass is considered). The meaning of the symbols and color code are the same of Fig. \ref{Mod1_3rel}.  \textsf{Left Panel}: The $I_e$ versus $R_e$ plane. \textsf{Middle Panel}:  the $L_V$ versus $\sigma$ plane. The dashed line is the best fit of the the previous Mod-A for comparison.   \textsf{Right Panel}: The $R_e$ versus $M_s$ plane. The filled circles are the original seed data.}
              \label{Mod3_3rel}
    \end{figure*}

\begin{figure*}         
   \centering
   {
   \includegraphics[scale=0.28]{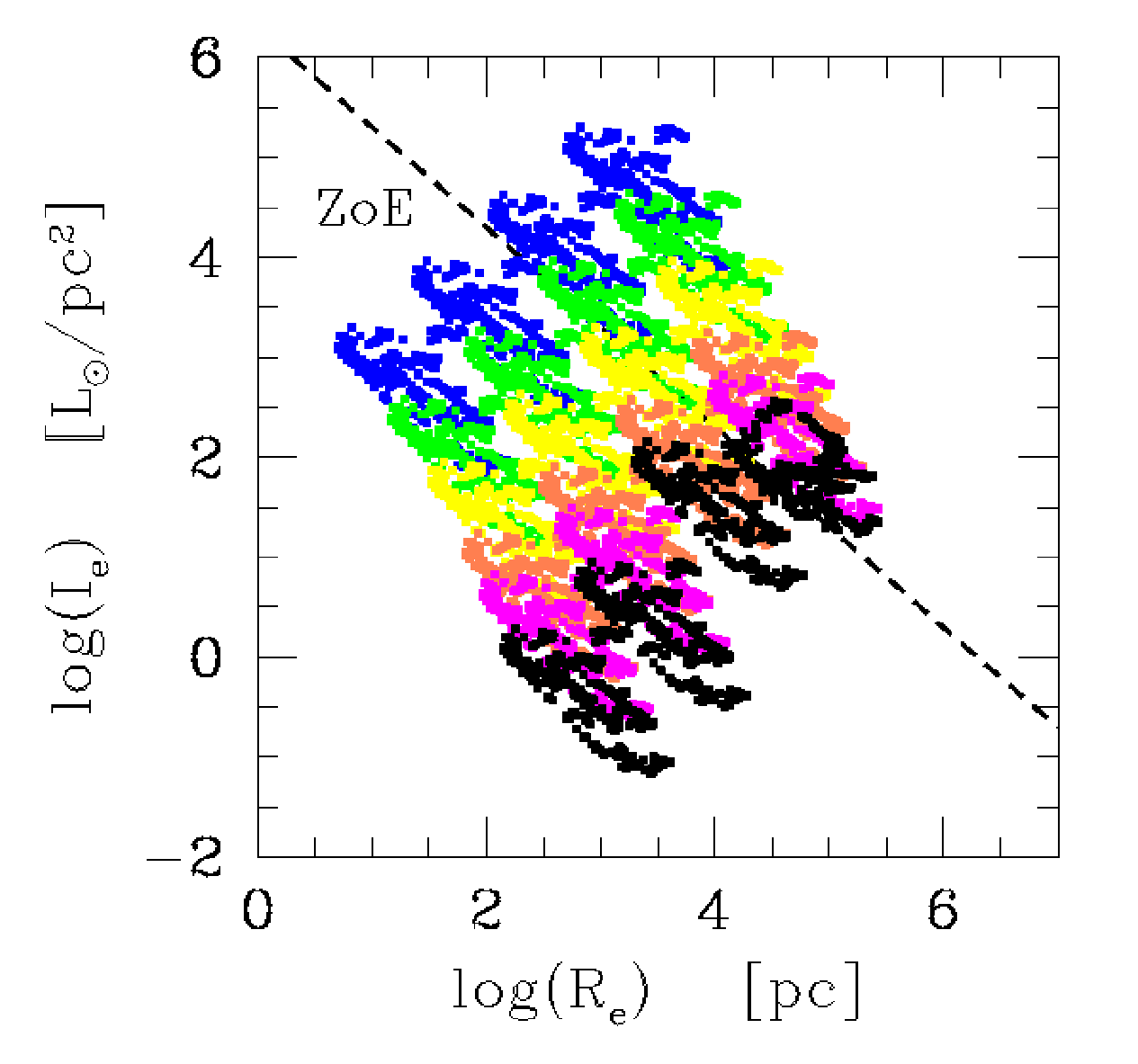}
   \includegraphics[scale=0.28]{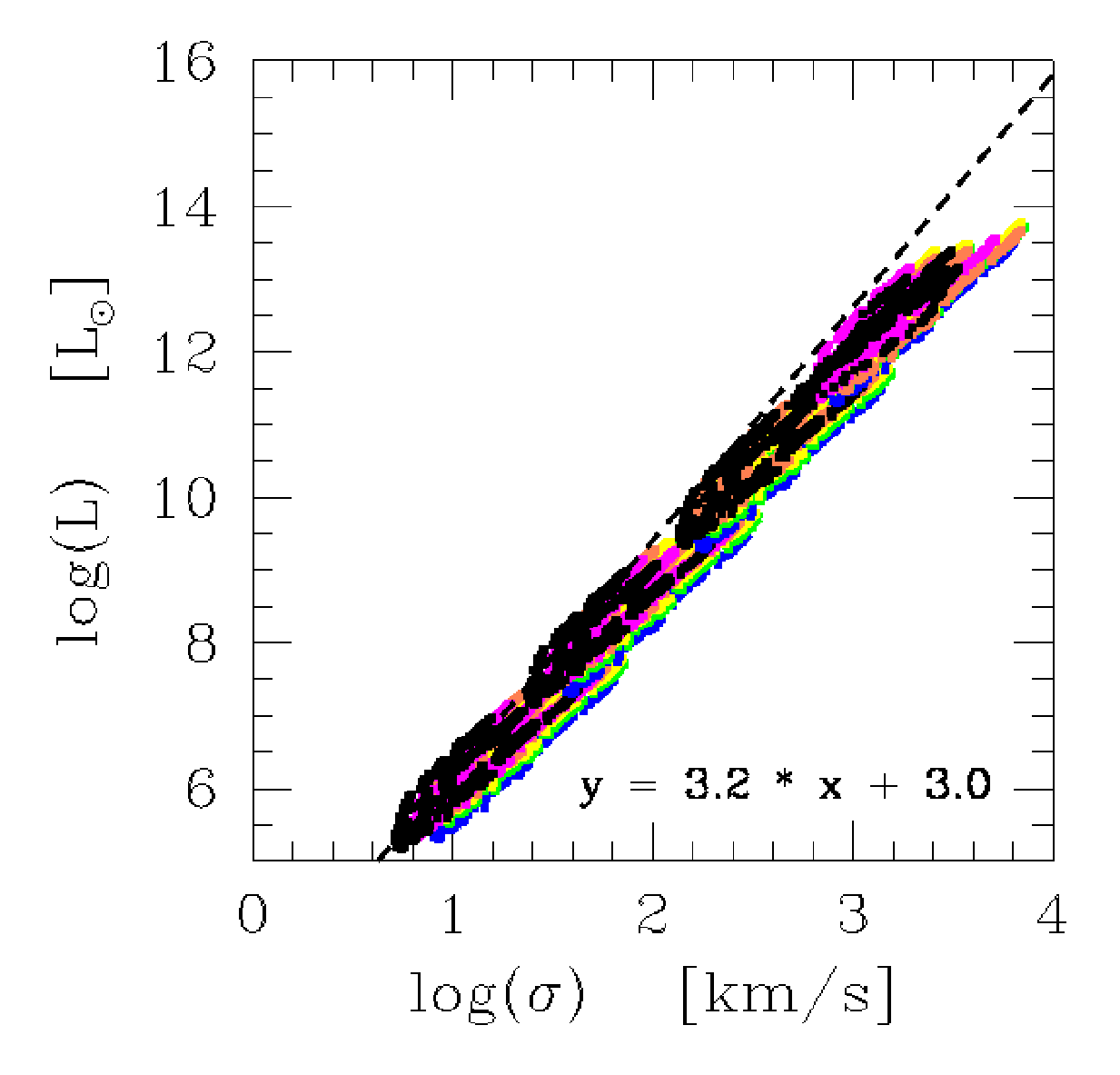} 
   \includegraphics[scale=0.28]{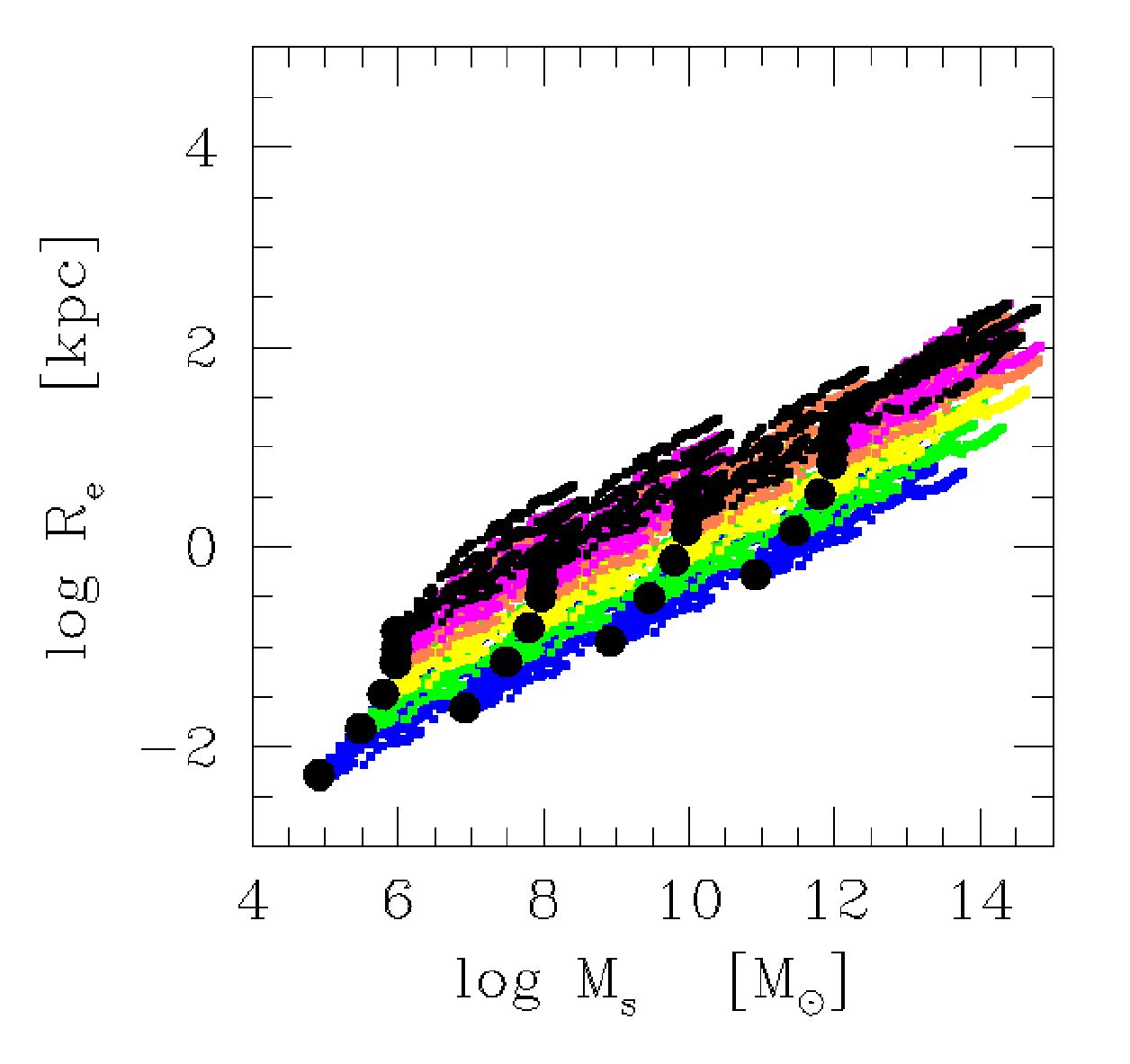}
   }
   \caption{Basic structural planes for Mod-B described by eqns. (\ref{Var_Ref_Gal_1}) to (\ref{Var_Ref_Gal_6}) in which  the  effect of the relative kinetic energy of the incoming object is much stronger than in the case of Fig. \ref{Mod3_3rel}. The parameters $\eta$,  $\epsilon$, and $\lambda$ are randomly varied in the intervals $0.1 \leq \eta \leq 1 $ and $0.1 \leq \epsilon \leq 1$ and summed up N-times. The parameter $\lambda$ is varied in the range  $0.1 \leq \lambda \leq 2$ and finally $1 \leq \theta \leq 3$. The meaning of the symbols and color code are the same of Fig. \ref{Mod1_3rel}.  \textsf{Left Panel}: The $I_e$ versus $R_e$ plane. \textsf{Middle Panel}:  the $L_V$ versus $\sigma$ plane. The dashed line is the best fit of the  previous Mod-A for comparison.   \textsf{Right Panel}: The $R_e$ versus $M_s$ plane. The filled circles are the original seed data.}
              \label{Mod4_3rel}
    \end{figure*}

In Figs. \ref{Mod1_3rel}, \ref{Mod2_3rel}, \ref{Mod3_3rel}, and \ref{Mod4_3rel} we show a few key results for Mod-A (two cases) and Mod-B (two cases), respectively. In the panels of these figures we note four groups of models corresponding to the four infall models with asymptotic mass $M_G$ out of which the seeds are taken. In each group in turn, we note models with different color code according to the age (i.e. redshift) at which the seeds are taken (see the caption of the figure). The seeds for all the cases are listed in Table \ref{Tab_2}.  

   \begin{table}
    \begin{center}
    \caption{Initial Seeds of the merger sequences. The meaning of the columns is as follows: column (1): T age in Gyr; column (2): z redshift; column (3):  logarithm of the stellar mass $M_s$in solar units; column (4): the logarithm of the effective radius $R_e$  in Kpc; column (5):  the logarithm of the velocity dispersion $\sigma)$ in km/s; column (6):   the logarithm of the luminosity $L_v$ in the V band in solar units; column (7): the  logarithm of the specific  intensity $I_{e,V}$ in  $L_\odot/ pc^2$; column (8): the  mass to light ratio $\frac{M_s}{L_V}$ in solar units.} 
    \label{Tab_2}
    \begin{tabular}{|r c  r c c  r c c|}
    \hline
 T   &   z    & $M_s$  &  $R_e$  & $\sigma$  &$L_V$   & $I_{e,V}$   & $\frac{M_s}{L_V}$ \\ 
\hline 
\multicolumn{8}{|c|}{$M_G = 10^6 M_\odot$ }\\
 \hline
  0.37 &  4.0  & 4.92 &  -2.28 &   0.90 &   5.27  &    3.03 &  -0.35 \\
  1.03 &  3.0  & 5.49 &  -1.81 &   0.95 &   5.49  &    2.32 &  -0.00 \\
  2.16 &  2.0  & 5.80 &  -1.47 &   0.94 &   5.55  &    1.69 &   0.25 \\
  4.74 &  1.0  & 5.96 &  -1.16 &   0.86 &   5.44  &    0.97 &   0.52 \\
  7.42 &  0.5  & 5.99 &  -1.02 &   0.80 &   5.31  &    0.56 &   0.68 \\
 12.47 &  0.0  & 5.99 &  -0.84 &   0.71 &   5.13  &    0.02 &   0.86 \\    
\hline
\multicolumn{8}{|c|}{$M_G = 10^8 M_\odot$ }\\
 \hline
  0.37 &  4.0  & 6.92 &  -1.61 &   1.57 &   7.27  &    3.70 &  -0.35 \\
  1.03 &  3.0  & 7.49 &  -1.15 &   1.62 &   7.49  &    2.99 &  -0.00 \\
  2.16 &  2.0  & 7.80 &  -0.80 &   1.60 &   7.55  &    2.36 &   0.25 \\
  4.74 &  1.0  & 7.96 &  -0.50 &   1.53 &   7.44  &    1.64 &   0.52 \\
  7.42 &  0.5  & 7.99 &  -0.35 &   1.47 &   7.31  &    1.22 &   0.68 \\
 12.47 &  0.0  & 7.99 &  -0.18 &   1.38 &   7.13  &    0.69 &   0.86 \\
\hline
\multicolumn{8}{|c|}{ $M_G = 10^{10} M_\odot$ }\\
 \hline
  0.37 &  4.0  & 8.92 &  -0.95 &   2.23 &   9.27  &    4.37 &  -0.35 \\ 
  1.03 &  3.0  & 9.46 &  -0.51 &   2.24 &   9.48  &    3.70 &  -0.02 \\
  2.16 &  2.0  & 9.80 &  -0.14 &   2.27 &   9.55  &    3.03 &   0.25 \\
  4.74 &  1.0  & 9.96 &   0.17 &   2.20 &   9.44  &    2.31 &   0.52 \\
  7.42 &  0.5  & 9.99 &   0.31 &   2.14 &   9.31  &    1.89 &   0.68 \\
 12.47 &  0.0  & 9.99 &   0.49 &   2.05 &   9.13  &    1.35 &   0.86 \\
\hline
\multicolumn{8}{|c|}{$M_G = 10^{12} M_\odot$ }\\
 \hline
  0.37 &  4.0  &10.92 &  -0.28 &   2.90 &  11.27  &    5.03 &  -0.35 \\
  1.03 &  3.0  &11.46 &   0.15 &   2.95 &  11.48  &    4.37 &  -0.02 \\
  2.16 &  2.0  &11.80 &   0.53 &   2.94 &  11.55  &    3.69 &   0.25 \\
  4.74 &  1.0  &11.96 &   0.83 &   2.86 &  11.44  &    2.97 &   0.52 \\
  7.42 &  0.5  &11.99 &   0.97 &   2.80 &  11.31  &    2.56 &   0.68 \\ 
 12.47 &  0.0  &11.99 &   1.15 &   2.71 &  11.13  &    2.02 &   0.86 \\
\hline 
\end{tabular}
\end{center}
\end{table}

Inspection of the results  presented in this study and many others not shown here for the sake of brevity allow us to say that:

i) First of all, we emphasize that even if the four cases presented here look alike as far as the general trend of the various relationship, in reality there are important differences that depend on the choice of values assigned to each parameter. Since there are no specific arguments for preferring a particular value with respect to others, we expect  a large variety of  results to be  possible. This is indeed what happens in nature where mergers occur among  galaxies of different mass, radius, ages, chemical properties, luminosities, energy budget, evolutionary state,  and so forth.   In any case, our choices are made following the general criterium  of minimal variations.
 
ii) We note that along the lines corresponding to the same  galaxy mass $M_G$ from which the  seeds are derived, the mass $M_f$ increases with $N$ according to the rule $M_f{(N)} = M_i\times (1+ N \eta)$. Depending on $\eta$ (typically $\simeq 0.2$) and maximum value $N=1000$, we get the  maximum mass  $M_f(1000) \simeq  200 M_i$. 
This implies that starting from a initial stage belonging to an object of low mass, say $M_G \simeq 10^ 6 M_\odot$, successive mergers cannot increase the total mass up to  the value typical of high mass galaxies, say a few $10^{12} M_\odot$. Indeed, they hardly go beyond the above limit. One can envisage the following picture: after spontaneous collapse of initial bodies made of baryons and dark mass that are nicely described by the infall models  and whose mass may span up to two or three orders of magnitude, at a certain time of their life, mergers with similar bodies of smaller ($\eta < 1$) or equal mass ($\eta =1$) may start; the case with $\eta > 1$ is similar to the first one in which the roles of the two merging objects are interchanged. For obvious reasons, the case with $\eta \ge 1$ are statistically less frequent. The process goes on so that bigger and bigger galaxies are in place. In this picture the most advantageous cases are mergers of equal mass. 

iii) The variations of luminosity caused by a merger are more difficult to predict. In a galaxy the luminosity is the sum of the partial luminosities emitted by stars of different generations and different chemical composition. Each generation contributes in a way proportional to its mass, age, and chemical composition. The mass of a stellar generation is proportional to the rate of star formation at the time of its birth. In a single stellar generation, the luminosity decreases with the age because of its natural fading as the turnoff mass gets smaller and smaller. {For the aims of this study, we can simplify 
this complicated situation assuming that the mix of stellar populations of different chemical compositions and age can be approximated to a single stellar population of suitable mean chemical composition (indicated here as <SSP>, the mass of which equals the total stellar mass of the galaxy. The chemical composition is indicated in the usual notation by means of the mass abundance of Hydrogen (X), Helium (Y) and all other heavy elements lumped together (Z). By definition, X, Y and Z satisfy the condition X+Y+Z = 1. In our case, we can further assume that the mean chemical composition we refer to is close to the solar value, that is [X=0.700, Y=0.28, Z=0.02]. This <SSP> is used to evaluate the time evolution of the luminosity of a  galaxy by natural fading caused by the aging of its stellar content. More details on the source of SSPs can be found in \citet[][and references]{Donofrio_Chiosi_2024}.
From the data of Fig. \ref{masslum} {(left panel)}, we note a fast decline during the first 1.5 to 2 Gyr after the start of star formation, followed by a smooth decline during the rest of the life by a factor of $\theta=3$ to 5. We adopted $\theta=3$.  }

At the time of merger, the two galaxies may have different mass, age, and mean chemical composition,  and therefore different luminosities. In absence of additional star formation at the merger of galaxies A and B, the resulting luminosity will be $L = (M_{s,A} L_A + M_{s,B} L_B) /(M_{s,A} + M_{s,B})$. In analogy, if a new generation of stars is created at the merger with total mass $M_{s,C}$ and luminosity $L_C$, the resulting new luminosity will be  $L = (M_{s,A} L_A + M_{s,B} L_B + M_{s,C} L_C)  /(M_{s,A} + M_{s,B} + M_{s,C})$. In our {present models}, the  bursting star formation at the merger is missing. This is a feature of the model to be improved. 

iv) In the $I_e$ vs $R_e$ plane all the groups of models draw lines that are nearly parallel to the ZoE, confirming that they correspond to loci along which virial equilibrium is obtained (imposed). The scatter of the models along each line is due to the scatter of the parameters $\eta$, $\epsilon$, $\lambda$, and $\theta$ when active.  

v) In general, variations of $\epsilon$ and $\lambda$ shift the lines for assigned $M_G$ along the direction perpendicular to the ZOE. An  increase of these parameter corresponds to a shift from the bottom left  corner to the  upper right one. It may also happens that  a slight variation of the slope of the virial loci toward the ZoE  occurs in the region of massive galaxies.

vi) The $L$ versus $\sigma$ plane is more intriguing. As expected the gross behavior is that the luminosity increases with the mass. However, looking at the particular history of a galaxy through its merger sequence, the possibility  arises that the luminosity increases while the velocity dispersion decreases. This is the result of the energy exchange among successive  mergers. Similar result is found and discussed by \cite{Naab_etal_2009}. The issue will be examined  in more detail later on. In any case a large dispersion in this relationship is expected.

vii) The $R_e$ versus $M_s$ relation is expected to increase with the mass both from galaxy to galaxy and also along the merger sequence of each object. 

viii) The results for each case depend on the input parameters $\eta$, $\epsilon$, $\lambda$, and $\theta$ but the general trend is preserved at least for reasonable choices of their values.

ix) In general, there is not much variation of these relationships  with  the age of the simulated galaxies and implicitly  the redshift at which galaxies are observed. 

x)  The theoretical models reasonably correspond to and reproduce the observational data. However, to be used for this purpose, one has to know the relative number of objects of different mass $M_G$ and an estimate of the number of mergers that each object of a certain mass can undergo.  It is reasonable to assume that in the low mass galaxies  only a limited number of mergers can occur, while in the most massive ones  a larger number of mergers is plausible. Based on this argument, we suppose that the number of merger increases from 10 for a $10^6 M_\odot$ to 1000 for a $,10^{11} M_\odot$. With aid of this we randomly populate the three planes above by a number of simulated galaxies seen at $z=0$ after their last merger episode.    

xi) By construction,  models of this type cannot deal with the temporal history of galaxy assembling and star formation  because there is no time in the equations governing their structure. This drawback of the models could be cured by estimating the probability that in the course of time  mergers can occur among objects of typical masses and sizes confined in certain volume of space. The issue is beyond the aims of this study.

To conclude, the model we are proposing can be considered as a proxy of the merger mechanisms of galaxy formation as far as masses and sizes and other quantities associated to these are concerned. In the following, we will compare real galaxies to these mock galaxies to interpret various diagnostic planes such as those examined in this study. To this aim, we compare the results of the analytical model to the same  observational samples of galaxies (MaNGA and WINGS at $z=0$) as we did for the EAGLE and TNG100 simulations. What we learn from using the analytical merger model and the comparison with the observational data is that even a simple model of galaxy mergers can be useful to investigate the deep physical reasons determining the observed behavior of real galaxies in the diagnostic planes, projections of the FP, in terms of a unified description.

\section{Comparison of our model with the observational data}\label{sec:9}  

\begin{figure*}            
   \centering
  { \includegraphics[scale=0.45]{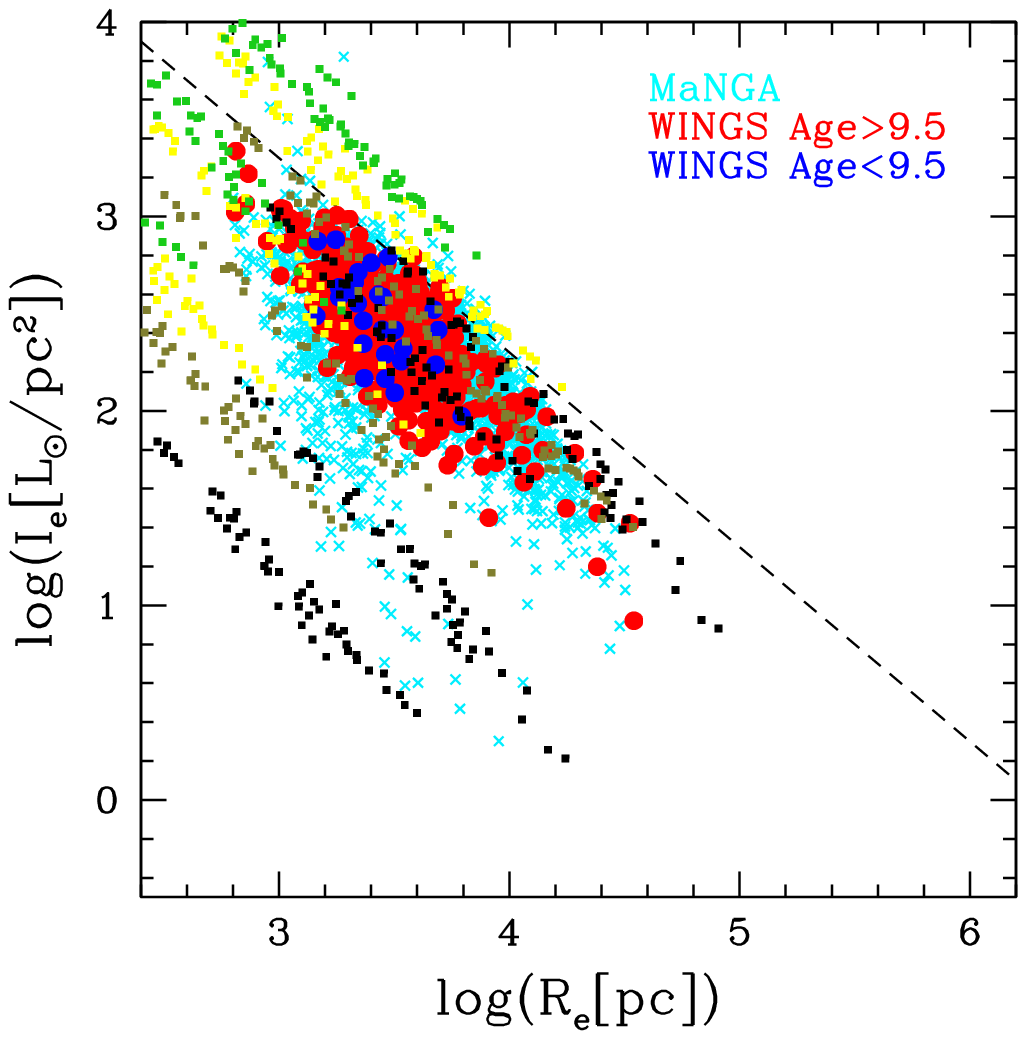}
     \includegraphics[scale=0.45]{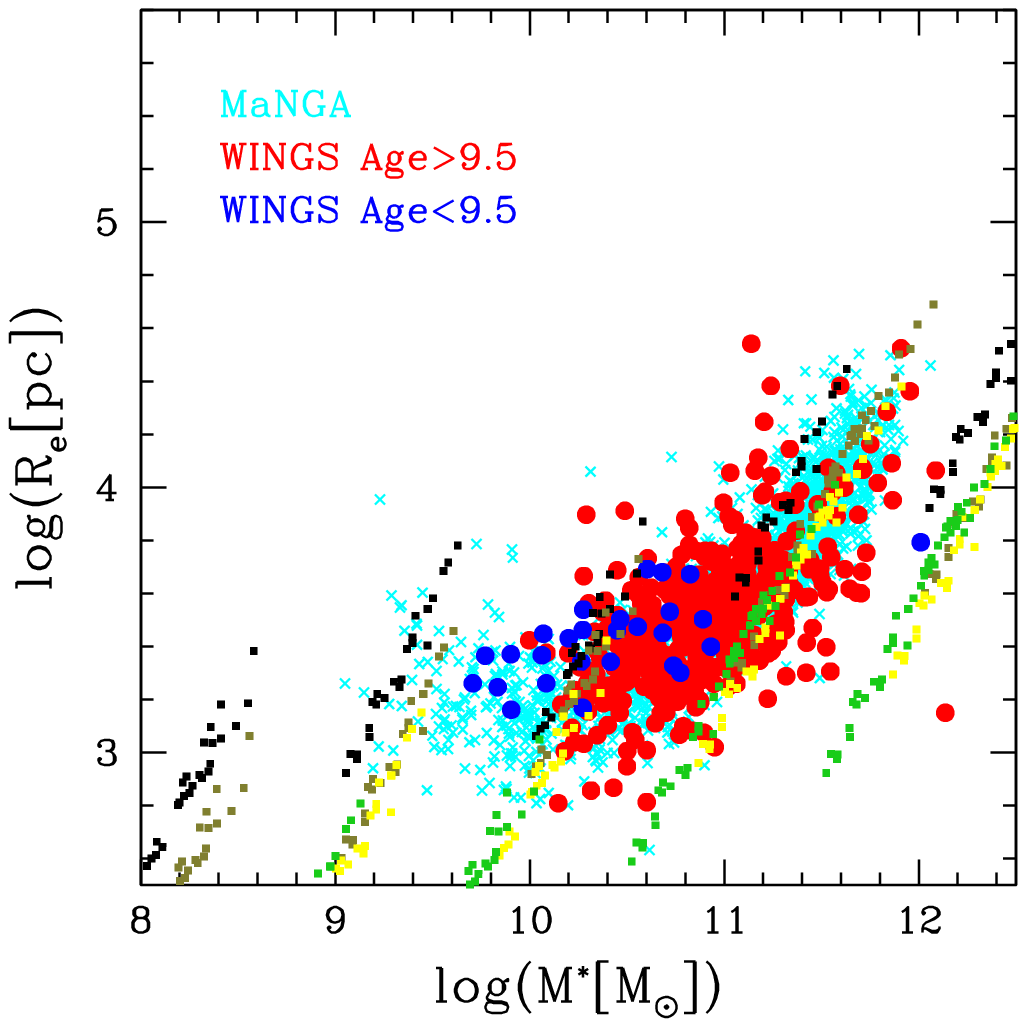} }
   \caption{
      Comparison of simulated galaxies of Mod-B described by eqns. (\ref{Var_Ref_Gal_1}) to (\ref{Var_Ref_Gal_6} with the observational data of MaNGA \citep[see][and references therein]{Blanton_etal_2017} and WINGS, both are described in the text. Only E and S0 galaxies are considered here.
       \textsf{Left Panel}:  the \IeRe\ plane where the sky-blue dots are the MaNGA data, while red and blue filled circles the WINGS galaxies of different ages as indicated. The black squares are our simulated mergers starting at z=0. Similarly, the colored dots mark mergers at increasing redshift: coral ($z=1$), yellow ($z=2$) and green ($z=3$).
       The dashed line is the ZoE.
       \textsf{Right Panel}: the \MRa\ plane of the same data. In the two panels,
       the parameters $\eta$,  $\epsilon$, and $\lambda$ are randomly varied in the intervals $0.1 \leq \eta \leq 0.3 $ and $0.1 \leq \epsilon \leq 0.3$, $0 \leq \lambda \leq 0.3$. In addition to this, the expression for the luminosity contains the parameter $\theta $ which is randomly varied at each step in the interval $1 \leq \theta \leq 3$. The value of N changes for the massive and less massive galaxies.  The meaning of the symbols and color code are as in Fig. \ref{Mod1_3rel}}
              \label{Fig:9}
    \end{figure*}

Fig. \ref{Fig:9} shows  in the left panel the \IeRe\ plane already displayed in Fig. \ref{Fig:1} with superposed our mock galaxies of type Mod-B in which  a series of mergers from initial seeds of different ages and mass are considered. The parameters of this models are $0.1 \leq \eta \leq 0.3$, $0.1 \leq \epsilon \leq 0.3$, $0 \leq \lambda \leq 0.3$, and  $0.1 \leq \theta = 1.0$. 
This means that in the luminosity only the effect of the increasing mass is considered while that of the age and star formation at the merger are neglected. There is no particular reason for this choice of the parameters. Other similar values would yield similar results. The value of N decreases  from the massive to less massive galaxies, to more closely mimic the different merger history of galaxies with different mass.
The black dots indicate objects whose seed is taken at at the present time while the colored dots indicate models whose seeds are taken at younger ages (i.e. higher redshift). The color code is as follows: coral ($z=1$), yellow ($z=2$) and green ($z=3$). The observational data are the MaNGA sample (sky-blue dots) and the WINGS sample split according to the age (red for $\log Age > 9.5$ and blue for 
$\log Age < 9.5$). 

The figure indicates that the hierarchical scenario in which galaxies {are  built} up by successive mergers is compatible with the observational data. The $\Lambda$-shaped distribution of the galaxies is well reproduced, in particular the long tail formed by the massive galaxies and the clouds of low mass objects. This result can be obtained by limiting the number of possible mergers in galaxies of different mass:  a large number of mergers across the whole galaxy age for the more massive objects and fewer  mergers for the low mass ones. In this way  it is possible to get a very good reproduction of the distribution in the \Ie\ plane. The choice of the number of mergers is based on simple theoretical arguments of statistical nature; we simply say that the more massive galaxies can merge many more galaxies than the less massive objects.

Similarly, the right panel of Fig. \ref{Fig:9} indicates that the same simulations can in principle reproduce also the  \MRa\ plane. They indeed  easily reproduce the tail of the massive galaxies and the cloud-shaped distribution  of the smaller ones. The large dispersion of the observational data can be  also accounted  for. 

\begin{figure*}       
   \centering
  { \includegraphics[scale=0.45]{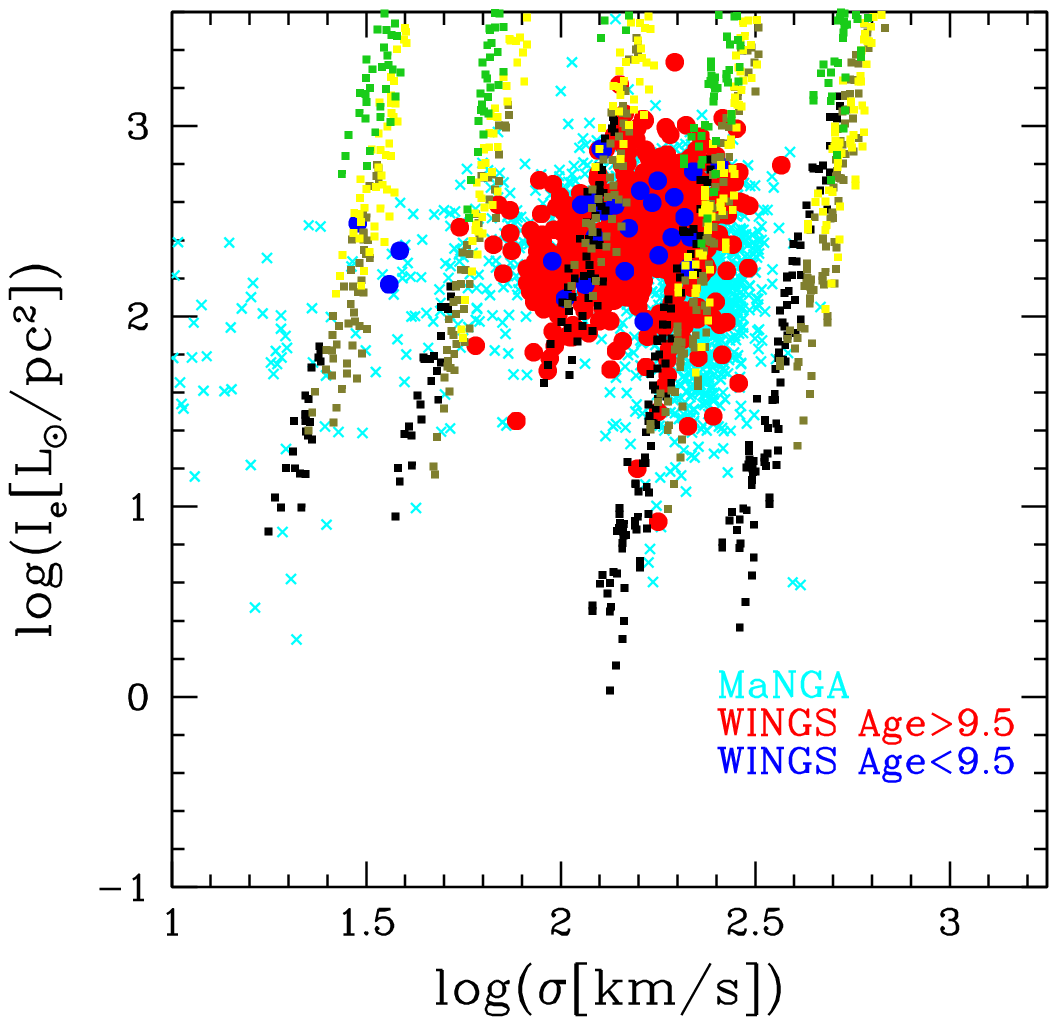}
      \includegraphics[scale=0.45]{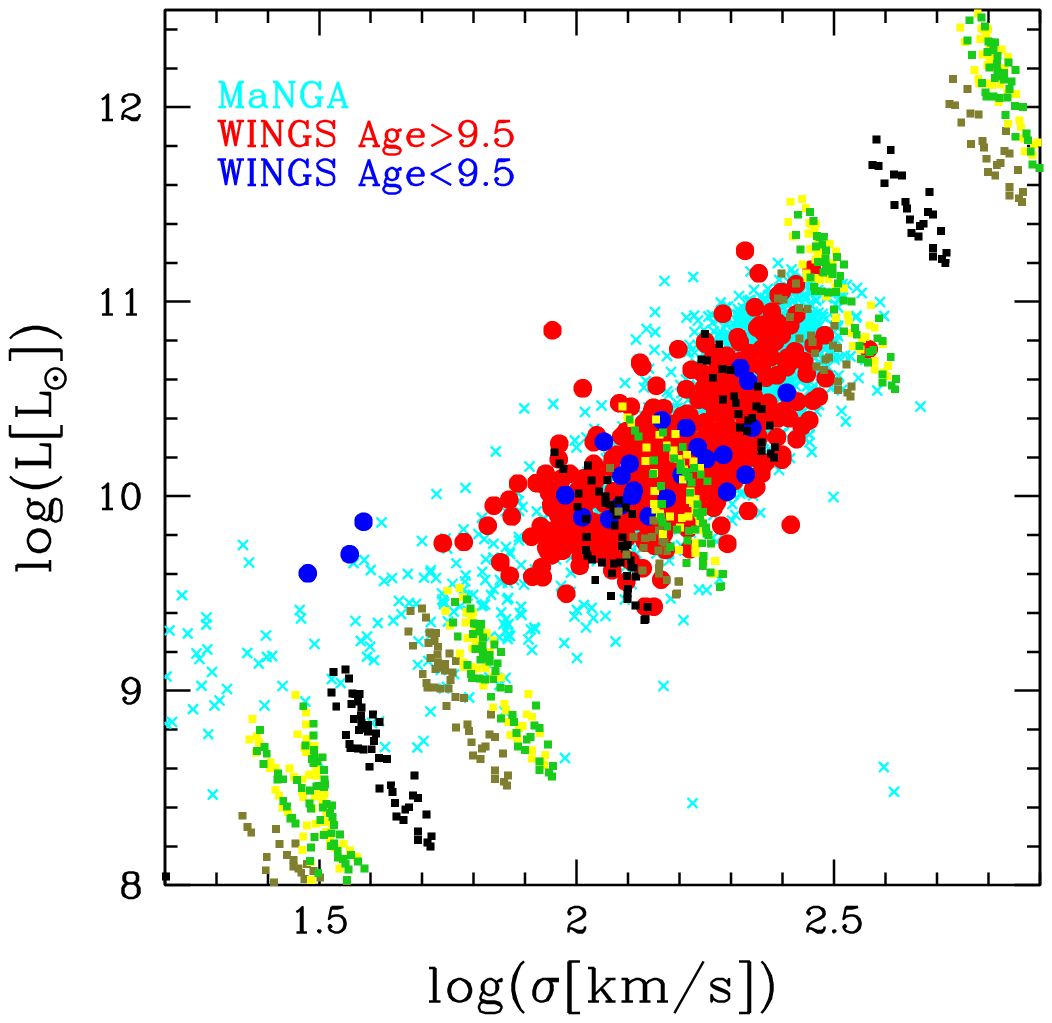} }
   \caption{
   Comparison of simulated galaxies of Mod-B described by eqns. (\ref{Var_Ref_Gal_1}) to (\ref{Var_Ref_Gal_6}) with the observational data of MaNGA \citep[see][and references therein]{Blanton_etal_2017} and WINGS both described in the text. Only E and S0 galaxies are considered.
   \textsf{Left Panel}:  the \IeSig\ plane.  \textsf{Right Panel}: The \Lsig\ plane with the MaNGA and WINGS data, and our artificial galaxy mergers. In the two  panels, the color code and meaning of symbols are  the same of Fig. \ref{Fig:9}.
   The parameters $\eta$, $\epsilon$, $\lambda$, and $\theta$ have the same values as in Fig. \ref{Fig:9}.   }
              \label{Fig:10}
    \end{figure*}

The same can be said for the relationships shown in the panels of  Fig. \ref{Fig:10}, that  is the \IeSig\ plane (left panel) and the \Lsig\ plane (right panel). Here, the same limits in the number of mergers has been applied. The values for the parameter $\eta$, $\epsilon$, $\lambda$ and $\theta$ are the same as in Fig.\ref{Fig:9}.  

The \Lsig\ plane (and also the $M_s -\sigma$ plane not shown here) deserves some comments and explanations. As already mentioned when describing the  \Lsig\ plane of Figs. \ref{Mod1_3rel},
\ref{Mod2_3rel} for two Mod-A, and \ref{Mod3_3rel} and \ref{Mod4_3rel} for the two Mod-B, the luminosity is expected to increase with the velocity dispersion $\sigma$ as amply confirmed by the observational data, see the classical Faber-Jackson relation, \Lsiga\ with $L_0$ and $\alpha$ positive. The velocity dispersion is derived from the VT in which the radius $R_e$  can be replaced by $R_e \propto M_s^{0.33}$ \citep[see][]{Tantaloetal1998};  to a first approximation, it follows that $\sigma^2 \propto M_s^{0.67}$. According to a typical mass to luminosity ratio for old galaxies, $L \propto M_s$. Therefore $L \propto \sigma^{3}$. The luminosity should only increase with $\sigma$. The real slope derived from observational data is about 3.5. The results plotted in Figs. \ref{Mod1_3rel},
\ref{Mod2_3rel},  \ref{Mod3_3rel} and \ref{Mod4_3rel} show that the situation is somewhat more   complicated, that is two regimes are possible: at increasing mass the luminosity increases. However, for  a system in mechanical equilibrium, that is subjected to the VT, the velocity dispersion may either decrease, if there is not enough energy to disposal, or increase when the amount of the brought-in energy is large enough. In our models, given the mass of the captured object ($\eta$), there are two possible sources of energy expressed by the parameters, $\epsilon$ (the amount of internal energy brought-in by the captured object ) and $\lambda$ (the amount of kinetic energy brought-in by the captured object).  In general, the radius of the composite system is larger or equal to that of the hosting object. If there little energy to disposal (typical of small $\epsilon$ and/or $\lambda$), in order to reach the equilibrium configuration, part of the      
energy must be taken from the dispersion velocity which decreases. If $\epsilon$ and $\lambda$ are large enough, the energy required to reach the equilibrium state is taken from the kinetic energy of the infalling object and the velocity dispersion may increase. It goes without saying that all intermediate situations are possible. This effect explains  the large dispersion seen in the observational \Lsiga\ relationship. 

Similar results are obtained using our Mod-A simulations, indeed they  produce the same trends. What does  change is only the number of mergers required to obtain a given mass.

The comparison of Mod-A and Mod-B with the real observational data  shows that our simple model is able to obtain a distribution of galaxies in the FP projections very similar to that obtained with the more sophisticated numerical simulations. The main features of the ScRs are reproduced, such as slopes, scatter and ZoE. 

\section{Our model in the context of the $\beta-L_0'$ theory}
\label{sec:9_beta_theory}

{There is a last step to undertake here. In a series of papers,  \citet{Donofrioetal2017,Donofrioetal2019,Donofrioetal2020, DonofrioChiosi2021,Donofrio_Chiosi_2022,Donofrio_Chiosi_2023a,Donofrio_Chiosi_2023b, Donofrio_Chiosi_2024} proposed a new way of reading the diagnostic planes, projections of the FP of galaxies, in terms of two parameters, $\beta(t)$ and $L'_0(t)$ of the \Lsigbtempo\ relation which was already presented in the introduction and is now repeated here for the sake of completeness

\begin{equation}
L(t)  = L'_0(t) \sigma(t)^{\beta(t)}. 
\label{lum_sig}
\end{equation}

The parameters $L'_0(t)$ and $\beta(t)$ were found to describe the distribution of galaxies in the various diagnostic planes and to predict the path of a galaxy in any of these planes in the course of evolution, by the natural changes of the various physical quantities defining a galaxy in the FP. } We refer to it as the I$\beta-L_0'$ theory. In this section first we refresh the foundations of the $\beta-L_0'$ theory,  derive the $\beta$ and $L'_0$ for the galaxies of the MaNGA and WINGS samples as well as for our model mergers, and finally compare these results to check  whether they are mutually compatible.

In the following we drop the explicit notation of time dependence for the sake of  simplicity. The two equations representing the VT and the \Lsigb\ law are:

\begin{eqnarray}
 \sigma^2 &= & \frac{G}{k_v} \frac{M_s}{R_e}   \\
 \sigma^\beta &= & \frac{L}{L'_0} = \frac{2\pi I_e R^2_e}{L'_0}. 
\label{eqsig}
\end{eqnarray}

The first equation is the VT itself, where $G$ is the gravitational constant, and $k_v$ a term that gives the degree of structural and dynamical non-homology. The presence of $k_v$ allows us to write $M_s$ (stellar mass) instead of the total mass $M_T$. $k_v$ is  a function of the S\'ersic index $n$, that is $k_v=((73.32/(10.465+(n-0.94)^2))+0.954))$ \citep[see][for all details]{Bertinetal1992, Donofrioetal2008}. In the two equations, all other symbols have their usual meaning. 
In these equations, $\beta$ and $L'_0$ are time-dependent parameters that depend on the peculiar history of each object. 

From these equations one can derive all the mutual relationships existing among the parameters $M_s$, $R_e$, $ L$, $I_e$, $\sigma$ characterizing a galaxy as a function of the $\beta$ parameter. Although they may be of some interest also in the context of this paper,  they are not repeated here. They can be found in \citet{Donofrio_Chiosi_2024}.
It suffices to recall that in all these relationships,  the slopes   depend  on $\beta$ while the proportionality coefficients, in addition to other structural parameters characterizing a galaxy, depend also on $L'_0$. This means that when a galaxy changes its luminosity $L$,  and velocity dispersion $\sigma$, and has a given value of  $\beta$ (either positive or negative), the effects of this change in the \Lsig\ plane are propagated in all other projections of the FP.  In these planes the galaxies cannot move in whatever directions, but are forced to move only along the directions (slopes) predicted by the $\beta$ parameter in the above equations. In this sense the $\beta$ parameter is the link between the FP space and the observed distributions in the FP projections.
In addition to this, it is possible to derive an equation holding for each single galaxy that looks like the classical FP. Combining eqs. (\ref{eqsig}), it is possible to write a FP-like equation involving the parameters $\beta$ and $L'_0$:

\begin{equation}
    \log R_e = a \log\sigma + b <\mu>_e + c
    \label{eqfege}
\end{equation}

\noindent
where  $<\mu_e>$ is the mean surface brightness $<I_e>$ expressed in magnitudes and the coefficients:

\begin{eqnarray}
a & = & (2+\beta)/3 \\ 
b & = & 0.26 \\ 
c & = & -10.0432+0.333(-\log (G/k_v) - \log (M/L)  \nonumber \\
  &   & -2\log (2\pi)-\log (L'_0)) 
\end{eqnarray}

\noindent
are written in terms of $\beta$ and $L'_0$. We note that this is the equation of a plane whose slope depends on $\beta$ and the zero-point on $L'_0$. The similarity with the FP equation is clear. The novelty is that the FP is an equation derived from the fit of a distribution of real objects, while here each galaxy independently follows an equation formally identical to the classical FP, but of profoundly different physical meaning. In this case, since $\beta$ and $L'_0$ are time dependent, the equation represents the instantaneous plane on which a generic galaxy is located in the FP space and consequently in all its projections.

Finally, the equation system (\ref{eqsig}) allows us  to determine the values of $\beta$ and $L'_0$, the two basic evolutionary parameters. Let us  write the above equations in the following way:

\begin{eqnarray}
\beta [\log(I_e)+\log(G/k_v)+\log(M_s/L)+\log(2\pi)+\log(R_e)] +  \nonumber   \\
    + 2\log(L'_0) - 2\log(2\pi) - 4\log(R_e) = 0 \\ 
 \beta\log(\sigma) + \log(L'_0) + 2\log(\sigma) + \log(k_v/G) - \log(M_s) +  \nonumber  \\
 - \log(2\pi) - \log(I_e) - \log(R_e) = 0. 
\label{eqbet}
\end{eqnarray}

\noindent
Posing now: 

\begin{eqnarray}
A  & = & \log(I_e)+\log(G/k_v)+\log(M_s/L)+\log(2\pi)+ \\ 
   &   & \log(R_e)  \\ 
B  & = & - 2\log(2\pi) - 4\log(R_e)  \\ 
A' & = &  \log(\sigma)  \\ 
B' & = & 2\log(\sigma) - \log(G/k_v) - \log(M_s) - \log(2\pi) -   \nonumber \\
   &   & \log(I_e) - \log(R_e)  
   \label{eq3}
\end{eqnarray}
we obtain the following system:

\begin{eqnarray}
A\beta + 2\log(L'_0) + B = 0 \\
A'\beta + \log(L'_0) + B'= 0
\label{eqsyst}
\end{eqnarray}
\noindent 
with solutions:

\begin{eqnarray}
 \beta & = & \frac{-2\log(L'_0) - B}{A} \\ 
 \log(L'_0) & = &\frac{A'B/A - B'}{1-2A'/A}.
 \label{eq2}
\end{eqnarray}

\begin{figure*}    
   \centering
   {
   \includegraphics[scale=0.43]{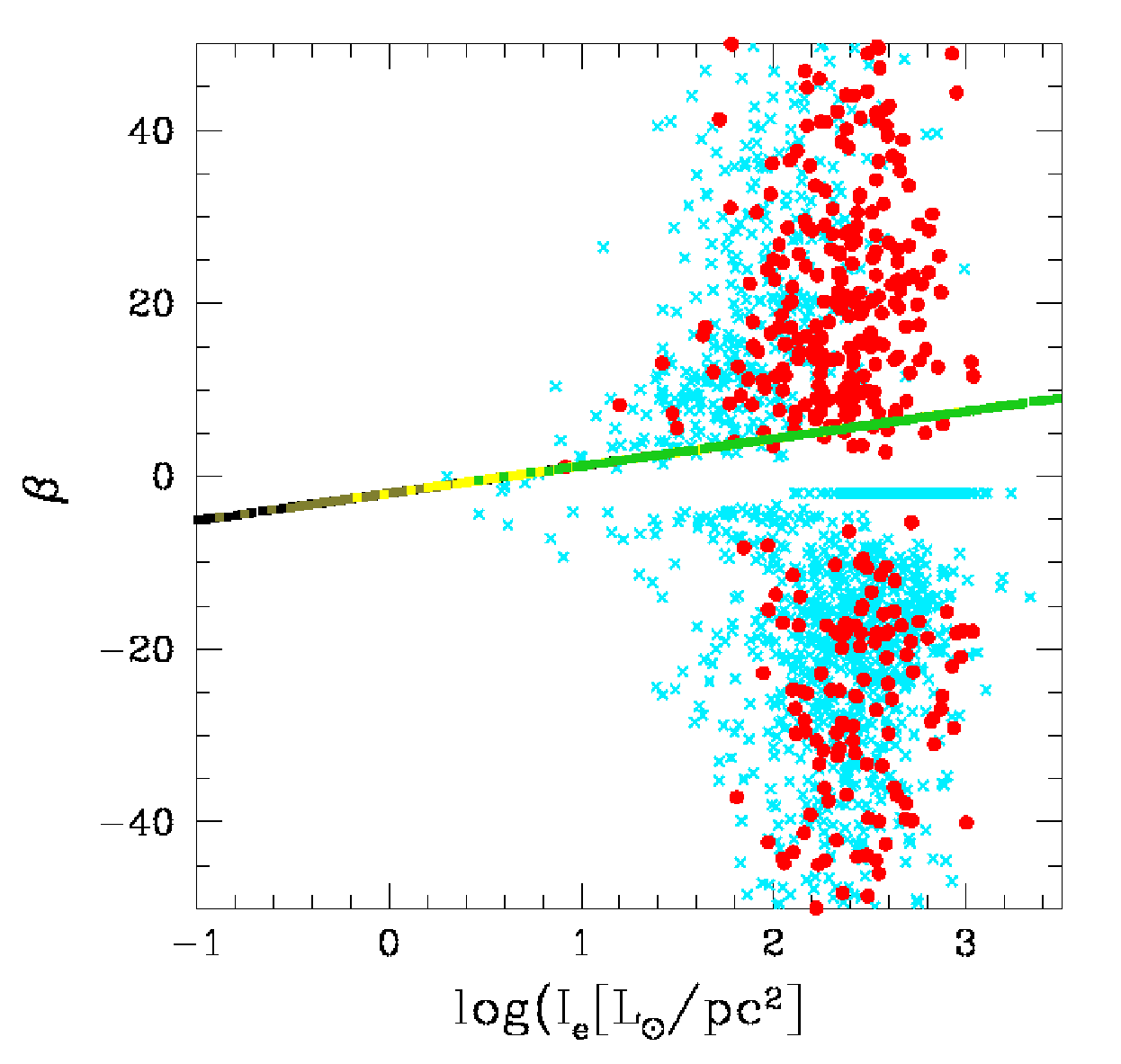}
   \includegraphics[scale=0.43]{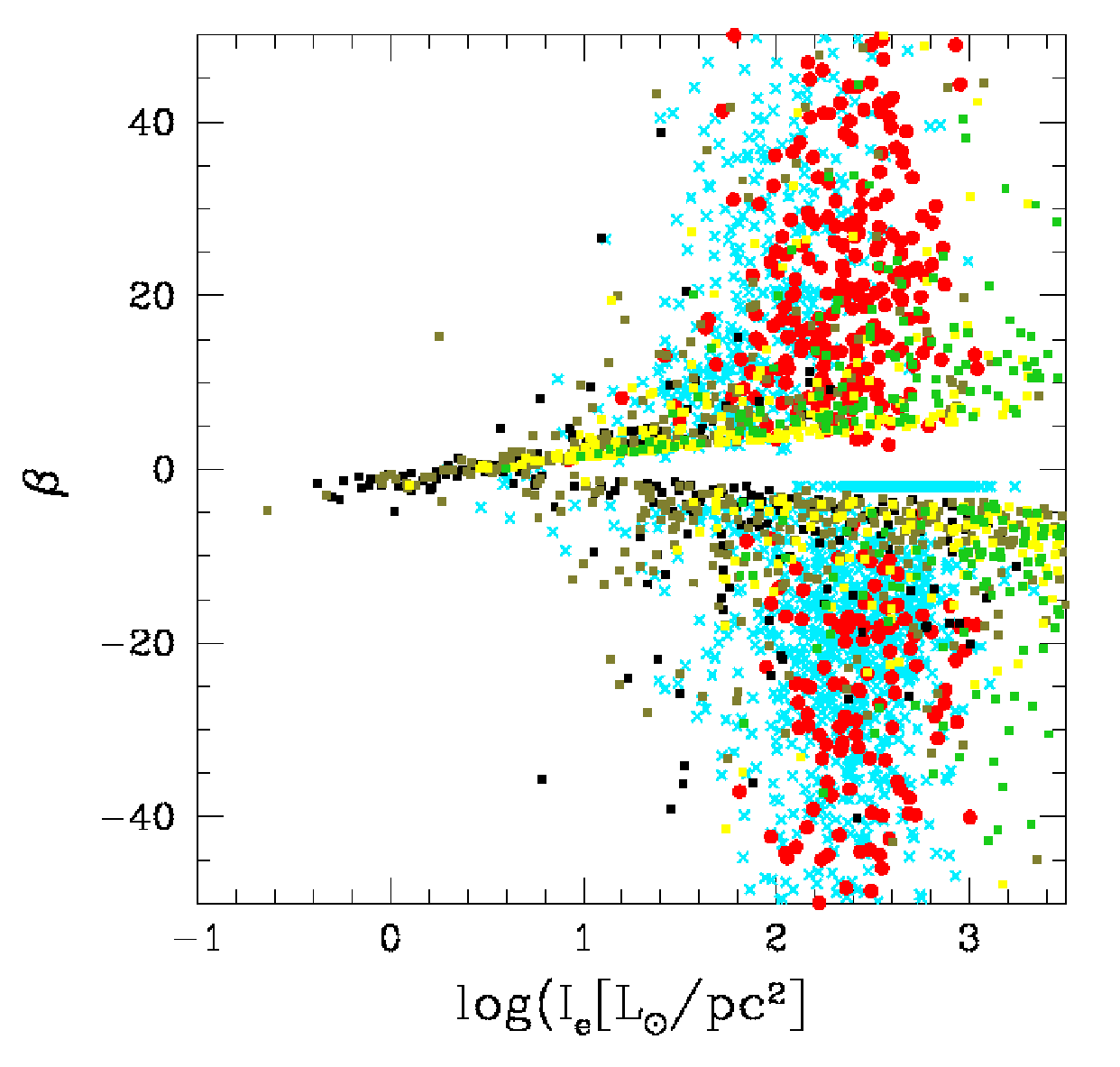} }
   \caption{ Parameters $\beta$  for the MaNGA and WINGS data and the models of type Mod-B. \textsf{Left Panel}: the $I_e-\beta$ plane for the MaNGA data (sky-blue crosses) and WINGS (red dots) as well as for our merger models of  type Mod-B (filled squares of different colors according to the age (i,e. redshift) at which the seeds are taken: blue (z=4), green (z=3), yellow (z=2), coral (z=1), and black (z=0), i.e. the same color-code {of Fig. \ref{Mod1_3rel}. All the models fall along the same straight line which coincides with low $\beta$-limit of data in the positive semi-plane. } \textsf{Right Panel}: the same plane after perturbing the theoretical data (see text). The meaning of the symbols and color code are the same as in the left panel.}
              \label{Fig:11}
    \end{figure*}

The key result is that the  parameters $L$, $M_s$, \re, \Ie\ and $\sigma$ of a galaxy fully determine its evolution in FP that is encoded in the parameters $\beta$ and $L'_0$.
    
Figure \ref{Fig:11} (left panel) provides a comparison between the values of $\beta$ and $I_e$ for the MaNGA and WINGS galaxies and our artificial simulations already presented in Figs. \ref{Fig:9} and \ref{Fig:10}. It is clear that the trend shown by simulations does not fit the more general distribution of $\beta$ and $I_e$ shown by the observational data, in which $\beta$ can assume either positive and negative values. The parameter $\beta$ is very low at very low surface brightness and progressively acquires large values (both negative and positive) for the high surface brightness galaxies. In contrast all the models fall on a single straight line which runs along the border of the distribution of real galaxies in the positive semi-plane. Similar results was found by \citet{Donofrio_Chiosi_2023a} in the case of infall models.  

Before proceeding further, it is worth commenting on the gap of triangular shape void of galaxies falling in between the distribution on the positive and negative semi-planes. It is easy to check that this gap corresponds in the $I_e$-$R_e$ plane to objects of relatively low surface brightness $I_e$ (and also luminosities) and intermediate radii $R_e$ that do not exist in our observational sample. They should correspond to the so-called diffuse galaxies that are difficult to identify (and perhaps are rare in the sky). So the gap is not a problem here and is left aside. 

\begin{figure}    
   \centering
   {
   \includegraphics[scale=0.40]{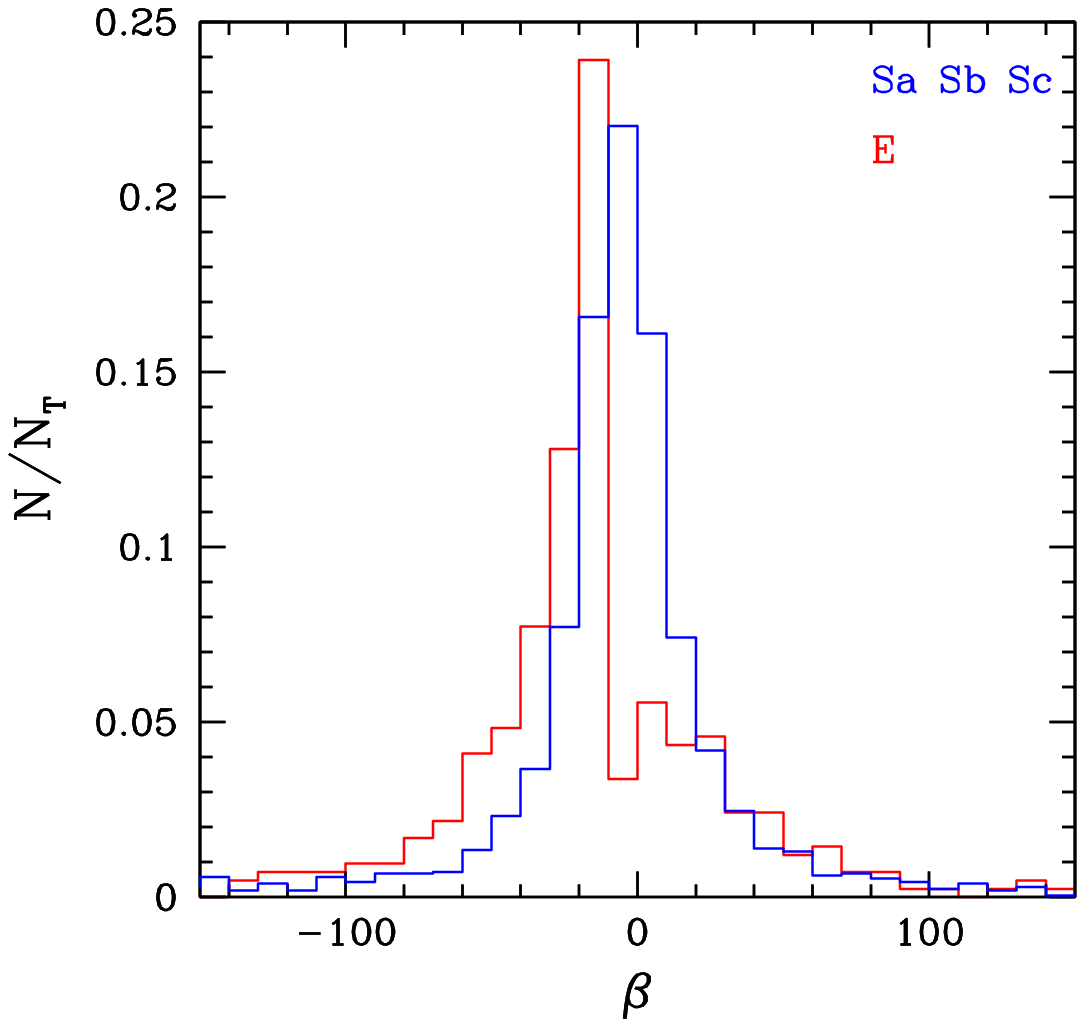} }
   \caption{ Histogram of the percentage  distribution of $\beta$'s of the the MaNGA galaxies. The red line shows the ETGs (only E galaxies, $N_T=414$), the blue line the LTGs (Sa+Sb+Sc, $N_T=2075$) .}
              \label{Fig:17}
    \end{figure}

To understand the causes of the above disagreement a few remarks may be helpful. The correct comparison should take into account the following facts: (i) the luminosity of galaxies of our models does not consider the variations  of the $M/L$ ratios present among   real galaxies. The luminosity of a galaxy primarily depends on its mass, but it also depends on other factors, such as  star forming episodes, star formation quenching, natural fading, and so forth). Taking all of them in account correctly is a cumbersome affair; (ii) the errors on the observed structural parameters that can be as high as 20\%; (iii) the radii evaluated by our simulations that do not strictly correspond to the real effective radii measured in the data.
In order to somehow include such effects in our simulations of mergers, we decided to apply a random variation in the \re\ and \Ie\ parameters of the model galaxies. The maximum  variation of \re\ is  $\simeq 0.4$ (in log units), and the maximum variation of \Ie\ is $\simeq 1$ (in log units). These variations on $R_e$ and $I_e$ soon propagate to $\beta$ and $L'_0$. The results are  shown in the right panel of Fig. \ref{Fig:11} limited to $\beta$. Now the theoretical values of $\beta$  spread out over the whole $I_e-\beta$ plane in that  closely mimicking the observational data. Therefore at least part of the scatter in the $\beta$-$I_e$ can be attributed to uncertainties in the determination of $I_e$ and $R_e$.

Another point to consider is related to the definition of $L'_0$ and $\beta$.
Relations (\ref{eq2}) show that $\beta$ depends on $logL'_0$ which in turn contains the quantity $1- 2A'/A$ at the denominator, where $A'$ is $2log\sigma$ while the $A$, expressed by means of $I_e$, $G/k_v$, $M_s$, $L$ and $R_e$, can also be reduced to $2log\sigma$. Therefore,  $\beta$ should diverge when the denominator  tends to zero.  So high values of $\beta$ and $L'0$ should be the signature that the two derivation of $\sigma$ are in mutual agreement and implicitly that the system is mechanical equilibrium.   Unfortunately, $I_e$,  $M_s$, and $R_e$, are all derived from the luminosity and its profile in the central region so that their are likely affected by some uncertainty. 
The quantity $k_v$ is an indirect evaluation of the morphology of the galaxy, which is unknown in most cases,  at least in the samples under examination, so that $k_v$ is also highly uncertain. 
Catching the right combination of all these parameters is highly unlikely.  To cast light on this issue we looked at the relative number of galaxies as a function of $ \beta$. In our observational sample (only 
MaNGA to avoid contamination by other sources,) we count 414 ETGs (only E galaxies, S0 are left aside) and 2075 LTGs (Sa+Sb+Sc lumped together). These are the same galaxies already shown in Fig.\ref{Fig:11}.  Their number distribution as a function of  $\beta$ is shown in Fig. \ref{Fig:17}. The histograms displays the two populations with different colors (red for ETGs and blue for LTGs) in steps of 10 for $\beta$. The ETGs population is centered at small values of beta ($-20 \leq \beta -10$), while LTGs  peak in between $-20 \leq \beta \leq 10$). In brief  ETGs are more shifted to the negative domain while LTGs are more shifted towards the positive domain. Finally both ETGs and LTGs have tails towards high values of $|\beta|$ on both semi-planes. Assuming as tail all galaxies falling below $\beta=-40$ and above $\beta=40$,  the estimated total  fractions in the tails and peaks amount to 0.15 (left), 0.45 (peak), 0.27 (right)  for ETGs and 0.09 (left), 0.43 (peak) and 0.09 (right) for LTGs. Apparently the great majority of galaxies have parameters that cannot be used to infer their dynamical conditions. However, the typical dynamical timescale for a galaxy is about 0.5 Gyr. This is also the timescale required to  lose memory of any dynamical perturbation like a merger. On the other hand, if the merger induces star formation accompanied by a variation in total luminosity, about 3 Gyr is the timescale required to lose memory of the luminosity perturbation (see the data in Fig. \ref{masslum}). It follows that while the memory of the dynamical event is quickly lost, that of the luminosity variations lasts longer. Therefore, all quantities that directly or indirectly derive from the luminosity keep memory of the event for longer time and therefore the dynamical conditions of the system cannot be safely ascertained by this kind of arguments.  

To conclude, 
mergers alone cannot explain the whole story encrypted in the ScRs. Luminosity effects should also be taken into consideration in a much better  way than what has been done so far  in order to explain the scatter present  in the $\beta$-$I_e$ plane (observational data and theoretical models). Work is in progress along this line.

\section{Conclusions}\label{sec:10_concl}

Three are the aims of this study: i) To analyze in detail the observational sample of galaxies MaNGA and to compare it with two theoretical databases of galaxy models in the $\Lambda$CDM hierarchical scenario of structure formation EAGLE and Illustris-TNG100; ii) To propose and validate a simple analytical model of galaxy formation in the hierarchical scheme aimed to be a proxy of real galaxies and at the same time of the galaxy models obtained by the large scale cosmo-hydro-dynamical simulations; iii) To analyze the various projection  planes of the galaxies FP under the light of the theoretical results to our disposal (large scale simulations and the simple analytical model)  in a unitary conceptual framework. In the following we briefly comment on the three topics.

(i) There is a satisfactory agreement between the MaNGA data and the results of the large scale theoretical simulation EAGLE and Illustris-TNG100. There are of course some differences, but they do not invalidate the general agreement. More precisely, the radii $R_e$ predicted by EAGLE are a bit greater than those of MaNGA while those of Illustris-TNG100 are a bit smaller.
There is the important result that the radii $R_e$ increase at decreasing redshift, in other words galaxy size increases as they become older. The same for the galaxy mass that is found to increase as the redshift decreases. The relation $R_e - M_s$ varies with time. At  high redshift the relationship seems to disappear. Objects with mass in excess of  $10^{10} M_\odot$) are rare and the bulk of galaxies have lower masses and their radii  span a large range of values with mean value either nearly constant or weakly increasing with the mass. It is only  at redshift $z\leq 2 - 1.5$, when massive galaxies are numerous that the standard \MRa\ relations gets in place with positive slope and small scatter around it. 
Finally,  the rate of star formation either decreases or is nearly zero as the  redshift tends to zero. This is particularly true for ETGs, while remains active at moderate levels of activity  in the late type objects. Dwarf galaxies belong to  a separate group that is  not included in our analysis.

(ii) 
In order to better clarify the role played by mergers in shaping the above trends and results without being forced to intensive numerical computing with the standard N-Body hydro-dynamical codes, we proposed a simple model of mergers. To avoid misunderstanding, it is better to clarify that our model mergers are not intended to compete with those from detailed large-scale numerical simulations. They must be considered as a simple-minded tool to get quick understanding of the physical causes why galaxies populate the diagnostic planes, projections of the FP, in the way they do. The model stands on  the VT, a physical condition in which galaxies are during most of their life, makes use of parameters to quantify the mass of the captured object, the energy injection, and the energy sharing,  the effect on the radius, velocity dispersion, luminosity, and specific intensity. The formalism in use is simple and allows to get results at no cost in terms of computing time and resources. A large number of simulations is made at varying the parameters, with the aid of these, the usual diagnostic planes are derived.  The results are found to be fully compatible with both observational data and extensive N-Body hydro-dynamical simulations.  We limit ourselves to highlight here that the model provides a simple explanation for the larger dispersion shown by the data in the diagnostic planes (for instance the $R_e-M_s$ and \Lsig\ planes), they also cast light on some unexpected sub-trends noticed in the $L-M_s$, $L-\sigma$ and $M_s-\sigma$ planes, they can explain the $\Lambda$-like distribution in the $I_e-R_e$ plane. Above all, they better clarify the role of mergers in shaping all the relation found in the diagnostic planes. At present the model cannot deal with or predict the behavior of star formation. Work is in progress to cope with it. {Despite its simplicity, the model of mergers we are proposing  can complement or
even guide more complex simulations}.

To conclude, we can say that the method and results presented in this study can be taken as the starting point of more sophisticated analyses, in which the histories of star formation and luminosity in turn are explicitly taken into account. Thanks to its simplicity, our approach offers some advantages   with respect to  numerical simulations. However, it cannot  yield the precise predictions  of these latter, but yet it can predict the effects of different physical phenomena on the observed ScRs. 

\begin{acknowledgements}
      M.D. and  F.T. thank the Department of Physics and Astronomy of the Padua University for the financial support. C.C. thanks the Padua University for the hospitality. 
\end{acknowledgements}

   \bibliographystyle{aa} 
   \bibliography{ScRs.bib} 

\end{document}